\documentclass[reprint,amsfonts, amssymb, amsmath,  showkeys,prx, superscriptaddress, twocolumn,longbibliography,nofootinbib,aps]{revtex4-2}

\usepackage{graphicx} 
\usepackage{float}
\usepackage[caption=false]{subfig}
\usepackage{amsmath}
\usepackage[left=0.7in, right=0.7in, top=0.9in, bottom=0.9in]{geometry}
\usepackage{times} 
\usepackage[colorlinks=true, linkcolor=blue, citecolor=blue, urlcolor=blue, linktoc=all]{hyperref}
\usepackage{orcidlink}

\usepackage{mathtools}

\usepackage{calligra}
\usepackage{tikz}
\newcommand*\circled[1]{\tikz[baseline=(char.base)]{
            \node[shape=circle,draw,inner sep=1pt] (char) {#1};}}
\newcommand{\fSim}{\mathrm{fSim}}
\newcommand{\iSWAP}{\mathrm{iSWAP}}
\newcommand{\CPHASE}{\mathrm{CPHASE}}

\usepackage{enumitem}

\begin{document}
 
\title{Tensor Networks with Belief Propagation Cannot Feasibly Simulate Google's \\ Quantum Echoes Experiment}

\author{Pablo~Bermejo\orcidlink{0000-0002-7467-4951}}
\email{pablo.bermejo@dipc.org}
\affiliation{Google Quantum AI, Santa Barbara, CA 93111, USA}
\affiliation{Donostia International Physics Center, Paseo Manuel de Lardizabal 4, E-20018 San Sebasti\'an, Spain}
\affiliation{Department of Applied Physics, Gipuzkoa School of Engineering, University of the Basque
Country (UPV/EHU), Plaza Europa 1, 20018 San Sebastián, Spain}

\author{Benjamin~Villalonga\orcidlink{0000-0002-3299-7226}}
\affiliation{Google Quantum AI, Santa Barbara, CA 93111, USA}

\author{Brayden~Ware\orcidlink{0000-0002-3321-3198}}
\affiliation{Google Quantum AI, Santa Barbara, CA 93111, USA}

\author{Guifre~Vidal\orcidlink{0000-0003-4513-9252}}
\affiliation{Google Quantum AI, Santa Barbara, CA 93111, USA}

\author{Aaron~Szasz\orcidlink{0000-0002-1127-2111}}
\email{aszasz@google.com}
\affiliation{Google Quantum AI, Santa Barbara, CA 93111, USA}

\begin{abstract}
In the recent quantum echoes experiment~\cite{google2025observation}, Google Quantum AI showed that out-of-time-order correlators (OTOCs) for random-circuit time evolution can be measured using a quantum processor more than 10,000x faster than they can be computed to similar accuracy via classical computation. 
This claim was substantiated by comparison with a variety of state-of-the-art classical simulation methods.  One classical simulation method that was not explicitly tested was tensor networks with belief propagation (TNBP).  TNBP should be poorly suited to simulating Google's echoes experiment: the states involved are highly entangled, a challenge for tensor network states; and the Willow chip has dense 2D connectivity, a challenge for belief propagation.  Here we confirm, via a combination of theoretical scaling arguments and explicit numerical simulation, the intuition that TNBP is unable to simulate the quantum echoes experiment.  We show that the OTOC circuits generate enough entanglement that they are largely incompressible, implying that other approaches in which OTOCs are computed by evolving a tensor network state in the Schr\"{o}dinger picture will also fail.   Our results further reinforce the claim that the quantum echoes experiment cannot be reproduced by classical computation. 
\end{abstract}

\maketitle

\section{Introduction}
\label{sec:introduction}

The promise of quantum computing is that it will solve important problems that are beyond the reach of any realistic classical computer.  Starting in 2019, numerous experiments on quantum chips have been claimed to show beyond-classical performance, meaning that reproducing the results with a classical computer would take many orders of magnitude more time~\cite{Arute2019,USTC2020,USTC2021a,USTC2021b,USTC2021c,Xanadu2022,morvan2024phase,USTC2023,IBM2023,DWave2024,GoogleAnalog2025,QuantinuumRuszlan2025,QuantinuumIsing2025,USTC2025,google2025observation}.  Improvements in classical algorithms have since led some of these claims to be refuted~\cite{tindall2024efficient, beguvsic2024fast, larose2024history, zhao2025leapfrogging}, but many still stand.  However, of the beyond-classical claims that remain unchallenged, most are for contrived problems with no practical application and with outcomes being statistical properties of ensembles of samples rather than a single reproducible value.  These experiments still leave open the question: can present-day quantum chips perform an experiment (1) with a reproducible result (2) that cannot be found by classical simulation and (3) that has an application to an important real-world problem?

One candidate for satisfying all three properties is the set of experiments on out-of-time-order correlators (OTOCs) performed last year on Google Quantum AI's Willow processor~\cite{google2025observation, OTOC_NMR}.  In the first experiment, known as the quantum echoes experiment and presented in~\cite{google2025observation}, a ``second-order OTOC'' was measured on 65 qubits and found to be more than 13,000x slower to classically compute to comparable accuracy using a contemporary supercomputer.  In the second paper~\cite{OTOC_NMR}, OTOCs on small systems were shown to be useful for interpreting nuclear magnetic resonance spectra to determine the geometrical structure of molecules.  Taken together, these results suggest a route to near-future beyond-classical applications of quantum computing to a real-world application.

In this paper, we provide further evidence that the quantum echoes experiment in~\cite{google2025observation} achieves goals (1) and (2).  Let us be more precise about what these goals entail, \textit{i.e.} what we mean when we say that a reproducible experimental result cannot be found classically.  Specifically consider the OTOC for concreteness.  As we explain in depth later in the paper, the true value is given by the expectation value of a local observable in the output state of a quantum circuit; this is a well-defined numerical value, even when it cannot be computed in practice.  The corresponding experiment, in which the circuit is implemented on a quantum computer and the local observable is measured on the device, is subject to noise and therefore returns an approximation to the OTOC.  We can define the experimental error as the difference between the experimental result and the true value of the OTOC.  Likewise, a classical simulation may not return the exact value of the OTOC, and the difference defines the classical simulation error.  If a classical computation achieves a lower error than does the experiment, we say that the classical method can simulate the OTOC experiment.  Conversely, if no classical computation can achieve as low an error as the experiment, or if to do so would take orders of magnitude more time, we say that the experiment is beyond classical.

The beyond-classical claim in~\cite{google2025observation} was substantiated by comparing runtimes for the experiment with those of numerous state-of-the-art classical simulation methods (see Ref.~\cite{google2025observation} for a detailed explanation). These included variants of exact tensor network contraction that aimed at exploiting the structure of the experiment, as well as a new family of Monte-Carlo methods.  From the latter family, Tensor Network Monte Carlo (TNMC) was able to efficiently reproduce the standard OTOC at moderate sizes (up to $N=40$ qubits) and is expected to reproduce large instances (up to $N=95$ qubits) with realistic but substantial computational resources. However, no method could reproduce the second-order OTOC, or OTOC$^{(2)}$, on 65 qubits with feasible computational cost. 

One key classical approach that was not explicitly tested is tensor networks (TNs) with belief propagation (BP)~\cite{yedidia2003understanding,alkabetz2021tensor}.  This method approximately computes the OTOC in two steps. In the first step, one evolves a wavefunction in the form of a projected-entangled pair state (PEPS). In order to keep computational cost under control, the size of the PEPS tensors after each step of the time evolution is reduced by making local truncations using a gauge specified by a BP fixed point.  In the second step, one extracts the OTOC from the final wavefunction using boundary matrix product state (bMPS) contraction; this step also requires approximation in order to be computationally feasible on large systems.  We refer to the full simulation method, including both the evolution step and the extraction step, as TNBP. 

TNBP has been used to refute past claims of beyond-classical performance in large-scale experiments on quantum devices~\cite{tindall2024efficient,tindall2025dynamics}, making it natural to ask whether the method also presents a challenge to the quantum echoes experiment.  Indeed, there are good reasons to believe that TNBP could outperform other methods based on tensor network evolution: compared with methods based on one-dimensional matrix product states, PEPS-based methods can more naturally capture the entanglement structure of two-dimensional systems; and compared with other PEPS-based methods, the truncations in TNBP are cheap, allowing for larger tensors that can capture more entanglement.  The approximations of TNBP are most accurate in locally tree-like lattices~\cite{sahu2022efficienttensornetworksimulation,tindall2023gauging,rudolph2025simulating}, leading to the successful simulations~\cite{tindall2024efficient} of the experiment in~\cite{IBM2023} on the heavy-hex lattice. 
However, TNBP has also successfully simulated~\cite{tindall2025dynamics} a large-scale experiment on a square lattice geometry~\cite{DWave2024}, showing that truncating a PEPS with the BP approximation can still be effective in lattices with small loops. Thus it is not \textit{a priori} clear how effective TNBP will be in simulating the quantum echoes experiment.

In this paper, we provide strong evidence that the TNBP method in fact cannot feasibly simulate the quantum echoes experiment.  Under the assumption that random circuit evolution is incompressible, as defined below, we show based on the geometry of the OTOC circuits that TNBP simulation must have cost scaling exponentially in system size.  We then confirm the hypothesis of incompressibility via explicit numerical simulation both in one and two spatial dimensions.  In the 1D simulations, we observe precisely the exponential cost scaling predicted by the geometric argument.   In the 2D simulations, we show that TNBP run with moderate computational resources already fails to accurately reproduce the OTOC for two systems with only 23 qubits that were studied in the experiment in~\cite{google2025observation}.  We further confirm incompressibility in 2D by running TNBP on new OTOC circuits we construct for a range of system sizes up to 28 qubits.  We also numerically confirm that TNBP \emph{could} efficiently compute the OTOC if the circuits were compressible; in particular, for OTOC circuits up to 28 qubits, we replace the highly-entangling two-qubit gates with weakly entangling gates and show that TNBP can accurately compute the OTOC at low cost.  

From the observed incompressibility, we conclude that computing the OTOC with TNBP, or indeed with any other method based on approximate tensor network evolution in the Schr\"{o}dinger picture, will be more expensive than performing an optimized exact tensor network contraction.  Thus TNBP cannot simulate the quantum echoes experiment any more effectively than the methods already considered in~\cite{google2025observation}, providing further confirmation that the experiment is beyond classical.

This paper is organized as follows. In Sec.~\ref{sec:background},
we formally introduce out-of-time-order correlators, and we review how the OTOC is implemented as a quantum circuit for the quantum echoes experiment.  We also define the figure of merit used to evaluate the quality of simulations, and we give an overview of tensor networks with belief propagation. Sec.~\ref{sec:results} contains the results.  We make a theoretical argument for exponential scaling of TNBP simulation cost with linear system size, and we back up this argument with extensive numerical evidence in small 1D and 2D systems.  Based on these results, we argue that TNBP simulation of the quantum echoes experiment is more expensive than the exact tensor network contraction already considered in~\cite{google2025observation}.  We conclude with a summary and future perspectives in Sec.~\ref{sec:conclusions}.


\section{Background and setup}
\label{sec:background}

\subsection{Reproducible quantum advantage with out-of-time-order correlators (OTOCs)}
\label{sec:otoc_background}

Out-of-time-order correlators (OTOCs) are correlation functions employed in many-body quantum physics to diagnose how quickly quantum information spreads in a quantum system ~\cite{hashimoto2017out, haehl2019classification, garcia2022out}. They serve as fingerprints of quantum chaos and information scrambling, providing a powerful framework for understanding how local perturbations evolve under complex quantum dynamics, spread across degrees of freedom, and ultimately become inaccessible to local probes. 
As shown in~\cite{google2025observation}, ``higher-order'' OTOCs are also promising for realizing a reproducible quantum advantage.  It is this aspect of OTOCs that we focus on in this paper.  

In this section, we give the reader an overview of the definition of the OTOC in \ref{subsubsec:OTOC_def}, explain how we evaluate the quality of an experiment or simulation of an ensemble of OTOC circuits in \ref{subsec:snr_def}, provide details of our OTOC circuit constructions in \ref{subsec:circuit_design}, and discuss why OTOCs are promising for reproducible quantum advantage in \ref{subsec:reproducible_advantage}.

\subsubsection{Definitions: first- and second-order OTOCs}
\label{subsubsec:OTOC_def}

The standard OTOC is defined as 
\begin{equation}
    C^{(2)} \equiv \langle B(t) M B(t) M\rangle
    \label{eq:OTOC_def_original}
\end{equation}
for two local operators $M$ and $B$, with $B(t)=U^\dagger(t) B U(t)$ where $U$ is the unitary operator generated by a quantum circuit.  The expectation value is traditionally taken in the maximally mixed state, in which case $C^{(2)}$ is equal to the trace of $B(t) M B(t) M$, divided by $2^N$ where $N$ is the total number of qubits in the system.  However, as was done in~\cite{google2025observation}, we use an alternative definition of the OTOC where we take the expectation value in the product state $|0\rangle^{\otimes N}$.  Given a large-dimensional Hilbert space and highly scrambling time-evolution $U$, the resulting value of the OTOC will be approximately the same. 

\begin{figure}
\includegraphics[width=1\linewidth]{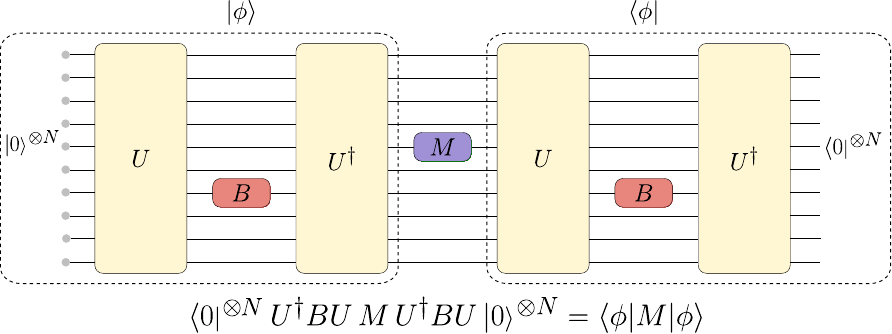}
\caption{\textbf{OTOC as a local measurement.} The OTOC, despite involving repeated forwards and backwards time evolution, can still be viewed as the expectation value of a local operator in a time-evolved state.  We define $|\phi\rangle$ as $U^{\dagger}BU|0\rangle^{\otimes N}$, then just measure $M$ in $|\phi\rangle$.  Note that the circuit diagram follows the usual convention, with the input state (ket) at the left, and with successive gates applied from left to right. }
\label{fig:OTOC_scheme}
\end{figure}

We refer to $M$ as the ``measurement operator'' and $B$ as the ``butterfly operator.''  To understand the name of $M$, as in~\cite{google2025observation} we take $M$ to be the Pauli $Z$ operator on some site; we will continue to assume $M=Z$ throughout the paper.  Then $M$ acts trivially on $|0\rangle^{\otimes N}$, $M|0\rangle^{\otimes N}=|0\rangle^{\otimes N}$, so that (if $B$ is Hermitian)
\begin{equation}
\begin{aligned}
C^{(2)} 
&= \langle 0 |^{\otimes N} \, B(t)\, M \, B(t)\, M| 0 \rangle^{\otimes N} \\
&= \langle 0 |^{\otimes N} \, B(t)\, M \, B(t)\, | 0 \rangle^{\otimes N} \\
&= \langle \phi | M | \phi \rangle
\end{aligned}\label{eq:OTOC definition}
\end{equation}
for the state $|\phi \rangle \equiv  U^\dagger B U |0\rangle^{\otimes N}$.  Thus to evaluate the OTOC, we can simply create the state $|\phi\rangle$, then measure the expectation value of $M$ in the state.  
In Fig.~\ref{fig:OTOC_scheme}, we visually show how the OTOC can be broken down into $\langle\phi|$, $M$, and $|\phi\rangle$.

\begin{figure}
  \centering
  \includegraphics[width=1\linewidth]{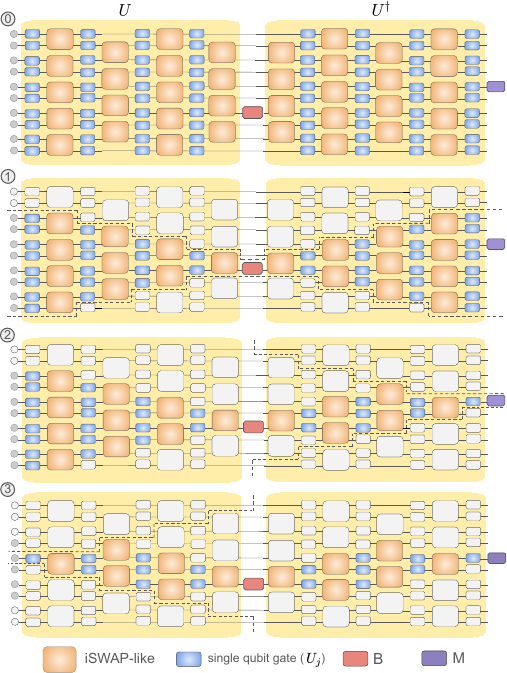}
\caption{\textbf{Construction of the time evolution circuit for the OTOC state $|\phi\rangle$.} Although the quantum echoes experiment was run with two-dimensional circuits, for simplicity we illustrate the circuit construction in 1D.  Our goal is to find the simplest circuit giving a state $|\phi\rangle$ on which a measurement of $M$ gives the OTOC.  We begin with $|\phi\rangle = U^\dagger B U |0\rangle^{\otimes N}$, schematically illustrated in Fig.~\ref{fig:OTOC_scheme}, and then remove any gates that do not affect the final observable in order to achieve the highest possible accuracy.  (0) We first construct the full time evolution operator $U$ on all qubits, consisting of alternating layers of two-qubit entangling gates ($\iSWAP$-like, tan) and single-qubit rotations (blue), then make the full circuit for $|\phi\rangle$ by acting first with the circuit for $U$, then with the butterfly operator $B$ (red), then with the circuit for $U^\dagger$.  In the figure we also show the location of the measurement of $M$ (purple). (1) We then remove gates from $U$ and $U^\dagger$ that are outside the lightcone of $B$, since these exactly cancel with each other. Grayed-out gates are the ones that have been removed from the circuit, and the dashed line shows the edge of the lightcone.  (2) We remove gates from $U^\dagger$ that are outside the lightcone of the measurement; these exactly cancel with gates in $\langle \phi|$.  (3) Finally, we remove gates outside the lightcone of the location of $M$ in the initial state.  Considering the original definition of the OTOC in Eq.~\eqref{eq:OTOC_def_original} where the expectation value is taken over the maximally mixed state and there is a second copy of $M$, these gates would cancel due to the cyclic property of the trace.  Here they do not actually cancel, but we remove them following the convention in~\cite{google2025observation}.  In (1)-(3), the dark gray qubits on the left are the only ones on which gates act after taking the lightcones into account.  We remove all other qubits from the simulation; the reported system size $N$ for each circuit in the paper corresponds only to the non-removed qubits.
}
    \label{fig:OTOC circuit}
\end{figure}

In this picture, we can also understand why $B$ is called the butterfly operator.  Imagine that we replaced $B$ by the identity operator.  Then the forwards and backwards evolution by $U$ and $U^\dagger$ will cancel, so $B(t)$ becomes the identity, and we end up with just $\langle M \rangle$ in the initial state $|0\rangle^{\otimes N}$.  Recalling that $M=Z$, the value of the OTOC is then exactly 1.  On the other hand, with $B$ present, there is a ``geometric lightcone'' of gates in $U$ and $U^\dagger$, spreading out from $B$, that no longer cancel; we use the word ``geometric'' because the lightcone depends only on where gates are located in the circuit, not on the contents of the gates.  The lightcone is illustrated in Fig.~\ref{fig:OTOC circuit}, panel \circled{1}.  The grayed-out gates outside the lightcone still cancel between $U$ and $U^\dagger$.  
If the measurement $M$ is placed outside the lightcone, for example on the uppermost qubit in Fig.~\ref{fig:OTOC circuit}, none of the remaining gates act on the measured qubit and the OTOC still gives exactly 1.  
But when $M$ is measured within the lightcone of $B$, we see a ``butterfly effect''---the wavefunction is scrambled inside the lightcone by the time evolution $U^\dagger$, leading to a value of the OTOC with sizable fluctuations that sensitively depend on the details of the evolution.  

We can also define higher-order OTOCs.  For example, in~\cite{google2025observation}, experiments were also performed to measure the second-order OTOC, or $\text{OTOC}^{(2)}$,
\begin{equation}
\begin{aligned}
    C^{(4)} &\equiv \langle 0|^{\otimes N} B(t) M B(t) M B(t) M B(t) M|0\rangle^{\otimes N}\\
    &\equiv \langle 0|^{\otimes N} B(t) M B(t) M B(t) M B(t)|0\rangle^{\otimes N}\\
    &\equiv \langle \varphi|M|\varphi\rangle
\end{aligned}
\label{eq:OTOC2}
\end{equation}
for $|\varphi\rangle \equiv B(t) M B(t)|0\rangle^{\otimes N}$.

\subsubsection{Evaluating the quality of simulation results or experimental measurements on an ensemble of OTOCs}
\label{subsec:snr_def}

Once the measurement operator $M$, butterfly operator $B$, and time evolution $U$ are specified, the value for the OTOC, $\mathcal{C}^{(2)}_\text{exact}$, is given by Eq.~\eqref{eq:OTOC definition}.  In this paper, we consider two ways of obtaining an approximation to $\mathcal{C}^{(2)}_\text{exact}$.  First, the OTOC can be approximated by running an experiment on a quantum computer, in which we evolve by $U$, $B$, and $U^\dagger$ to get the state $|\phi\rangle$, then measure $M$; the average over many shots is the experimental estimate of the OTOC, $\mathcal{C}^{(2)}_\text{exp}$.  Second, the OTOC can be estimated by numerical simulation; we call the result $\mathcal{C}^{(2)}_\text{sim}$.  
The quality of the approximation to the OTOC derived from experiment or simulation can be evaluated with respect to the exact value $\mathcal{C}^{(2)}_\text{exact}$.  

A natural metric is the absolute error in the OTOC, or $|\mathcal{C}^{(2)}_\text{sim} - \mathcal{C}^{(2)}_\text{exact}|$ for approximate simulation and likewise for experiment.  
However, if our goal is not to evaluate a particular OTOC, but rather to assess the ability of a quantum processor to accurately evaluate OTOCs more generally, achieving lower error in experiment than in simulation for one specific OTOC instance is not sufficient.  For example, we may accidentally pick a random time evolution $U$ that happens to be particularly hard or particularly easy to simulate classically. 

We therefore want to evaluate the performance of a simulation method, or of experiment, on an \emph{ensemble} of OTOCs. We define an OTOC ensemble to be a fixed butterfly operator $B$ and measurement operator $M$ (both the choice of operator and the location are fixed), together with a distribution over possible time evolutions $U$.  For example, we could consider time evolutions $U$ drawn from the Haar distribution.  The specific distribution of $U$ used in~\cite{google2025observation} and in this paper is defined in Sec.~\ref{subsec:circuit_design} below.  

We would like to diagnose whether a quantum experiment can reproduce the OTOCs in an ensemble more accurately than an approximate classical simulation can reproduce them.  Standard ensemble metrics, such as mean and standard deviation across instances, cannot be used here because they are easy to simulate classically~\cite{google2025observation, mi2021information, khemani2018operator, nahum2018operator} and hence cannot reveal the power of the quantum experiment.  Instead, we measure how accurately the experiment or simulation captures instance-to-instance fluctuations in the OTOC value, \textit{i.e.} the butterfly effect.

For this purpose we use the signal-to-noise ratio (SNR) defined by 
\begin{equation}
    \text{SNR} = 
    \frac{1}{\sqrt{\,\,\rule{0pt}{3.6ex} \overline{\left( \mathcal{C}^{(2,s)}_{\text{exact}} - \mathcal{C}^{(2,s)}_{\mathrm{approx}} \right)^2}}} 
    \label{eq:snr}
\end{equation}
where $\mathcal{C}^{(2,s)} \equiv \frac{\mathcal{C}^{(2)} - \overline{\mathcal{C}^{(2)}}}{\sigma(\mathcal{C}^{(2)})}$ measures how far the OTOC value for a given instance, $\mathcal{C}^{(2)}$, is from the mean over the full ensemble, $\overline{\mathcal{C}^{(2)}}$, normalized by the standard deviation $\sigma$ that gives the typical size of instance-to-instance fluctuations.  $\mathcal{C}^{(2,s)}$ can be thought of as a normalized version of the OTOC $\mathcal{C}^{(2)}$, whose mean and standard deviation (size of typical fluctuations) over the OTOC ensemble are 0 and 1, respectively.  This SNR can be defined for any approximate method of evaluating the OTOC, including both simulation and experiment; $\mathcal{C}^{(2,s)}_{\mathrm{approx}}$ is the normalized OTOC for the approximate method, while $\mathcal{C}^{(2,s)}_{\mathrm{exact}}$ is the normalized OTOC for the theoretical exact evaluation.  

In practice, the average over the ensemble is defined using a finite number of instances, $m$, typically on the order of 50 to 100.  In the limit of no error,  $\mathcal{C}^{(2,s)}_{\mathrm{approx}}$ and $\mathcal{C}^{(2,s)}_{\mathrm{exact}}$ are equal, so SNR goes to infinity.  In the opposite limit, where the simulation or experiment returns random results completely uncorrelated with the exact OTOC values, the SNR estimate using $m$ instances follows the inverse $\chi_{m-1}$ distribution, $\mathrm{SNR}^{\text{uncorrelated}} = (\sqrt{m}\,\Gamma(m-1))/(2\,\Gamma(m - 1/2))$; as $m\rightarrow\infty$, SNR in the case of no correlation approaches $1/\sqrt{2}$~\cite{google2025observation}. 
In general, the higher the SNR that an experiment or simulation achieves on the OTOC ensemble, the better it is able to reproduce the true value of the OTOC across instances for that ensemble.  

This, finally, gives us a metric for comparing the quality of experiment and of classical simulation.  
We say that an experiment on an OTOC ensemble is beyond classical if it achieves a higher SNR than is possible by any feasible classical simulation.  Conversely, we say that a classical method successfully simulates an experiment if it achieves an SNR equal to or higher than the SNR achieved by the experiment.

Note that we can define an analogous SNR for the second-order OTOC as well, just replacing $\mathcal{C}^{(2)}$ by $\mathcal{C}^{(4)}$.

\subsubsection{OTOC ensemble and quantum circuit construction}\label{subsec:circuit_design}

We construct OTOC ensembles, and from them quantum circuits that we simulate with TNBP, following a similar procedure to \cite{google2025observation}.   We first design a distribution of time-evolution circuits $U$ that are (typically) hard to simulate classically.\footnote{Note that classical hardness of $U$ does not automatically imply hardness of the OTOC because, for instance, in the OTOC, gates in $U$ will cancel with gates in $U^\dagger$.}  We then construct each individual ensemble by picking the locations of the measurement and butterfly operators, $M$ and $B$, to achieve a large quantum signal (size of fluctuations in the OTOC) while maintaining classical hardness.  Finally, we simplify the resulting circuits, removing gates that do not contribute to the value of the OTOC; the last step improves the SNR for both simulation and experiment.

While we cannot in practice draw unitaries $U$ from the $N$-qubit Haar distribution, as this would require exponential-depth circuits, we still draw $U$ from a set of highly-entangling random evolutions.  Following~\cite{google2025observation}, entanglement is generated by uniform two-qubit $\fSim$ gates which can be decomposed into an $\iSWAP$ gate [$\exp(i(\pi/4)(XX+YY))$] followed by a $\CPHASE$ gate with a conditonal phase of $\approx$ 0.35  rad.  Circuit-to-circuit randomness is provided by single-qubit gates of the form $\exp\left(i\theta\left(\cos(\phi)\,X + \sin(\phi)\,Y\right)\right)$, where $\theta/\pi \in \{0.25, 0.5, 0.75\}$ and $\phi/\pi$ is uniformly sampled at random from the interval $[-1,1]$.  The full time evolution $U$ is constructed by alternating layers of entangling two-qubit gates and random single-qubit gates in a brickwall structure, as shown in Fig.~\ref{fig:OTOC circuit}, panel \circled{0}.  Moreover, following~\cite{google2025observation} we take $M$ to be a single-qubit Pauli $Z$ and $B$ a single-qubit Pauli $X$.\footnote{Some circuits from~\cite{google2025observation} use a tensor product of $X$ on a few sites for $B$.  However, we do not perform simulations of those ensembles in this paper.}

With this choice for the distribution of time evolutions $U$, measurement operator $M$, and buttefly operator $B$, each OTOC ensemble is determined by: the depth of the brickwall time evolution circuit; the location of $M$; and the location of $B$.  The choice of the locations for $B$ and $M$ relative to each other in space is of critical importance.  This choice determines both the size of the OTOC signal and the classical hardness of simulating it.  

\begin{figure}
    \centering
    \includegraphics[width=\linewidth]{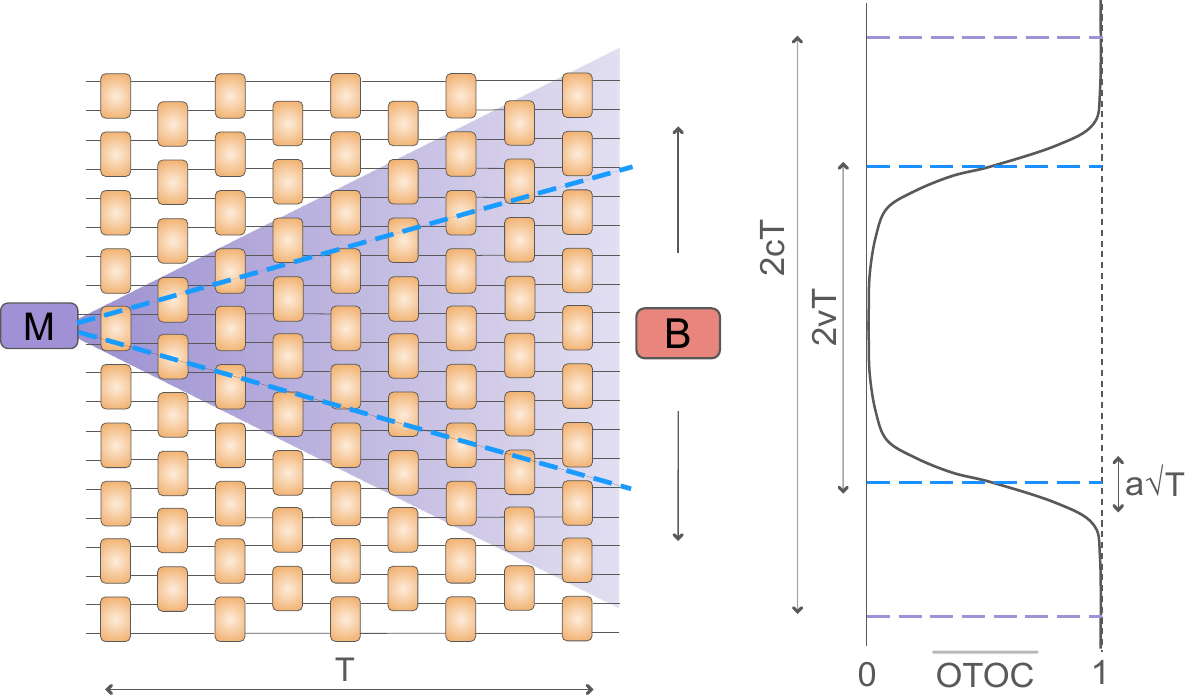}
    \caption{\textbf{Dependence of OTOC value on relative positions of measurement and butterfly operators.} The measurement operator $M$ has a geometric lightcone, the region shaded purple, given by the brickwall structure of the circuit.  The edge of the lightcone travels at speed $c$, so after time $T$ the extent of the lightcone in space is $2cT$.  The operator also has a physical lightcone, bounded by the dashed blue lines, determined by the speed at which information actually spreads through the system, $v$.  The average value of the OTOC over an ensemble depends on the location of the butterfly operator $B$ relative to the lightcone edges (indicated by the vertical arrows above and below $B$ in the figure).  As shown on the right, the ensemble average is exactly 1 outside the geometric lightcone, approximately 1 far outside the physical lightcone, and approximately 0 deep inside the physical lightcone.  At the edge of the physical lightcone, the value is close to 1/2 and has large instance-to-instance fluctuations.  In 1D, these fluctuations give the physical lightcone a diffusive front; the value of the OTOC goes from 1 to 0 over a distance that scales as the square root of the circuit depth, $a\sqrt{T}$ for some constant $a$.  In 2D, the width of the physical lightcone front scales as $T^{1/3}$ rather than $\sqrt{T}$.~\cite{nahum2018operator}.  See App.~\ref{app:2D_lightcone_geometry} for further information on the 2D lightcone geometry.      
    Notice that the OTOC is defined such that it vanishes inside the physical lightcone and is 1 outside; this is in contrast to disconnected two-point correlators with $M$, which vanish outside the lightcone and obtain their largest values inside.}
    \label{fig:OTOC_vs_B_loc}
\end{figure}

At one limit, if $M$ is outside the geometric lightcone of $B$ (see Fig.~\ref{fig:OTOC circuit}, panel \circled{1}), the OTOC is exactly 1. Furthermore, for any choice of specific entangling gates, actual quantum correlations spread more slowly than in principle allowed by the circuit structure of $U$, giving rise to a narrower ``physical lightcone,'' as shown in Fig.~\ref{fig:OTOC_vs_B_loc}.  Outside of the smaller physical lightcone, the OTOC remains approximately 1.  In the other limit, if $M$ is on the same qubit as $B$, the OTOC is close to 0 for all random instances in the ensemble.  The dependence of the ensemble-averaged OTOC on the position of $B$ is shown for a one-dimensional system in Fig.~\ref{fig:OTOC_vs_B_loc}.  The 2D case is shown in App.~\ref{app:2D_lightcone_geometry}.

When the ensemble-averaged OTOC is close to either 1 or 0, the instance-to-instance fluctuations in the OTOC value are small, so there is little signal to be measured in either experiment or simulation.  
We therefore choose $M$ to be near the edge of the physical lightcone of $B$, where we observe a maximum signal size.  The precise locations for $B$ are chosen to maximize an estimate of classical hardness while retaining the large fluctuations, as described in detail in App.~\ref{section:obtain_lightcone}.

Once $U$, $M$, and $B$ are determined, we simplify the resulting circuits by removing gates that do not contribute to the value of the OTOC.  
We begin with a time evolution circuit that acts uniformly on the full space of qubits in the system, as shown in Fig.~\ref{fig:OTOC circuit}, step \circled{0}.  
After constructing each circuit instance, we ``prune'' irrelevant gates.  Gates from $U$ and $U^\dagger$ outside the lightcone of $B$ exactly cancel and are removed (step \circled{1}).  Additionally, gates outside the backward lightcone of $M$ cannot affect the final measurement, hence we also remove them (step \circled{2}).  

Finally, following~\cite{google2025observation}, we remove some gates in a way that slightly modifies the definition of the OTOC.  If we used the original OTOC definition from Eq.~\eqref{eq:OTOC_def_original}, we could also exactly cancel gates outside the lightcone of the $M$ operator that acts on the initial maximally mixed state, $\text{Id}/2^N$, thanks to the trace implicit in Eq.~\eqref{eq:OTOC_def_original}.  While this is no longer true once we replace the maximally-mixed state with the $|0\rangle^{\otimes N}$ state, as in \cite{google2025observation} we still remove all of these gates and redefine our OTOC circuit to be the result of this step.
This is illustrated in Fig.~\ref{fig:OTOC circuit}, step \circled{3}.  

One subtlety is that, after pruning, some qubits no longer have any gates acting on them and hence are irrelevant to the value of the OTOC.  We simply remove those qubits from the circuit. As a result, the pruning procedure may reduce the system size.  This is shown in Fig.~\ref{fig:OTOC circuit} via the color of qubits on the left: after each step, the dark gray qubits are the ones still included in the circuit. 

We provide further details on the construction of our new OTOC circuits in App.~\ref{section:obtain_lightcone}.  We explain our precise method for selecting the locations of the $M$ and $B$ operators. We also show how gate pruning makes the effective system size dependent on circuit depth. Finally, we illustrate some of the specific circuit layouts generated by this procedure, with systems sizes ranging from $N=8$ to $N=28$ qubits.

\subsubsection{Reproducible quantum advantage with OTOCs}\label{subsec:reproducible_advantage}

A major step towards useful applications of quantum computers is reproducible quantum advantage.  A quantum computer accomplishes a computational task with a reproducible quantum advantage if:
\begin{enumerate}
    \item Quantum advantage:    
    a classical computer would require far more time or memory to complete the computational task as accurately as the quantum computer, and    
    \item Reproducibility: the output of the task must be an observable whose value is the same (to within a certain accuracy) every time the same experiment is run.
\end{enumerate}
Achieving both quantum advantage and reproducibility simultaneously has proven difficult.  Many of the existing demonstrations of quantum advantage involve 
sampling bitstrings from the final output state of a highly scrambling evolution, \textit{i.e.} random circuit sampling~\cite{morvan2024phase, decross2406computational}.  However, if any of these experiments is run again, a very different set of bitstrings will be drawn, even if in both cases the distribution of bitstrings is consistent with the expected evolution.  In this sense, the outcome of the experiment (namely, the set of bitstrings) is not reproducible.  
On the other hand, reproducible quantities are generally averages over many repetitions of a local measurement on the final state, and several previous experiments measuring potentially beyond-classical local observables~\cite{IBM2023,DWave2024,QuantinuumIsing2025} have proven to be classically simulable~\cite{tindall2024efficient,Kechedzhi2024effective,beguvsic2024fast,tindall2025dynamics,mandra2025}. 

OTOCs with highly scrambling evolution $U$, for example generated using the circuits described in \ref{subsec:circuit_design} above, have features that make them strong candidates for achieving both a large reproducible value and classical hardness simultaneously.  Classical hardness comes from deep random circuits for $U$ and $U^\dagger$.  If we evolved only forward in time, this would lead to local expectation values becoming 0, so there would not be a reproducible signal.  However, when $M$ is near the edge of the physical lightcone of $B$, the OTOC value is around 1/2, with large instance-to-instance fluctuations that can be reliably reproduced when an accurate simulation or experiment is run again for the same random circuit instance.  There is a remaining question: can the same structure of forwards and backwards evolution that gives rise to the reproducible signal be exploited in simulation to reduce classical hardness, removing the potential for quantum advantage?  This question must be answered empirically.

The answer, as demonstrated in \cite{google2025observation}, is that the current experiments on the first-order  OTOC $\mathcal{C}^{(2)}$ are actually \emph{not} clearly beyond classical.  The tensor network Monte Carlo (TNMC) method exploits precisely the mirrored forward and backward evolution to dramatically reduce the complexity of the simulation relative to the naive cost for simulating random circuits of the same total depth.  In~\cite{google2025observation}, TNMC obtained higher SNR than the experiment up to intermediate-size instances up to 40 qubits, and it is expected to perform similarly well over the largest instances with 95 qubits.   
On the other hand, the second-order OTOC $\mathcal{C}^{(4)}$ \emph{is} beyond the reach of any classical method tested in~\cite{google2025observation}, including TNMC.

\subsection{Tensor networks with belief propagation}
\label{sec:bp}

As discussed in Sec.~\ref{sec:otoc_background} above, in an experiment we can find the value of the OTOC by first time-evolving the initial state $|0\rangle^{\otimes N}$ to get the state $|\phi\rangle$ defined in Eq.~\eqref{eq:OTOC definition}, then measuring the value of the local observable $M$.  We follow the same procedure when computing the OTOC by simulation using TNBP.  
In the ``evolution step,'' we simulate the time evolution to find a classical representation of $|\phi\rangle$ as a tensor network, specifically a projected entangled pair state (PEPS)~\cite{verstraete2004renormalization}.
In the ``extraction step,'' we compute the expectation value of $M$ in this PEPS. With finite classical computational resources, for large systems both steps may have to be done approximately.

Let us first consider the evolution step.  We begin with a brief review of PEPS.  A PEPS is specified by a collection of high-rank tensors, one for each qubit.  The tensor has one (``bond'') index, or leg, for each nearest neighbor of the corresponding qubit, plus one additional (``physical'') index of dimension 2 to encode the local qubit basis states.  The dimension of the largest auxiliary index among all PEPS tensors is called the ``bond dimension,'' which we denote by $D$.  When $D=1$, the PEPS encodes a product state with no entanglement.  As $D$ increases, the PEPS can faithfully represent increasingly entangled states. However, generic states such as those produced by random time evolution are highly entangled, and often they can only be approximately represented as a PEPS for the bond dimensions $D$ achievable in simulations with practical computational resources.  

In our OTOC simulation, we begin with the all-zero state, $|0\rangle^{\otimes N}$, a product state which we represent as a PEPS with $D=1$.  We then evolve the state by applying gates according to the OTOC circuit.  In an exact simulation, when we apply an $\iSWAP$-like two-qubit entangling gate between neighboring qubits, the dimension of the bond index between them typically grows by a factor of 4.  The result is that, by the end of an exact simulation of the circuit, the PEPS bond dimension becomes exponential in the number of gates applied on each bond, making the simulation computationally intractable.

Thus we must perform the evolution step approximately.  Specifically, when the dimension of any bond in the PEPS grows beyond some allowed maximum $D$, which controls the degree of approximation, we ``truncate'' the bond back to dimension $D$.  Many schemes have been developed for truncating while attempting to maintain a good approximation to the true evolution.  These range from the simple update scheme~\cite{jiang2008accurate}, which is computationally inexpensive but also not very accurate because it does not take into account enough information about the rest of the system, to the full update scheme~\cite{lubasch2014algorithms, jordan2008classical}, which is much more expensive but takes into account long-ranged correlations to improve accuracy.

A recent addition to the stable of PEPS truncation schemes is belief propagation~\cite{tindall2023gauging}, which was originally developed for machine learning and graphical models~\cite{yedidia2003understanding}.  BP improves upon simple update, taking into account more information about the rest of the system, without significant computational overhead; details about the specific version of the BP truncation scheme we use are given in App.~\ref{app:TNBP}.  The cost of evolving a PEPS on an $N$-site square lattice, with BP-based truncation to a maximum allowed bond dimension of $D$, scales as $\mathcal{O}(N\,D^5)$~\cite{tindall2023gauging}.

We now turn to the extraction step.  To measure a local observable $M$ in a PEPS, we must ``contract'' the tensor network consisting of the PEPS, $M$, and the complex conjugate of the PEPS; this is called a double-layer tensor network because every tensor in the PEPS is ``doubled''---paired with its complex conjugate.  If performed exactly, with no approximation, the contraction of the double-layer tensor network has computational cost scaling with $D^2$ raised to the power of the smaller of the two linear dimensions of the system. 

For many of the small systems we study in this paper, up to $N=28$, and for the PEPS bond dimensions $D$ on the order of 35 that we reach, this exact contraction is still possible.  In those cases, we perform such exact contractions, in order to isolate the effect of the approximation made during the evolution step by limiting $D$. However, our ultimate goal is to understand whether TNBP can scalably simulate large systems, for which exact contraction will be impossible.  We therefore also need approximate methods for measuring the value of observables in a PEPS.

The approximate extraction method we consider here is boundary matrix product state (bMPS) contraction~\cite{lubasch2014algorithms,gonzalez2024random}.  In this method, we create a one-dimensional tensor network, a matrix product state (MPS), for the PEPS tensors and their complex conjugates along the longer edge of the system.  We then move inwards, merging more PEPS tensors into the MPS.  With no truncation, the bond dimension $\chi$ of the MPS would grow exponentially in the smaller linear system size; we would reproduce the exact contraction, which we therefore think of as the $\chi\rightarrow\infty$ limit of bMPS contraction.  To keep computational cost under control, we limit $\chi$ and perform truncations as we incorporate additional tensors from the PEPS.  The cost of contracting a PEPS with bond dimension $D$ using a bMPS of bond dimension $\chi$ scales as 
\begin{equation}
    \mathcal{O} (N (D^4\chi^3 + D^6\chi^2)). \label{eq:bMPS_scaling}
\end{equation}

Thus the final estimated value of an observable such as the OTOC simulated with TNBP depends on two approximations: truncation during the evolution step to a maximum PEPS bond dimension $D$ and truncation during the extraction step to a maximum bMPS bond dimension $\chi$.  As we will show in Sec.~\ref{sec:results} below, to accurately compute the OTOC, both $D$ and $\chi$ must be large.  When $D$ is too small, truncations during time evolution will discard large parts of the wavefunction $|\phi\rangle$, leading to an incorrect value for the OTOC $\langle \phi|M|\phi\rangle$.  When $\chi$ is too small, contributions to the OTOC from long-ranged correlations in $|\phi\rangle$ will be lost during the extraction step, again leading to an incorrect value for the OTOC.

In this work, for each computation we use an a2-ultragpu-8g or a c2-standard-60 CPU node on Google Cloud, limiting us to bond dimensions up to $D=35$ and $\chi=30$. 
With the same resources, we could also use a larger $D$ at the cost of smaller $\chi$, or vice-versa, but our results (below) suggest that $D=35$, $\chi=30$ gives a close-to-optimal performance for the given resources.  
These bond dimensions are bigger than the largest ones used in other state-of-the-art studies~\cite{tindall2024efficient, tindall2025dynamics, rudolph2025simulating}, which allowed for successful simulation of notable experiments by IBM~\cite{IBM2023} and D-Wave~\cite{DWave2024}.  We show that these bond dimensions are not sufficient to accurately compute the OTOC even on a small 23 qubit system.


\section{Results}
\label{sec:results}

We now argue that TNBP cannot feasibly simulate the most challenging experiments from~\cite{google2025observation}.  Through a combination of theoretical arguments and numerical experiments, we demonstrate that the OTOC circuits are largely incompressible via tensor network states and that incompressibility leads to a TNBP simulation cost that is exponential in system size.  The projected cost is already beyond the realm of feasibility for the largest circuits in~\cite{google2025observation}, namely OTOC on 95 qubits and OTOC$^{(2)}$ on 65 qubits.  In particular, we argue that the cost of TNBP simulation is lower-bounded by the cost of an optimal exact tensor network contraction with some gates removed from the circuits, and it was already shown in~\cite{google2025observation} that exact contraction with gate removal cannot reproduce the experiment.

By TNBP, we specifically refer to the classical algorithm described in Sec.~\ref{sec:bp} above: in the evolution step, we evolve a PEPS using BP-based truncations to get the state $|\phi\rangle$ defined in Eq.~\eqref{eq:OTOC definition}; then in the extraction step, we compute the OTOC $\langle\phi|M|\phi\rangle$ with boundary MPS contraction.  However, the argument for computational hardness ultimately does not depend on the truncation scheme used during the evolution step, so our conclusion also applies if we evolve the PEPS using, \textit{e.g.} simple update or full update. 

We proceed in six steps:
\begin{enumerate}[label=\Alph*.]
    \item \hyperref[subsec:IIB_scaling_theory]{\textbf{Theoretical scaling argument}} We argue, based on incompressibility of typical random circuits and based on the geometry of the circuit layouts, that the PEPS bond dimension $D$ needed for accurate simulation of OTOC or OTOC$^{(2)}$ scales exponentially in the circuit depth.  This translates to a cost exponential in the system size $N$ in 1D, and exponential in $\sqrt{N}$ in 2D.  
    
    The fact that the physical lightcones are narrower than the geometric lightcones can lead to compressibility in some parts of the system, but we show that this compressibility does not affect the asymptotic cost scaling.

    \item \hyperref[subsec:1D_exp_scaling_empirical]{\textbf{Empirical numerical demonstration in 1D}} We demonstrate the exponential scaling empirically for 1D systems simulated with matrix product states, confirming the hypothesized incompressibility in 1D.
    
    \item \hyperref[sec:comparison with experiment]{\textbf{Empirical numerical demonstration in 2D, experimental circuits from~\cite{google2025observation}}} We explicitly show that TNBP cannot easily simulate even 23-qubit circuits from~\cite{google2025observation}.  We confirm that the PEPS cannot be significantly compressed during time evolution without losing accuracy in computing the OTOC.  We furthermore show that a large boundary MPS bond dimension $\chi$ is needed to accurately extract the OTOC from the final PEPS.
    
    \item \hyperref[sec:scaling_experiment]{\textbf{Empirical numerical demonstration in 2D, new circuits}} We construct new circuits analogous to those from the experiment in~\cite{google2025observation}, on a range of small system sizes in 2D from $N=8$ to $N=28$ qubits.  By including much smaller systems, we can fully converge TNBP simulations even with computational resources corresponding to a single compute node on Google Cloud, leading to a more complete picture of the computational requirements for accurate TNBP simulation in 2D.  We further confirm both incompressibility of the PEPS during time evolution (and therefore large $D$) in 2D and the necessity of a large $\chi$ for extraction.  

    \item \hyperref[sec:modified_gates]{\textbf{TNBP can simulate OTOC with weakly-entangling evolution}} For each of the new circuits we constructed for the empirical test of TNBP in 2D, with $N=8$ through 28 qubits, we create and simulate analogous OTOC ensembles with less-entangling gates.  We observe that when the evolution is sufficiently weakly entangling, TNBP \emph{succeeds} at efficiently simulating the OTOC.  This result suggests that if the OTOC circuits were compressible, TNBP could exploit the compressibility to reduce computational cost, and therefore that the previously observed high cost of TNBP simulation in 2D indicates fundamental incompressibility in the circuits. 

    \item \hyperref[subsec:Frontier_estimates]{\textbf{TNBP cannot simulate the quantum echoes experiment}} Finally, we argue that reproducing the largest experiments in ~\cite{google2025observation}, namely OTOC on $N=95$ qubits and OTOC$^{(2)}$ on $N=65$ qubits, and at the level of accuracy reached in the experiment, would take longer with TNBP than with exact tensor network contraction already considered in~\cite{google2025observation}.  Hence TNBP does not present a challenge to the claim that the experiments in~\cite{google2025observation} are beyond classical.
\end{enumerate}

We assess the accuracy of our numerical simulations (B. $-$ E.) in two ways.  First, we use the SNR as defined in Eq.~\eqref{eq:snr}.  With 50 random instances per OTOC circuit layout, with no signal we would see $\mathrm{SNR}^{\text{uncorrelated}} \approx  0.725$; perfect simulation with no noise would give $\text{SNR}\!\rightarrow\!\infty$.  The echoes experiment achieved SNRs on the order of 3 to 5.  Second, we use the global wavefunction fidelity between the state $|\phi\rangle$ and the finite-$D$ PEPS approximation to the state after the evolution step of TNBP.  We observe that the two metrics are correlated if we truncate the PEPS during the evolution step to a spatially uniform bond dimension $D$.

Each item A. through F. is the focus of one subsection below.

\subsection{Prediction of exponential scaling of TNBP cost with system size}
\label{subsec:IIB_scaling_theory}

\begin{figure*}
    \centering
    \includegraphics[width=\linewidth]{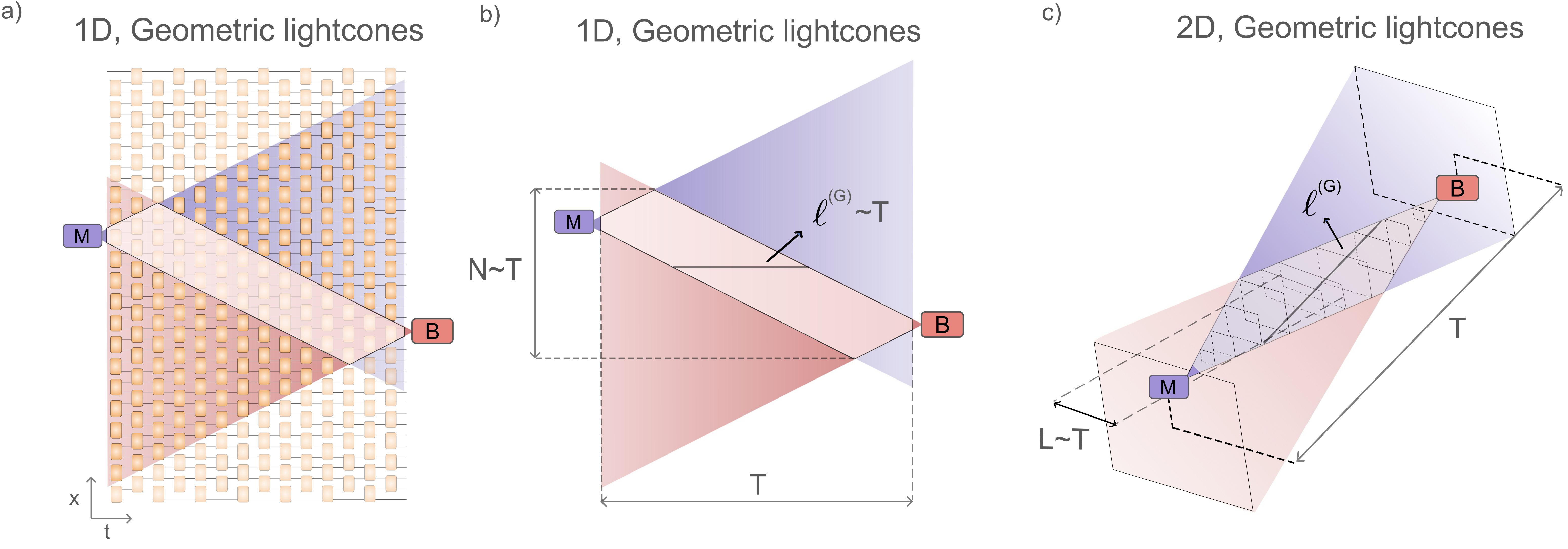}
    \caption{\textbf{Lightcone geometry determines entanglement growth.} We show geometrically the gates that contribute to the value of the OTOC and hence to entanglement generation that must be captured in simulation.  (a) The geometric lightcone for the measurement operator $M$ is shaded in purple, for the butterfly operator $B$ in red.  Only gates in the overlap between the two cones, in the white parallelogram, affect the value of the OTOC and are retained.  All other gates are pruned, as described in Sec.~\ref{subsec:circuit_design}.  (b) Abstract view of the overlap of geometric lightcones, with the gate structure of the time evolution circuit for $U$ not shown.  The number of qubits that remains after pruning, $N$, is proportional to the circuit depth $T$.  Furthermore, the number of gates applied to some bonds of a PEPS during Schr\"{o}dinger picture evolution scales as $\ell^{(G)}$, which is also proportional to $T$.  It follows that effective circuit depth is proportional to system size.  (c) The same structure also appears for the OTOC in two dimensions, except the geometric lightcones become three-dimensional square pyramids, and the number of qubits remaining after pruning scales as $N\sim T^2$; the linear size of the two dimensional qubit grid, $L\sim\sqrt{N}$, scales with $T$.}
    \label{fig:depth_scaling}
\end{figure*}

To show that TNBP cannot feasibly simulate the quantum echoes experiment in~\cite{google2025observation}, we start with a theoretical derivation of the expected asymptotic cost scaling with system size.  
Specifically, we argue that the required PEPS bond dimension $D$ should scale exponentially in the linear size of the system, meaning that at even moderate system sizes, classical simulation with TNBP would be out of reach.

Note that exponential PEPS bond dimension also implies exponential total cost for the TNBP simulation.  The bottleneck in the cost is the bMPS contraction in the extraction step, with scaling as given in Eq.~\eqref{eq:bMPS_scaling}.  The computation time scales at least as $D^6$, and the required bMPS bond dimension $\chi$ is also expected to increase with $D$.  The memory cost is also exponential, since the size of the tensors that must be stored scales polynomially in $D$. 

We first derive, in Sec.~\ref{subsec:scaling_with_incompressibility}, the exponential scaling under the assumption that the circuits are incompressible, as defined in App.~\ref{app:def_incompressibility}.  We then show in Sec.~\ref{subsec:scaling_with_compressibility} that, for well-chosen locations of the butterfly and measurement operators, the exponential scaling survives even despite limited compressibility inherent in the structure of the OTOC circuits. 

\subsubsection{Scaling under the assumption of incompressibility}\label{subsec:scaling_with_incompressibility} 

We begin by assuming incompressibility, as defined in App.~\ref{app:def_incompressibility}. Specifically, we assume that the PEPS representation of the final state, in which we measure $M$ to get the OTOC or OTOC$^{(2)}$, must have on each bond between adjacent qubits a bond dimension that scales as $D_g$ raised to the power of the number of gates applied on that bond in the circuit, where $D_g=4$ in 2D and $D_g=2$ in 1D.   Thus we estimate the required $D$ for an OTOC simulation from the scaling of number of entangling gates applied on each bond.

We are ultimately most interested in the cost scaling for 2D systems, since the experiments in~\cite{google2025observation} were performed on a 2D qubit grid. However, here we will also derive the cost scaling for 1D systems, as the circuit structures are simpler to visualize but the logic behind the derivation of cost scaling is similar.

\paragraph{Geometric derivation of exponential cost:}   
We consider an OTOC ensemble where the time evolution circuit $U$ has depth $T$, and where the measurement operator $M$ and butterfly operator $B$ are placed so that they are inside of each other's geometric lightcones (since otherwise the OTOC is exactly 1, as established in Sec.~\ref{subsec:circuit_design} above). 

The evolution step of the TNBP method as outlined in Sec.~\ref{sec:bp} consists of a wavefunction evolution in the Schr\"{o}dinger picture.  For OTOC, we obtain the state $|\phi\rangle$ by evolving under $U$ (with gates outside the geometric lightcones removed), then applying $B$, then evolving under $U^\dagger$ (again with gates removed).  For OTOC$^{(2)}$, we simply continue to evolve this state, under $M$, then $U$, then $B$, then $U^\dagger$, finally arriving at the state $|\varphi\rangle$.  To estimate the number of entangling gates applied on each bond during the evolution, we focus on the first part of the evolution step, evolution by the circuit for $U$. We then multiply this number of gates by 2 for OTOC and by 4 for OTOC$^{(2)}$.

We begin with the simpler case of 1D systems.  As illustrated in Fig.~\ref{fig:depth_scaling}(a), in 1D the gates that remain after pruning based on the geometric lightcones form a parallelogram with $M$ and $B$ at two opposite vertices. 
The number of gates on a given bond is proportional to the length (in the time direction) of the parallelogram at that qubit location.  We focus on the largest such length, denoted by $\ell^{(G)}_{ }$ in Fig.~\ref{fig:depth_scaling}(b),  which is the same as Fig.~\ref{fig:depth_scaling}(a) but with the background circuit structure of $U$ removed for clarity. ``G'' indicates that the length is determined by the geometric lightcones.  
$\ell^{(G)}$ is proportional to the circuit depth $T$, and therefore to the system size $N$.  

More precisely, let the ``speed of light'' for the geometric lightcone be $c$, and let the speed of propagation required to reach $B$ from $M$ in time $T$ be $v_{MB}$.  Then it can be seen that $N=cT$, and $\ell^{(G)} = (1-v_{MB}/c)T$.  Then as long as $B$ is inside the geometric lightcone of $M$, so that $v_{MB}/c < 1$, we find that the number of gates that must be applied on some bonds of the PEPS scales linearly in system size, resulting in an exponentially large bond dimension, $D\sim \exp\left(\ell^{(G)}\right)$ which implies $D\sim \exp(N)$.  Thus the total simulation cost is exponential in system size.

The same argument can be extended to 2D systems, as illustrated in Fig.~\ref{fig:depth_scaling}(c).  
We again keep just the gates in the intersection of the geometric lightcones, which become square pyramids as shown in the figure.  As in 1D, the maximum number of gates applied on any given bond is proportional to $\ell^{(G)}$, which itself is proportional to the circuit depth $T$.  However, in 2D it is not the full system size $N$ but rather the linear size $L\sim\sqrt{N}$ that scales linearly in $T$ after pruning gates outside the geometric lightcones; the precise scaling of $N$ with $T$ is complicated by the lack of rotational symmetry of the lightcones, as we explain in detail in App.~\ref{app:2D_lightcone_geometry}.  Thus $\ell^{(G)}$ scales as $L\sim\sqrt{N}$, and the required PEPS bond dimension scales exponentially in the square root of the system size, $D\sim \exp(\sqrt{N})$. 

\paragraph{More precise scaling for our specific OTOC circuits:} 
We have established that the PEPS bond dimension $D$ should grow exponentially in $N$ in 1D, and exponentially in $\sqrt{N}$ in 2D.  These bounds generically hold for incompressible evolution with a strict geometric lightcone whose edge travels at speed $c$, and for $B$ located within the geometric lightcone of $M$ and vice-versa.  
We now find more precise scaling specific to our OTOC circuits by considering the brickwall structure and the choice of gates in $U$.

We first consider 1D systems.  If we define $T$ to be the number of 2-qubit gate layers in the brickwall circuit for $U$, then $c=1$ because each layer expands the geometric lightcone by exactly 1 qubit both to the left and to the right.  Thus $N=T$.  The largest local circuit depth in $U$ on any given bond is $\ell^{(G)}=(1-v_{MB}/c)T$, measured in terms of numbers of two-qubit gate layers at that location.  However, in the brickwall circuit only half of layers apply a gate on any particular bond, so that during evolution by $U$, the largest number of gates applied on a given bond is $(1-v_{MB}/c)T/2$.  To find the OTOC, in the evolution step we apply first $U$ and then $U^\dagger$, thus doubling the number of gates applied on each bond, so the largest number applied on a given bond becomes $(1-v_{MB}/c)T$.  Using $N=T$, we predict that for OTOC the required bond dimension should scale as 
\begin{equation}
    D = A_1 \cdot 2^{(1-v_{MB}/c)N}. \label{eq:expected_scaling_1D}
\end{equation}
$A_1$ is a constant that depends on the desired level of accuracy.  For OTOC$^{(2)}$ the gate depth on any bond is doubled, and the scaling becomes $D = A_1'\cdot 2^{2(1-v_{MB}/c)N}$.

For concreteness, consider the situation where $M$ and $B$ are separated along the line with velocity $v_{MB}/c=3/5$.  We choose this specific value because when the time evolution $U$ for a 1D system is given by brickwall circuits made from Haar-random 2-qubit gates, $v/c=3/5$ is the velocity of the physical lightcone edge~\cite{nahum2018operator,khemani2018operator,xu2020accessing,zhou2020entanglement}, so setting $v_{MB}/c=3/5$ maximizes the size of fluctuations in the OTOC, \textit{i.e.} the signal to be measured.  The random circuits $U$ used in~\cite{google2025observation}, where the two-qubit entangling gates are iSWAP-like, are dual unitary circuits~\cite{Bertini2019dualUnitaries} where the physical lightcone is as wide as the geometric lightcone.  Then the speed of the physical lightcone edge is $v/c=1$, so $v_{MB}/c=3/5$ puts $B$ well inside the physical lightcone of $M$.  In both these cases (Haar-random two-qubit gates with $B$ on the edge of the physical lightcone of $M$, and iSWAP-like two-qubit gates with $B$ inside the physical lightcone of $M$), we predict a required bond dimension scaling as $D=A_1 \cdot 2^{2N/5}$.  In Sec.~\ref{subsec:1D_exp_scaling_empirical} below, we confirm this prediction in both cases via explicit numerical simulations.

We can likewise predict a more precise scaling for 2D systems.  The full geometric derivation is given in App.~\ref{app:2D_lightcone_geometry}, but when the locations of $M$ and $B$ are chosen to ensure large instance-to-instance fluctuations in the OTOC, then to maximize the hardness of exact tensor network contraction, the end result is that the required bond dimension for system size $N$ is 
\begin{equation}
    D \approx A_2 \cdot 4^{0.6\sqrt{N}} \label{eq:expected_scaling_2D_sep_x}
\end{equation}
where $A_2\leq 0.5$ is again a constant that depends on the desired accuracy, and it is upper-bounded by 0.5 because the first layer of gates is applied to a product state and therefore cannot increase the bond dimension by more than a factor of 2.  For OTOC$^{(2)}$ the scaling is increased to $A_2'\cdot 4^{1.2\sqrt{N}}$.

\subsubsection{Effect of compressibility outside of physical lightcones}\label{subsec:scaling_with_compressibility} 

\begin{figure*}
    \centering
    \includegraphics[width=1\linewidth]{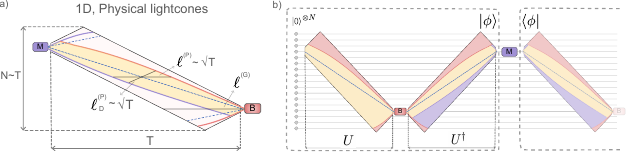 }
    \caption{\textbf{Compressibility outside of physical lightcones.}  (a) Gates inside the geometric lightcones of $M$ and $B$ but outside their physical lightcones might be compressible in some tensor network simulation methods and thus contribute less to computational complexity.  After taking into account the diffusive fronts of the physical lightcones, there remains an incompressible core (shaded yellow) which we expect cannot be compressed by any tensor network state method.  The width of the incompressible core in different directions, $\ell^{(P)}$ and $\ell^{(P)}_D$, scales with the square root of the circuit depth, compared with $\ell^{(G)}$ which scales linearly in $T$.  (b) When computing the OTOC with TNBP, in the evolution step no compression is possible during evolution by $U$.  During evolution by $U^\dagger$, gates in the red-shaded regions in $U$ and $U^\dagger$, outside the physical lightcone of $B$, approximately cancel, allowing for some compression. However the circuit depth remains $\ell^{(G)}$ in the vicinity of $B$, so the final bond dimension required in that part of the PEPS is not reduced.  In the extraction step, the approximate cancellation of gates in the purple-shaded regions allows for some compression in the bMPS contraction, allowing for a lower $\chi$ than would otherwise be required, but this still does not affect the overall cost scaling.
    }
    \label{fig:geometric_compressibility}
\end{figure*}

So far we have used the geometry of the circuits, in particular the geometric lightcones of $M$ and $B$, to derive the expected asymptotic cost scaling for computing OTOC and OTOC$^{(2)}$ with TNBP.  These derivations assumed the circuits to be incompressible. Here we show that the evolution actually \emph{is} compressible in some parts of the system, but that for good choices of the relative locations of $M$ and $B$, the required PEPS bond dimension remains exponential in linear system size.  We divide the possible locations of $M$ and $B$ into three cases:
\begin{enumerate}
    \item \label{MP_case1} If $B$ is on or near the edge of the physical lightcone of $M$, the scaling remains exponential.  This is the situation of most relevance to the experiments in~\cite{google2025observation}, since the instance-to-instance fluctuations are largest near the physical lightcone edges.

    \item \label{MP_case2} If $B$ is well inside the physical lightcone of $M$, the scaling remains exponential, and the effect of compressibility is minimal.

    \item \label{MP_case3} If $B$ is well outside the physical lightcone of $M$, the OTOC becomes highly compressible. 
\end{enumerate}
We will explain the compressibility while assuming case \ref{MP_case1}, then return to the other two cases briefly at the end of this section.

\paragraph{Compressibility in case 1:} 
To understand the origin of the compressibility, we consider the physical lightcones of $M$ and $B$ in addition to their geometric lightcones.  As shown in Fig.~\ref{fig:OTOC_vs_B_loc}, the OTOC is exactly 1 when $M$ and $B$ are outside of each other's geometric lightcones, and it remains close to 1 when the operators are inside each other's geometric lightcones but outside of each other's physical lightcones. In other words, gates outside of the physical lightcones of $M$ and $B$ do not substantially affect the value of the OTOC.  Thus it is reasonable to think that the portion of the circuit that lies outside the physical lightcones could be compressed without changing the value of the OTOC computed by TNBP.  In particular, gates outside of the physical lightcone of $B$ will partially cancel between $U$ and $U^\dagger$.  We now formalize this intuition.

The first question to answer is which gates should be viewed as outside the physical lightcones, in the sense that they do not cause the value of the OTOC to significantly deviate from 1 for any random instance in the OTOC ensemble.  Naively, the physical lightcone edge is the dashed blue line in Fig.~\ref{fig:OTOC_vs_B_loc}, indicating where the OTOC transitions from 0 inside to 1 outside. 
However, as seen in Fig.~\ref{fig:OTOC_vs_B_loc}, the transition from 0 to 1 is not sharp.  There are two reasons: information spreads at different speeds for different random instances of the evolution $U$, leading to wider and narrower physical lightcones for the different instances; and the OTOC value for any given instance fluctuates non-monotonically as the location of $B$ crosses the edge of the physical lightcone of $M$.  The result in 1D is that the physical lightcone front is diffusive---for time evolution $U$ with depth $T$, the OTOC value transitions from 0 to 1 over a range of $M$-$B$ location separations that scales as $\sqrt{T}$.
In 2D, there is also spreading of the physical lightcone front, with a width that scales as $T^{1/3}$~\cite{nahum2018operator}.

The physical lightcones are shown in the context of an OTOC circuit in Fig.~\ref{fig:geometric_compressibility}(a), specifically in case \ref{MP_case1}: $M$ and $B$ located on each other's physical lightcone edges.  The lightcone edges, defined as where the ensemble-averaged OTOC crosses 1/2, are shown as dashed blue lines.  At both $M$ and $B$ these bound a triangular region narrower than the geometric lightcone bounded by the black lines on the boundaries of the parallelogram.  Note that because we are assuming case 1, $M$ and $B$ share one physical lightcone edge, so only three dashed blue lines appear in the figure.  

Because the physical lightcone fronts have finite width, the gates that contribute to information spreading from $M$ are not just those between the blue dashed lines emanating from $M$ in Fig.~\ref{fig:geometric_compressibility}(a), but in fact all the gates between the two purple lines in the figure.  These gates define the ``dressed physical lightcone'' of $M$.  Likewise, the dressed physical lightcone of $B$ consists of the gates between the red lines in the figure.  The portion of the OTOC circuits outside of these dressed physical lightcones may be compressible.  The intersection of the two dressed physical lightcones, shaded yellow in Fig.~\ref{fig:geometric_compressibility}(a), we expect to be essentially incompressible.    

We now show that the TNBP algorithm can partially exploit the compressibility of the circuits outside the dressed physical lightcones, as illustrated in Fig.~\ref{fig:geometric_compressibility}(b).  We begin the evolution step by evolving the state $|0\rangle^{\otimes N}$ under $U$; this evolution is not compressible, and the required PEPS bond dimension for accurate simulation must scale exponentially in $\ell^{(G)}$.  Once we apply $B$ and begin to evolve under $U^\dagger$, some compression becomes possible.  Namely, gates outside the dressed physical lightcone of $B$, in the region shaded red in Fig.~\ref{fig:geometric_compressibility}(b), approximately cancel between $U$ and $U^\dagger$.  Thus in the parts of the system far from $B$, the effective circuit depth is reduced substantially.  The required PEPS bond dimension in that part of the system is reduced in tandem.  

However, in the vicinity of $B$, the local circuit depth is unchanged since the full circuit in that spatial region is inside the physical lightcone of $B$.  This is illustrated in Fig.~\ref{fig:geometric_compressibility}(a): at the location of $B$, the local circuit depth within the dressed physical lightcone of $B$ is still given by $\ell^{(G)}$.  The final bond dimension required for the PEPS representing $|\phi\rangle$ is determined by the deepest part of the circuit and hence is unchanged from Eqs.~\eqref{eq:expected_scaling_1D} (in 1D) or \eqref{eq:expected_scaling_2D_sep_x} (in 2D).  

Compressibility of the part of the circuit outside of the dressed physical lightcones could also lead to reduced computational cost for the extraction step.  When performing bMPS contraction to compute $\langle\phi|M|\phi\rangle$, the bMPS bond dimension $\chi$ required for accurate extraction is reduced because some of the entanglement in the final PEPS tensors comes from gates outside the dressed physical lightcone of $M$, in the region shaded purple in Fig.~\ref{fig:geometric_compressibility}(b).  In the double-layer tensor network representing $\langle\phi|M|\phi\rangle$, the effect of these gates approximately cancels between $|\phi\rangle$ and $\langle\phi|$, and thus the bMPS bond dimension $\chi$ required for an accurate contraction is only as large as what would have been needed if the gates in the purple regions had been removed from the circuit.  

To summarize, when simulating the OTOC with TNBP, some compression is possible because the circuit for $|\phi\rangle$ includes gates outside the dressed physical lightcones of $B$ and $M$.  In the evolution step, the required PEPS bond dimension $D$ can be reduced in the part of the system far from $B$ because gates outside the physical lightcone of $B$ approximately cancel between $U$ and $U^\dagger$.  However, the bond dimension near $B$ remains close to maximal, so the overall scaling of $D$ with system size is unchanged.  In the extraction step, the required bMPS bond dimension $\chi$ can be reduced because in the part of the system far from $M$ gates outside the dressed physical lightcone of $M$ approximately cancel between $|\phi\rangle$ and its adjoint in the double-layer tensor network contraction.  The latter could potentially result in a lower asymptotic cost scaling in 2D systems.

Similar compression based on partial cancellation of gates outside the physical lightcones is possible when computing OTOC$^{(2)}$ with TNBP.  In the evolution step, we can compress the part of the circuit outside of the dressed physical lightcones of both $B$ operators in $|\varphi\rangle$ (see Eq.~\ref{eq:OTOC2}) and the part of the circuit outside the dressed physical lightcone of the $M$ operator in $|\varphi\rangle$.  In the extraction step, we can use a lower $\chi$ due to cancellation between $|\varphi\rangle$ and $\langle\varphi|$ in the part of the circuit outside the physical lightcone of the measured $M$.  However, the required PEPS bond dimension remains exponential in linear system size.

\begin{figure*}[ht!]
    \centering 
    \includegraphics[width=1\linewidth]{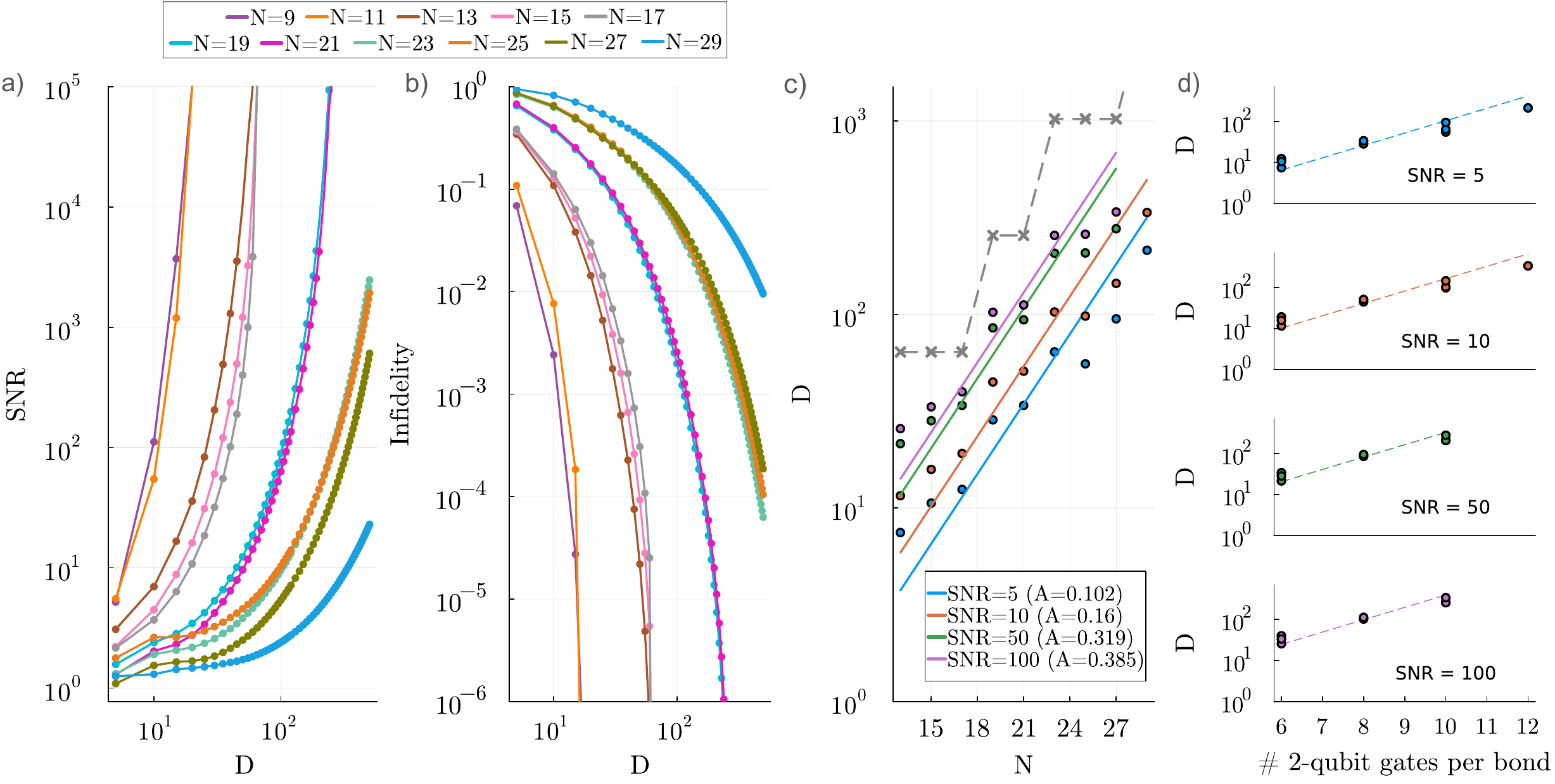}
     \caption{\textbf{Empirical demonstration of exponential scaling in 1D, Haar ensembles.}  
     To achieve fixed accuracy (SNR or fidelity), the required MPS bond dimension $D$ grows exponentially in system size.  
     (a) For each system size, we show the SNR computed using 50 instances of the corresponding OTOC ensemble where the evolution circuit $U$ is constructed from Haar-random two-qubit gates, for a range of bond dimensions.  For each size, we place $B$ on the edge of the physical lightcone of $M$, so $v_{MB}/c=3/5$.  We perform the simulations in single precision, so SNR values are accurate up to around $10^5$. (b) For each system size, we show the mean infidelity across instances for a range of bond dimensions.  (c) 
     For each system size, we find the bond dimension needed to reach a fixed SNR of 5, 10, 50, or 100.  We expect a scaling of the form $D = A\cdot 2^{(2/5)N}$ predicted in Eq.~\eqref{eq:expected_scaling_1D} when $v_{MB}/c=3/5$, where the coefficient $A$ depends on the target value of SNR.  Empirically, the data for each target SNR roughly match a curve of this form, but the required $D$ also shows a series of plateaus with increasing $N$.  The plateaus are expected, for the following reason.  Eq.~\eqref{eq:expected_scaling_1D} was derived by assuming the bond dimension to scale with 2 raised to the power of the maximum number of gates applied on any given bond, then finding in the large-system limit that the number of gates per bond scales as $2N/5$.  Lattice discretization leads to plateaus in [max $\#$ of gates/bond] vs $N$, and when we plot $D=2^{[\text{Max \# gates/bond}]}$ (X markers and dashed gray line), we see the same plateaus observed in the numerical results.  (d) To make the exponential scaling clearer, for each target SNR we plot required $D$ vs max \# of two-qubit gates per bond, grouping together system sizes $N$ that have the same gate depth.  Dashed lines show $D=A\cdot 2^{[\text{Max \# gates/bond}]}$ with $A$ taken from the best fit for the corresponding target SNR in panel (c).}
    \label{fig:empirical_scaling_1D_Haar}
\end{figure*}

\paragraph{Compressibility in cases 2 and 3:} 
For concreteness, we have focused on case \ref{MP_case1} for the relative locations of $M$ and $B$: they are positioned on each other's physical lightcone edges to maximize the instance-to-instance fluctuations in the value of the OTOC.  This case is the one most relevant to the quantum echoes experiment, where the locations of $M$ and $B$ were chosen such that the fluctuations are close to maximal.  We concluded that scaling should remain exponential in linear system size.

In case 2, where the operators are well inside of each other's physical lightcones, the required PEPS bond dimension is again exponential in linear size even after compression of the circuit outside the dressed physical lightcones.  Consider Fig.~\ref{fig:geometric_compressibility}(a).  If the $B$ operator is moved up, it remains the case that the local circuit depth within the physical lightcone of $B$ is $\ell^{(G)}$, which in fact increases in length as $v_{MB}$ decreases.  

On the other hand, in case 3 where the operators are outside of each other's physical lightcones, the circuit becomes highly compressible.  Although some incompressible parts of the circuit, near $B$, still have depth $\ell^{(G)}$ scaling with the linear system size (though $\ell^{(G)}$ is smaller when $v_{MB}$ is closer to $c$), all gates acting on the location of $M$ are compressible.  Hence the full simulation could be performed with a low bond dimension, and although that would produce a very bad approximation to the state away from $M$, in the vicinity of $M$ the state would be approximately correct and hence the OTOC values would remain accurate.  This prediction is compatible with~\cite{Xu2020otocOutside}, where the authors found that the OTOC outside the physical lightcone could be computed accurately with a small bond dimension.

\paragraph{Compressibility in the Heisenberg picture:} 
We note that further compression based on the physical lightcone geometry may be possible in a tensor network simulation in the Heisenberg picture.  We briefly describe a possible approach in App.~\ref{sec:Heis_picture}, which we expect to be asymptotically more efficient though still exponential.

\subsection{Empirical numerical demonstration in 1D: exponential scaling}
\label{subsec:1D_exp_scaling_empirical}

\begin{figure*}[ht!]
    \centering 
    \includegraphics[width=1\linewidth]{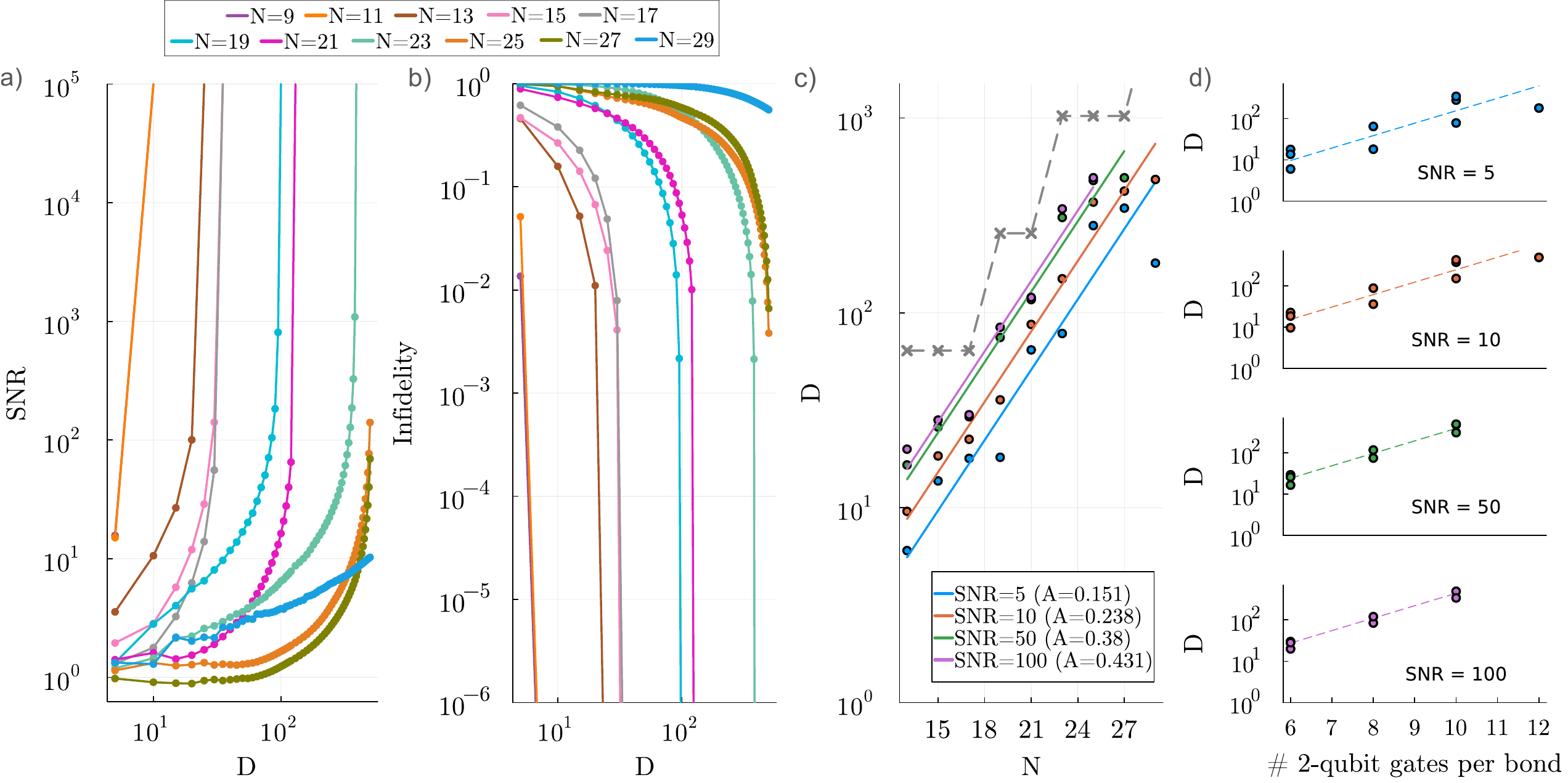}
     \caption{\textbf{Empirical demonstration of exponential scaling in 1D, iSWAP ensembles.}  
     Same analysis as in Fig.~\ref{fig:empirical_scaling_1D_Haar}, but for 1D OTOC ensembles where the time evolution $U$ is constructed from iSWAP-like entangling gates and random single-qubit gates as described in Sec.~\ref{subsec:circuit_design}.  We again use $v_{MB}/c=3/5$, which places $B$ well inside the physical lightcone of $M$. 
     (a) SNR computed using 50 instances for each system size, for a range of bond dimensions. 
     (b) Mean fidelity across instances for each system size, for a range of bond dimensions. 
     (c) For each system size, we find the bond dimension needed to reach a fixed SNR of 5, 10, 50, or 100.  Again the observed scaling of required $D$ with $N$ roughly matches the scaling $D = A\cdot 2^{(2/5)N}$ predicted in Eq.~\eqref{eq:expected_scaling_1D}, but with a series of rough plateaus due to lattice discretization.  The bond dimension required for an exact simulation ($\text{SNR}\rightarrow\infty$), given by $2^{[\text{Max \# gates/bond}]}$ and shown by the X markers and dashed gray lines, exhibits the same plateaus. (d) For each target SNR we plot required $D$ vs max \# of two-qubit gates per bond, grouping together system sizes $N$ that have the same gate depth.  Dashed lines show $D=A\cdot 2^{[\text{Max \# gates/bond}]}$ with $A$ taken from the best fit for the corresponding target SNR in panel (c).
     }
    \label{fig:empirical_scaling_1D}
\end{figure*}

We now demonstrate, via explicit numerical simulation, that TNBP simulation of the OTOC in 1D scales precisely as predicted in Eq.~\eqref{eq:expected_scaling_1D}. 

We study the scaling in two different settings.  In the first setting, we construct OTOC ensembles with the evolution $U$ given by a 1D brickwall circuit made from Haar-random two-qubit gates.  In 1D, these evolution circuits have a physical lightcone that spreads at speed $v/c=3/5$, and we place the butterfly operator $B$ on the edge of the physical lightcone of the measurement operator $M$, so $v_{MB}/c=3/5$ as well.  We use these ``Haar OTOC ensembles'' to test the scaling of required bond dimension when $B$ is on the edge of the physical lightcone of $M$.  The predicted scaling of bond dimension with system size is $D = A\times 2^{2N/5}$ for some constant $A$ depending on the desired level of accuracy, as derived in Sec.~\ref{subsec:scaling_with_incompressibility}.

In the second setting, we construct OTOC ensembles with the evolution $U$ given by a 1D circuit with iSWAP-like entangling gates and random single-qubit gates as described in Sec.~\ref{subsec:circuit_design}, analogous to the circuits in~\cite{google2025observation}.  In 1D, these evolution circuits are dual-unitary, so the physical lightcone is as wide as the full geometric lightcone.  We therefore use these ``iSWAP OTOC ensembles'' to test the scaling of required bond dimension when $B$ is \emph{inside} the physical lightcone of $M$.  We choose to place $B$ such that $v_{MB}/c=3/5$, again giving a predicted bond dimension of $D = A\times 2^{2N/5}$ for some constant $A$.  

In each of these two settings, a numerical experiment provides useful information regarding incompressibility and cost scaling for the types of OTOC circuits studied in~\cite{google2025observation}.  Numerical experiments showing the predicted exponential scaling for the iSWAP ensembles will confirm that evolution circuits of the type used in~\cite{google2025observation} are inherently incompressible with tensor networks.  Numerical experiments showing the predicted exponential scaling for the Haar ensembles will confirm that, as predicted in Sec.~\ref{subsec:scaling_with_compressibility}, compressibility due to the presence of gates outside the dressed physical lightcones of $B$ and $M$ will not reduce the scaling of bond dimension with system size.

For both settings, we construct ensembles for a range of circuit depths, such that pruning gates outside the geometric lightcones (see Sec.~\ref{subsec:circuit_design}) results in a range of system sizes from $N=3$ through $N=29$. For each ensemble, we first perform an exact state vector simulation to determine the final state $|\phi\rangle$ and the value of the OTOC.  We then perform TNBP simulations, where the PEPS representing the wavefunction reduces to an MPS, for each of a range of fixed cutoff bond dimensions $D$.  For each system size $N$ and bond dimension $D$, we compute the SNR and the mean fidelity across instances.  For the Haar OTOC ensembles, SNR is shown in Fig.~\ref{fig:empirical_scaling_1D_Haar}(a) and the mean fidelity is shown in Fig.~\ref{fig:empirical_scaling_1D_Haar}(b).  For the iSWAP OTOC ensembles, SNR is shown in Fig.~\ref{fig:empirical_scaling_1D}(a) and the mean fidelity is shown in Fig.~\ref{fig:empirical_scaling_1D}(b)

We then study how the required bond dimension to achieve a fixed accuracy (measured by SNR) scales with system size.  For a given target SNR, $\text{SNR}_0$, we find the required bond dimension as a function of system size $N$ by finding where the corresponding $\text{SNR}$ vs. $D$ curve in Fig.~\ref{fig:empirical_scaling_1D_Haar}(a) or Fig.~\ref{fig:empirical_scaling_1D}(a) crosses a horizontal line at $\text{SNR}_0$.  This step involves interpolation in $D$, which is quite accurate.  
We plot the resulting $D$ vs. $N$ for $\text{SNR}=5$, $10$, $50$, and $100$ in Fig.~\ref{fig:empirical_scaling_1D_Haar}(c) for the Haar ensembles and Fig.~\ref{fig:empirical_scaling_1D}(c) for the iSWAP ensembles.  

We observe that the $D$ vs $N$ curve for each $\text{SNR}_0$ roughly follows a straight line on a log-linear plot, in accordance with the predicted exponential scaling $D\sim\exp(N)$.  However, there are a series of rough plateaus in the required bond dimension as $N$ increases.  We can understand the origin of the plateaus by considering the derivation of Eq~\eqref{eq:expected_scaling_1D}.  We first predicted that (i) the required bond dimension should scale as 2 raised to the power of the maximum number of gates applied on any given bond during the evolution by $U$ and $U^\dagger$, and we then estimated geometrically that (ii) the maximum number of gates per bond is $(1-v_{MB}/c)N$, or $2N/5$ when $v_{MB}/c=3/5$ as in our numerical experiments.  However, on a lattice, results based on geometry in the continuum are only approximately correct; gate depth only increases in discrete steps and at certain values of $N$, leading to plateaus.  If step (ii) of the derivation being approximate is the origin of the plateaus, then a bond dimension predicted by step (i) alone should be more accurate.  In that case, the predicted bond dimension is $D=A\cdot 2^{[\text{Max \# gates/bond}]}$ where $A$ is a constant that depends on the desired level of accuracy (\textit{i.e.} the target $\text{SNR}_0$).  

To see that this prediction is correct, we first plot the bond dimension required for perfect accuracy ($\text{SNR}_0\rightarrow\infty)$, namely $D=2^{[\text{Max \# gates/bond}]}$, as a function of $N$ in Fig.~\ref{fig:empirical_scaling_1D_Haar}(c) and Fig.~\ref{fig:empirical_scaling_1D}(c).  This curve, shown with X markers and a dashed gray line, exhibits the same structure of plateaus as do our numerical simulation results.  Furthermore, in Fig.~\ref{fig:empirical_scaling_1D_Haar}(d) and Fig.~\ref{fig:empirical_scaling_1D}(d), we plot required $D$ vs the maximum number of two-qubit gates applied on a given bond, rather than vs system size.  For both the Haar and iSWAP OTOC ensembles, the data very closely match the predicted scaling.

In short, these results clearly demonstrate that, at least in 1D, the theoretical scaling argument presented in Sec.~\ref{subsec:IIB_scaling_theory} gives an accurate prediction of TNBP simulation cost.  We can further conclude that the assumption of incompressibility of the iSWAP-based evolution circuits is correct in 1D, and that the required bond dimension scales exponentially despite compressibility outside of the dressed physical lightcones as discussed in Sec.~\ref{subsec:scaling_with_compressibility}).

\subsection{Empirical numerical demonstration in 2D: Circuits from quantum echoes experiment}
\label{sec:comparison with experiment}

\begin{figure*}[t]
    \centering
    \includegraphics[width=0.8\linewidth]{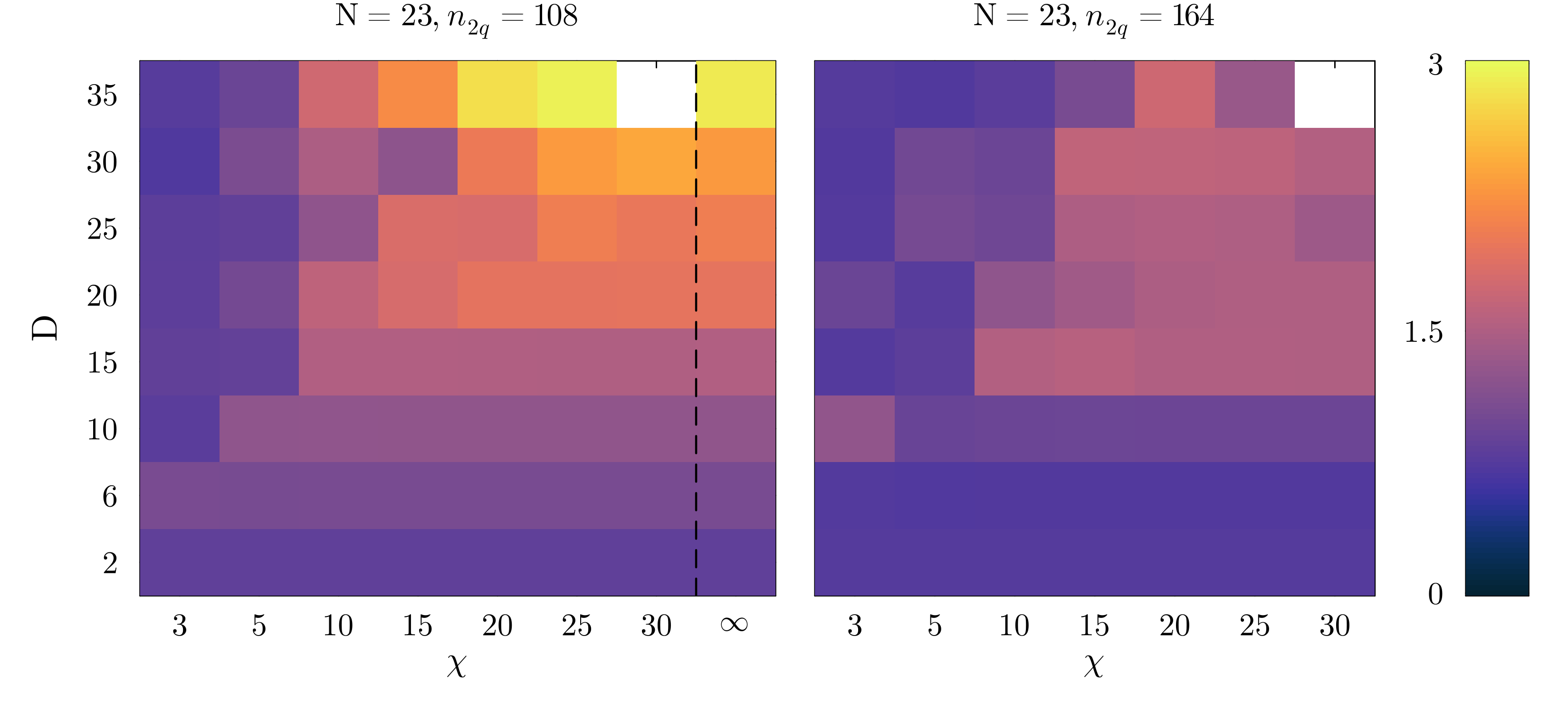}
    \caption{\textbf{SNR for the OTOC experiment using TNBP}. Performance of TNBP with boundary MPS contraction for the simulation of two 23-qubit experiments analyzed in \cite{google2025observation}. The left panel shows simulation of a shallower OTOC circuit with $n_{2q} = 108$ two-qubit gates.  We use a variety of PEPS and bMPS bond dimensions, $D$ and $\chi$, respectively.  The blank space for $D=35$, $\chi=30$ indicates that the simulation with these bond dimensions exceeds 80GB of RAM.  Furthermore, we perform an exact contraction of the final PEPS, equivalent to bMPS contraction with no truncation ($\chi=\infty$), to isolate the effects of truncation to $D$ during time evolution and to check whether bMPS contraction is converged as a function of $\chi$.  In the right panel, we show the same analysis for a deeper circuit with $n_{2q} = 164$ two-qubit gates}
    \label{fig:SNR_heatmap_original_circuit}
\end{figure*}

\begin{figure*}[t]
    \centering
    \includegraphics[width=0.6\linewidth]{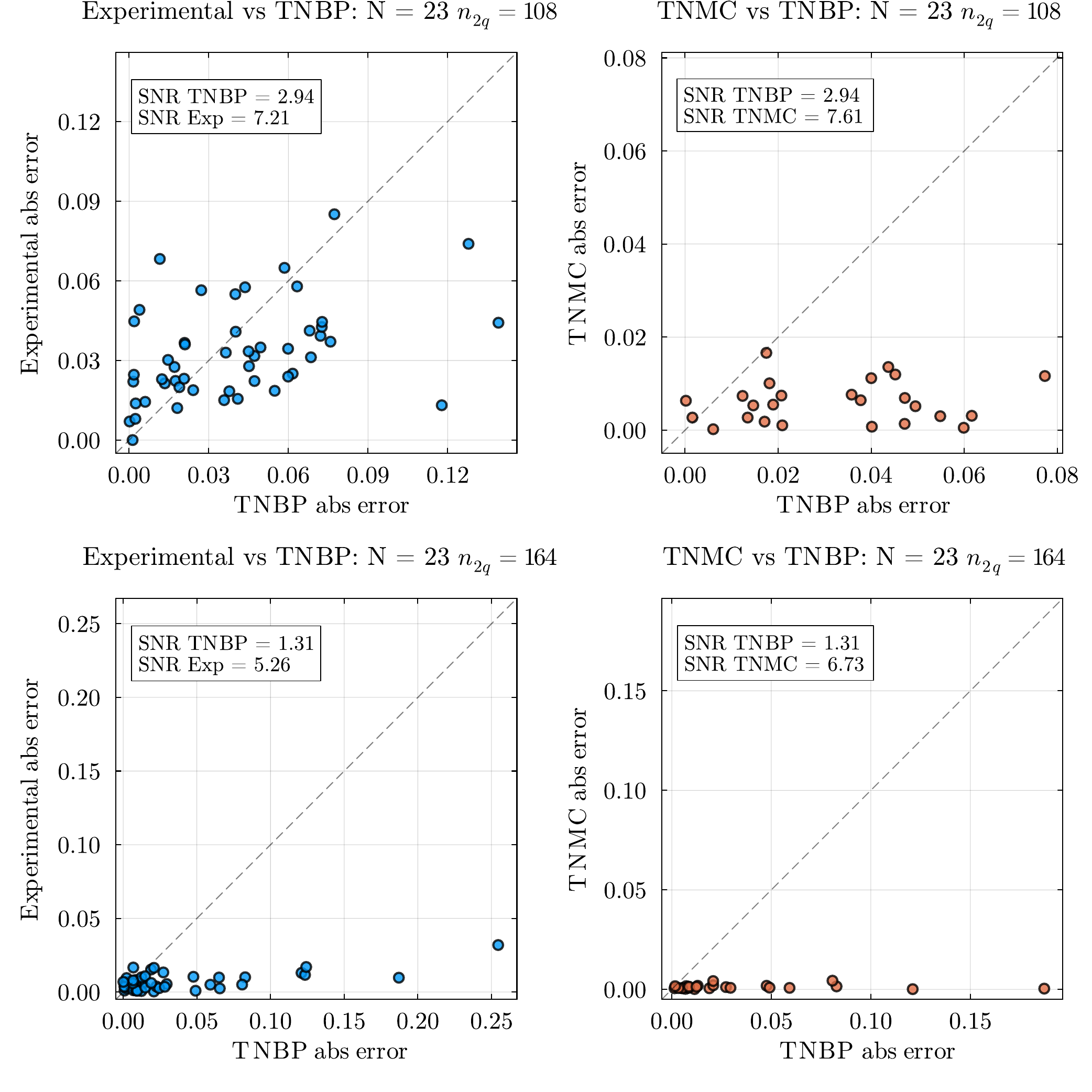}
    \caption{\textbf{Comparison of methods: absolute error OTOC value.}  For each random circuit instance for the shallow (top row) and deep (bottom row) 23-qubit OTOC circuits from~\cite{google2025observation}, we show the OTOC values measured in the experiment and computed using TNBP and TNMC.  We consider 50 circuit instances, each of which was run in the experiment and simulated with TNBP; 25 of them were simulated with TNMC.  For each circuit depth (108 gates vs 164), we make a scatterplot of error in experiment vs error in BP (left column) and of error in TNMC vs error in BP (right column).  Points below the 45$^\circ$ dashed line are instances with larger error for TNBP than for experiment/TNMC.}
    \label{fig:comparison_tnmc_experiment}
\end{figure*}

So far, we have argued based on the incompressibility and geometry of the OTOC circuits that TNBP simulation of both OTOC and OTOC$^{(2)}$ should have cost scaling exponentially in system size.  We demonstrated the expected scaling numerically for 1D systems.  We now turn to empirical demonstration in the more challenging setting of OTOCs in 2D.

Here we consider the 23-qubit circuits from the quantum echoes experiment (\cite{google2025observation} SM, page 29) with 108 and 164 two-qubit gates, respectively.  In both cases, we perform the evolution step with TNBP for a range of PEPS bond dimensions, $D$; for each $D$, we perform the extraction step with bMPS contraction, for a range of MPS bond dimensions $\chi$.  We go up to $D=35$ and $\chi=30$, chosen so that the simulation can be as accurate as possible while fitting within the memory of a single A100 GPU.  These bond dimensions are comparable to the values of $D$ and $\chi$ used in previous state-of-the-art TNBP simulations~\cite{tindall2024efficient, tindall2025dynamics}.  To isolate the effect of the truncation during time evolution, and to understand how close the bMPS contraction is to convergence in $\chi$, for the shorter-depth circuit (108 two-qubit gates) we also contract the final PEPS for each $D$ exactly ($\chi=\infty)$.\footnote{For the larger-depth circuit an exact contraction is not possible on a single n2-highmem-128 CPU node (employed for the exact contraction of 23 qubits and 108 two-qubit gates) due to memory constraints.}  The resulting SNR as a function of $D$ and $\chi$ is shown in Fig.~\ref{fig:SNR_heatmap_original_circuit}.

We find that TNBP achieves a lower SNR than do TNMC and the experiment on the Willow chip.  Even with the largest $D$ and exact ($\chi=\infty$) contraction, for the shorter-depth circuit with 108 two-qubit gates, TNBP achieves an SNR of only around 2.94, in comparison with 7.61 from TNMC (run with the same computational resources as TNBP) and 7.21 in the experiment.  On the deeper circuit, TNBP with the largest $D$ and $\chi$ gives an SNR of just 1.31, in comparison with 6.73 for TNMC and 5.26 for the experiment.  

The lower SNR for TNBP is partially explained by a small but significant fraction of circuit instances for which the method fails dramatically.  In Fig.~\ref{fig:comparison_tnmc_experiment}, we show scatterplots of the error in the OTOC relative to the ideal value computed by an exact simulation for each instance, with TNBP error on the horizontal axis and either experimental or TNMC error on the vertical axis.  Comparing TNBP with the experiment, we see that most instances have slightly larger error for TNBP, but a few have much larger error.  For TNBP with TNMC, most instances have significantly larger error for TNBP.

An important conclusion from these empirical tests is that, compatible with incompressibility of the OTOC circuits, both $D$ and $\chi$ must be large in order to achieve a high SNR.  For the shallower circuit, when the final wavefunction is contracted exactly (column labeled $\chi=\infty$ in Fig.~\ref{fig:SNR_heatmap_original_circuit}), the only approximation is the truncation to $D$ during the time evolution.  Looking at the exact-contraction data, we see that the SNR is still low at $D=35$, meaning that important information in the wavefunction is discarded during truncation after each gate application even with large $D$.  No increase in $\chi$ can compensate---a more accurate extraction of observables is not helpful if the information has already been lost.  
Likewise, the substantial signal that \emph{is} retained in the wavefunction in the case of $D=35$ is subsequently lost when the extraction step is inaccurate.  Moving along each row (with fixed $D$) in Fig.~\ref{fig:SNR_heatmap_original_circuit}, we see that the SNR only reaches its maximum value, so that all relevant information in the wavefunction is captured in the extraction step, when $\chi$ is comparable to $D$.  For the deeper circuit, while memory constraints preclude an exact contraction as a standard of comparison for convergence in $\chi$, we see the same trends: large $D$ is needed for the wavefunction to keep important information about the OTOC, and large $\chi$ is needed to accurately extract that information. 

We note that these results do not definitively show that $D=35$ is insufficient to represent the final state as a PEPS, only that repeated truncation to $D=35$ throughout the time evolution discards too much information from the wavefunction.  To explore this distinction, we also perform time evolution with no truncation for the circuit with 108 2-qubit gates, reaching $D'=72$,\footnote{The final bond dimension is not a power of 2 because numerically vanishing singular values below $10^{-14}$ are removed.} then do a single truncation at the end to smaller $D$.  We find that the final wavefunction contains no further relevant information than if we had truncated to $D$ throughout the evolution.  See App.~\ref{section:final_truncation} for details.

In the scaling argument in Sec.~\ref{subsec:IIB_scaling_theory} above, we assumed that the time evolution is close to incompressible: any significant truncation in $D$ would substantially degrade the accuracy of the final extracted OTOC value.  This assumption was demonstrated to be accurate in 1D in Sec.~\ref{subsec:1D_exp_scaling_empirical}.  We have now provided evidence that the assumption also holds in 2D.  Notably, for the shallower circuit (with $n_{2q}=108$ 2-qubit gates), a numerically exact PEPS simulation with no truncation reaches $D'=72$, and even compression by a factor of 2, to $D=35$, brings the SNR down to below 3. 

In light of this incompressibility, it is not surprising that TNBP performs substantially worse on the deeper circuit ($n_{2q}=164$) than on the shallower circuit ($n_{2q}=108$).  Entanglement builds up with successive circuit layers, requiring larger bond dimensions to capture faithfully.  Thus a fixed bond dimension of $D=35$ leads to less accurate evaluation of the OTOC and a smaller SNR when simulating the deeper circuit.

\subsection{Empirical numerical demonstration in 2D: new circuits with 8 to 28 qubits}
\label{sec:scaling_experiment}

\begin{figure*}[t]
    \centering
    \includegraphics[width=1\linewidth]{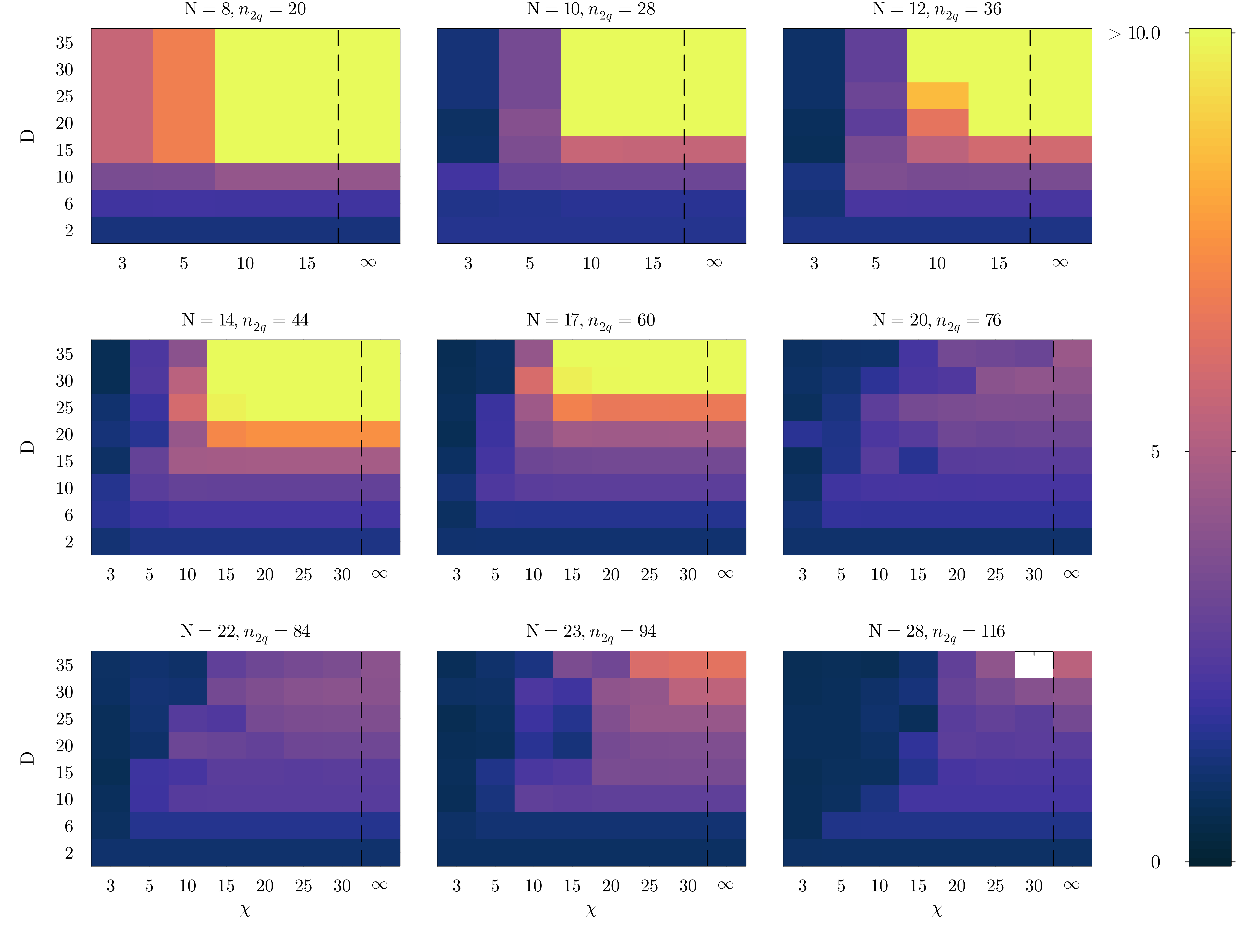}
    \caption{\textbf{SNR with TNBP for the scaling analysis circuits.} We show the same analysis as in Fig.~\ref{fig:SNR_heatmap_original_circuit}, but for newly generated circuits analogous to those from~\cite{google2025observation}, with a range of system sizes and circuit depths.  $D$ is the PEPS bond dimension and controls the accuracy of time evolution; $\chi$ is the boundary MPS bond dimension and controls the accuracy of extraction of the final OTOC observable.  We also perform exact contraction for each system size and PEPS bond dimension $D$, to isolate the effects of truncation during evolution.  The white square for $N=28$ indicates where bMPS contraction would exceed the available memory on one A100 GPU.}
    \label{fig:heatmap_scaling}
\end{figure*}

We have demonstrated the incompressibility of the tensor network representation of $|\phi\rangle$ in 1D (Sec.~\ref{subsec:1D_exp_scaling_empirical}) and for specific small circuits from~\cite{google2025observation} in 2D (Sec.~\ref{sec:comparison with experiment}).  However, even for the shallower of the 2D circuits, with the computational resources corresponding to a single node on Google Cloud we were only able to reach an SNR of around 3, still far from a fully accurate simulation.  We therefore still have an incomplete picture of (in)compressibility in 2D.  We would like to understand what bond dimension $D$ is really needed in 2D for a numerically exact simulation ($\text{SNR}\rightarrow\infty$), and how the SNR responds when we make only a very small truncation to smaller bond dimension.

To this end, we generate new OTOC ensembles, constructed analogously to the ones in~\cite{google2025observation} as described in Sec.~\ref{subsec:circuit_design} above, for a range of small system sizes from $N=8$ qubits to $N=28$ qubits.  For each OTOC ensemble, we find both the SNR and the global wavefunction fidelity as a function of bond dimensions $D$ and $\chi$.  
From our results, we reach two key conclusions:
\begin{enumerate}
    \item[] \textbf{Incompressibility of circuit evolution:} Across all the new ensembles we consistently find that to achieve an accurate value of the OTOC for each instance, and thus a high SNR, $D$ must be large.  In particular, for system sizes up to $N=14$, where we can reach large enough $D$ to achieve a numerically exact evaluation of the OTOC, even a 20\% decrease in bond dimension already brings the SNR down to the order of 5 to 10, confirming that the PEPS is almost completely incompressible.
    
    \item[] \textbf{High cost of OTOC extraction:} Additionally, we confirm across a range of system sizes that even if $D$ is large enough to achieve an accurate PEPS representation of the state $|\phi\rangle$, extracting an accurate value of the OTOC from the PEPS requires a large boundary MPS bond dimension $\chi$.  This makes the total simulation cost at least a high-degree polynomial in $D$.  
    
\end{enumerate}

Because we study a range of system sizes, one might expect that we could also numerically observe the scaling of cost with system size to confirm the exponential scaling predicted in Sec.~\ref{subsec:IIB_scaling_theory}.  For completeness, we show the results of such a scaling analysis in App.~\ref{app:empirical_scaling_2D}.  However, even at the scale of $N=28$ qubits the simulation cost is dominated by finite size effects, so we are not able to see the expected exponential scaling.  See the appendix for details.  In the appendix, we also discuss the observed relationship between SNR and global wavefunction fidelity.

\vspace{0.3cm}

\paragraph{Measured SNR vs $D$ and $\chi$:} 
We begin by computing SNR vs $D$ and $\chi$ for OTOC circuits with a number of qubits ranging from 8 to 28, each with 50 random instances.  The precise circuit construction is described in Sec.~\ref{subsec:circuit_design}, with additional details in App.~\ref{section:obtain_lightcone}. 
The results are shown in Fig.~\ref{fig:heatmap_scaling}.  The highest SNR achieved for each circuit layout, when restricting to $D\leq 35$ and $\chi\leq 30$, along with computation time and the mean and standard deviation of the OTOC across instances, is also printed in Table~\ref{table:OTOC_scaling}; computation times correspond to simulation on c2-standard-60 CPU nodes on Google Cloud.  For the circuits with $N\leq 14$, we are able to reach a perfect simulation giving numerically exact values for the OTOC. For $N = 17$ we achieve high SNR ($\text{SNR}>10$).  Starting from $N=20$, however, we can only reach SNR of around 5 for the largest accessible $D$ and $\chi$, or even with a large $D$ followed by exact contraction to extract the observable of interest.

\begin{table}[h]
\centering
\begin{tabular}{| c | c c c c c c |}
\hline
N & $\overline{\mathrm{OTOC}}$ & $\sigma(\mathrm{OTOC})$ & $D$ & $\chi$ & $t_{\mathrm{bMPS}}^{m} (h)$ & SNR  \\
\hline
8 & -0.052 & 0.241 & 14 & 15 & 0.02 & Exact \\
10 & 0.053 & 0.203 & 22 & 15 & 0.02 & Exact \\ 
12 & 0.019 & 0.202 & 24 & 17 & 0.03 & Exact \\
14 & 0.067 & 0.199 & 35 & 30 & 0.05 & Exact \\
17 & 0.055 & 0.113 & 35 & 30 & 0.77 & 16.52 \\
20 & 0.051 & 0.093 & 35 & 30 & 7.40 & 3.06\\
22 & 0.100 & 0.087 & 35 & 30 & 15.28 & 3.56\\
23 & 0.056 & 0.069 & 35 & 30 & 61.49 & 6.28\\
28 & 0.085 & 0.074 & 35 & 25 & 76.24 & 4.13 \\
\hline
\end{tabular}
\caption{\textbf{Summary of scaling analysis}. We show the bMPS computation time and SNR for different system sizes at the maximum values of $D$ and $\chi$ employed for each system size in the scaling analysis. Here, $t^m_{bMPS}(h)$ stands for median time, in hours, across 50 different instances.
We also include the mean value of the OTOC and its standard deviation over 50 instances. Contractions with bMPS were run on c2-standard-60 CPUs on Google Cloud. ``Exact'' means that the OTOC in each instance was computed accurately to single precision accuracy, with errors less than 10$^{-6}$.
\label{table:OTOC_scaling}
} 
\end{table}

\begin{figure}[h]
    \centering
    \includegraphics[width=1\linewidth]{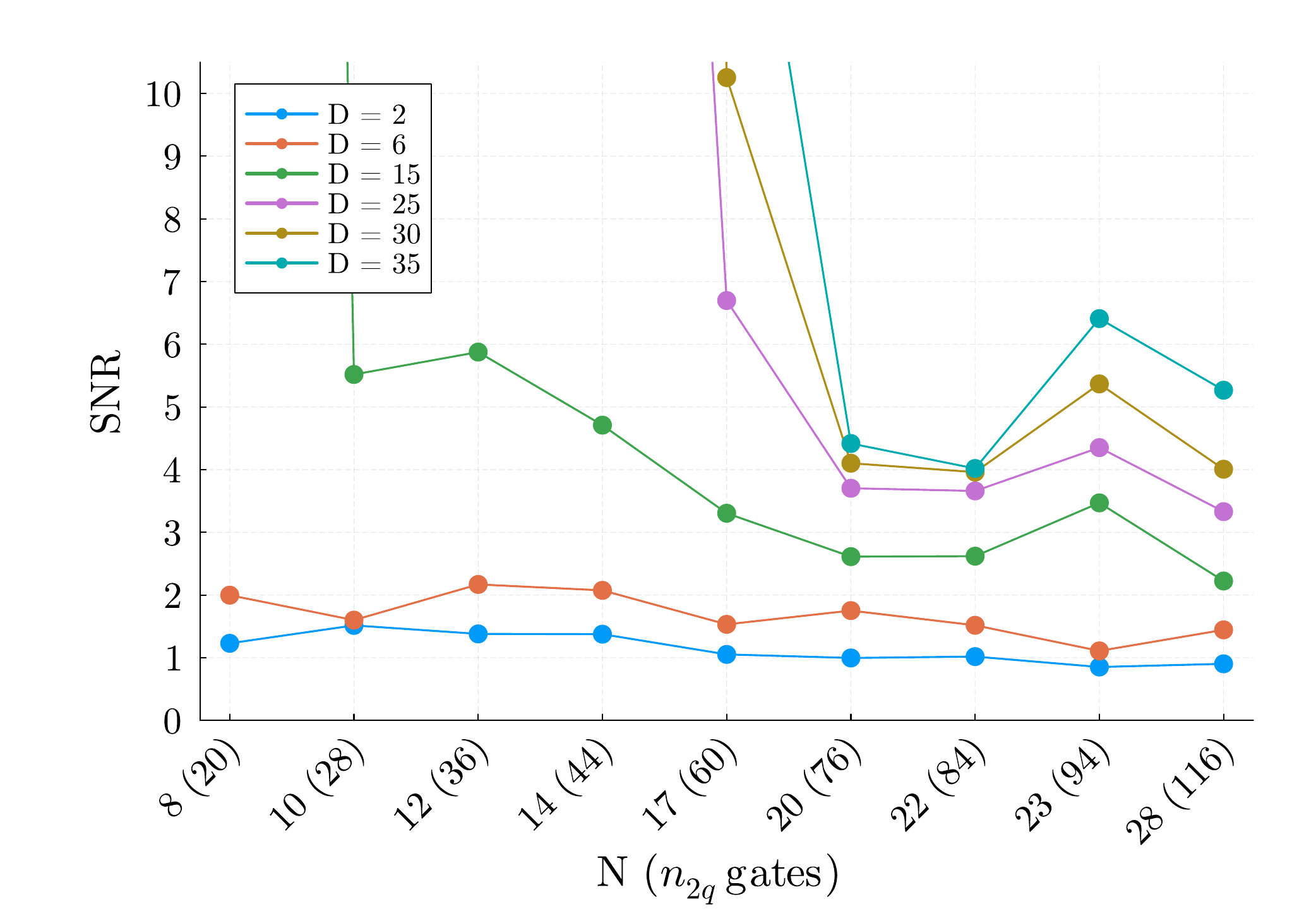}
    \caption{\textbf{SNR as a function of system size and $D$}. We isolate the effect of truncation during time evolution by performing an exact contraction of the final PEPS to extract the value of the OTOC.  Thus a numerically exact time evolution would give an infinite SNR.  In contrast, we find that even with large PEPS bond dimensions up to $35$, we still fail to achieve high SNR for moderate system sizes.  Each point on the $x$-axis is labeled by the number of two-qubit gates, with the number of qubits in parentheses.}
    \label{fig:SNR_vs_N}
\end{figure}

\paragraph{Isolating $D$ and $\chi$ approximations:}
As in our study of the 23-qubit circuits from the echoes experiment, we observe that both $D$ and $\chi$ must be large to accurately estimate the OTOC and get a high SNR.  To see this more clearly, we can isolate the two approximations.  First, we consider finite $D$ but with exact contraction, so that we only make approximations in the evolution step and not in the extraction step.  The resulting SNR as a function of number of two-qubit gates (and therefore of system size) and PEPS bond dimension $D$ is shown in Fig.~\ref{fig:SNR_vs_N}.  We see that even with perfect extraction of information, $D$ must be large and generally increasing with system size in order to achieve a high SNR.\footnote{We observe a non-monotonic trend in complexity vs system size.  The non-monotonicity is a result of the irregularly shaped circuit layouts produced by the construction described in App.~\ref{section:obtain_lightcone}.}  When $D$ is too small, relevant information for the final observable has already been lost to truncation during the evolution step and cannot be recovered during the extraction step, even if the extraction is exact.

We also check again whether a small $D$ could be sufficient to represent the final wavefunction faithfully enough to achieve a high SNR, if we were to truncate only once at the end of the evolution rather than after applying each gate.  As with the circuits from the experiment, we find that the problem with small $D$ is \emph{not} the repeated truncations, but rather that the final wavefunction itself requires a larger $D$.  See App.~\ref{section:final_truncation} for details.

\begin{figure}[t]
    \centering
    \includegraphics[width=0.9\linewidth]{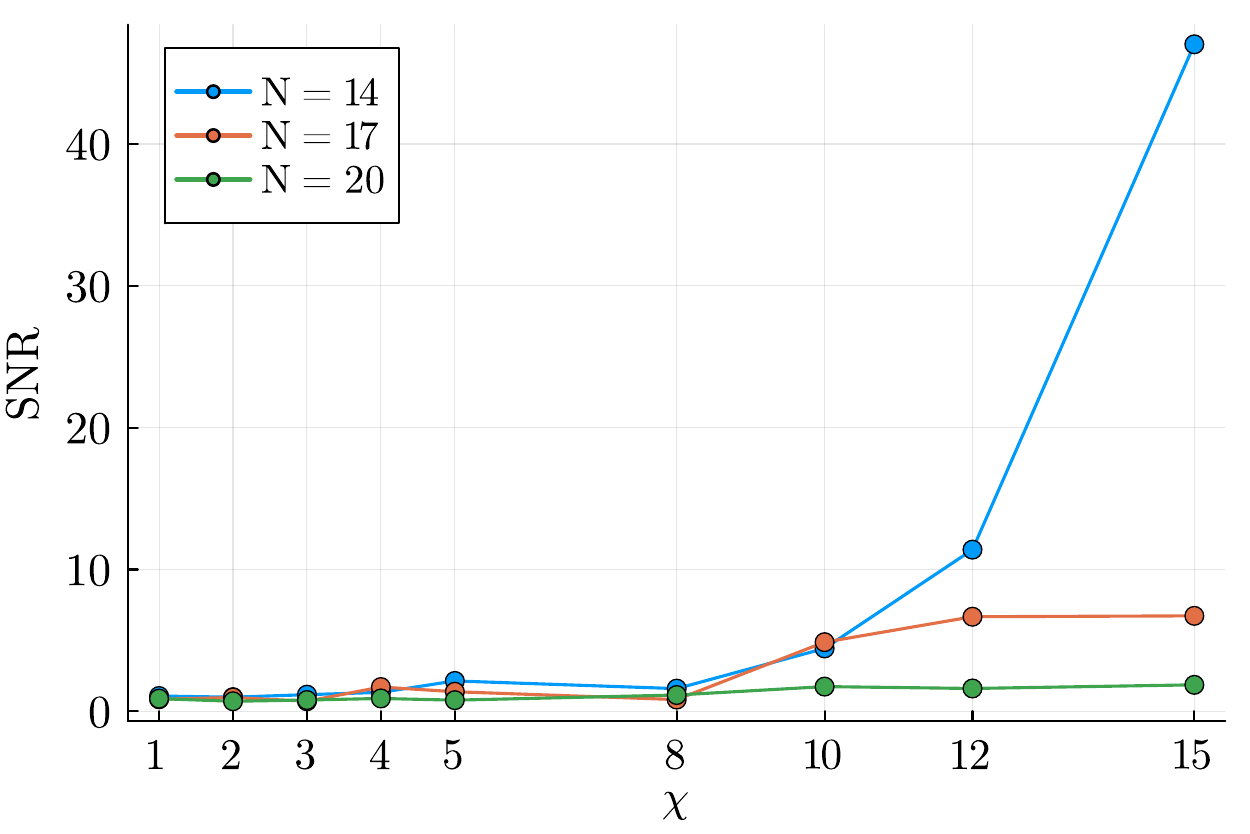}
    \caption{\textbf{SNR for \text{N}}$\boldsymbol{= 14,17,20}$ \textbf{at maximum bond dimension \text{D} $\boldsymbol{= 64}$}. We isolate the effect of approximate boundary MPS contraction on SNR by performing the PEPS time evolution with no truncation.  Exact contraction of the final PEPS would then give a numerically exact result and infinite SNR.  In contrast, a large $\chi$ is needed in the bMPS contraction in order to get a high SNR, even on small system sizes.  Here we limit $N$ to be at most 20 and $\chi=15$.}
    \label{fig:SNR_exact_time_ev}
\end{figure}

A large $\chi$ is also needed.  Even if we make no approximation during the time evolution, forgoing BP truncations and allowing $D$ to grow exponentially in time so that no information is lost in the final wavefunction, a small $\chi$ leads to low SNR.  This is shown for $N=14$, 17, and 20 in Fig.~\ref{fig:SNR_exact_time_ev}.  In all three cases, $D$ is allowed to grow to its maximum value of 64 for these circuits, and then information is extracted using bMPS contraction with different values of $\chi$.  For $\chi\lesssim 8$, there is no significant signal, and only when $\chi$ is a significant fraction of $D$ do we see SNRs on the scale of 10 or larger.  An exact time evolution cannot compensate for an insufficiently expressive boundary MPS contraction.

\subsection{Circuits with modified entangling gates}
\label{sec:modified_gates}

\begin{figure*}[t]
    \centering
\includegraphics[width=1\linewidth]{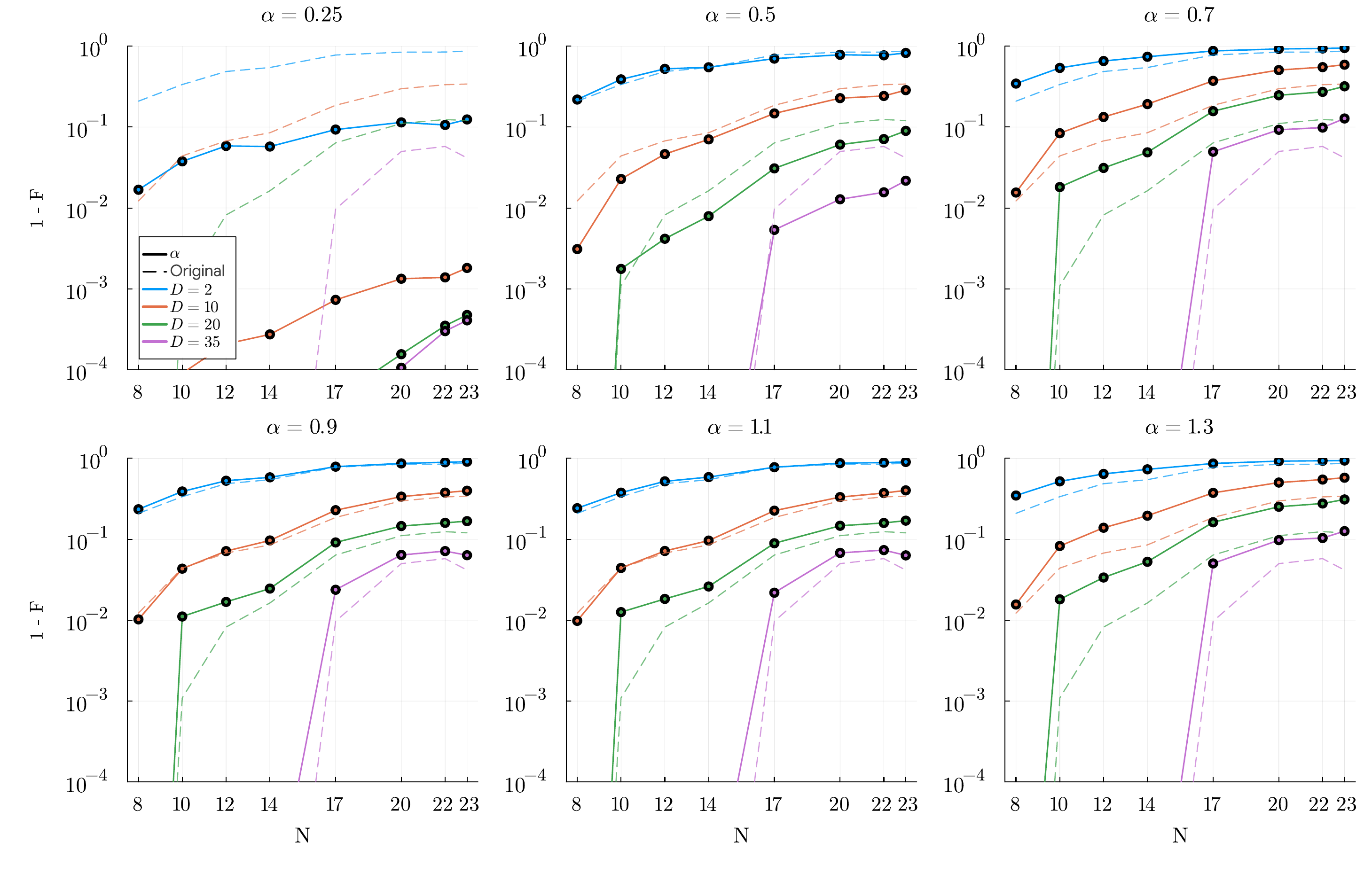}
    \caption{\textbf{Dependence of computational difficulty on rate of entanglement growth.}  We replace the $\iSWAP$-like entangling gates from the original OTOC circuits with modified gates \(\iSWAP(\alpha) = e^{\alpha \frac{\pi}{4}(XX + YY)}\).  The original gates have $\alpha=1$; non-entangling gates would have $\alpha=0$.  For small $\alpha$, we find that for every bond dimension (different line colors), the TNBP simulation fidelity is much higher for the modified, slower-entangling, circuits (solid lines) than for the original circuits with the same gate layout (dashed lines).  This advantage disappears for larger $\alpha$, and some values lead to circuits harder to simulate than the original ones.  For both the original and modified circuits, we compute fidelity by an exact contraction of the final wavefunction.}
    \label{fig:fidelity_alphas}
\end{figure*}

We have demonstrated that simulating the OTOC to high accuracy with TNBP has high computational cost, due to needing large bond dimensions $D$ and $\chi$ that grow rapidly with system size.  We furthermore argued that the large bond dimensions required follow from incompressibility of the circuit evolution.  Incompressibility for tensor network states is equivalent to the generation of large amounts of volume law entanglement.  

We therefore expect that, if the iSWAP-like entangling gates in $U$ and $U^\dagger$ in the OTOC circuits are replaced by more weakly-entangling gates, the evolution \emph{would} be compressible, and the random-circuit OTOCs could be efficiently simulated with TNBP. We now show that this is indeed the case.

We construct new OTOC circuits exactly as before, except we replace the $\iSWAP$-like entangling gate from the original experiment with a tunable $\iSWAP$ gate of the form
\(\iSWAP(\alpha) = e^{i\alpha \frac{\pi}{4}(XX + YY)}\).
The prefactor \(\alpha\) controls the rate of entanglement, with $\alpha=1$ giving a standard $\iSWAP$ gate.  Small values of $\alpha$ lead to slow entanglement growth.

For this numerical experiment, we evaluate the accuracy of TNBP simulation using the global wavefunction fidelity.  If the evolution is very compressible, TNBP should be able to return not just an accurate value for a local observable, but also the correct final state, efficiently.  
The global wavefunction infidelity is shown as a function of $\alpha$, $N$, and $D$ in Fig.~\ref{fig:fidelity_alphas}, averaged over $m=50$ instances in each case.  As in Sec.~\ref{sec:scaling_experiment}, fidelity is computed by exactly contracting ($\chi=\infty)$ the final PEPS after time evolution, then comparing the resulting state vector with the results of an exact simulation of the circuit.  

As can be seen in the top left panel of Fig.~\ref{fig:fidelity_alphas}, with $\alpha=0.25$, a slower rate of entanglement and scrambling leads to efficient TNBP simulations.  A relatively small bond dimension $D$ suffices to achieve high fidelity, even for some of the larger system sizes.  Starting from $\alpha$ around 0.5, the modified circuits are approximately as difficult to simulate with TNBP as the original OTOC circuits with $\alpha=1$.  Surprisingly, the $\alpha=0.7$ and 1.3 cases are actually harder than the original circuits to simulate with TNBP.

In short, we confirm that the precise choice of entangling gate, namely the fast-entangling $\iSWAP$ gate, is essential to the difficulty of simulating random OTOC circuits with TNBP, and we verify that the method \emph{can} be efficient for simulating OTOC circuits with more weakly scrambling time evolution.  This numerical experiment provides evidence that the observed incompressibility in the other 2D numerical experiments (Sec.~\ref{sec:comparison with experiment} and \ref{sec:scaling_experiment}) is indeed a feature of the OTOC circuits themselves, rather than being a failure of the TNBP method.

\subsection{TNBP cannot simulate the largest OTOC and OTOC$^{(2)}$ circuits from the echoes experiment} 
\label{subsec:Frontier_estimates}

We now argue that TNBP cannot simulate the most challenging experiments from~\cite{google2025observation}, OTOC on 95 qubits and OTOC$^{(2)}$ on 65 qubits.  We make two arguments, either of which is sufficient on its own.

\paragraph{PEPS bond dimensions are too large for any feasible simulation:} 
The circuits run in the experiments do not precisely follow the construction described in Sec.~\ref{subsec:circuit_design} and App.~\ref{section:obtain_lightcone}, with a single-site measurement operator and a single-site butterfly operator located so that each lies on the edge of the other's physical lightcone.  Therefore, the scaling of bond dimension with $N$ that we derived in Sec.~\ref{subsec:IIB_scaling_theory} does not directly apply.

However, we can keep one of the key underlying assumptions of that analysis: for evolution by a circuit that cannot be compressed, the required bond dimension for an accurate tensor network simulations scales as 4 raised to the power of the largest number of iSWAP-like entangling gates applied on any individual bond in the lattice.  Furthermore, we have provided extensive evidence in both 1D and 2D, including for 23-qubit circuits taken directly from the experiment, that the random circuit evolution in these OTOC circuits is indeed incompressible in at least some parts of the system.

Thus we can estimate the required bond dimension for TNBP simulation by simply counting the number of gates applied to each bond in the circuits in the experiment.  In the $N=95$ OTOC experiment, the largest number of entangling gates on any given bond is 8, giving an estimated bond dimension of $4^8 \approx 6.4\times 10^4$.  In the $N=65$ OTOC$^{(2)}$ experiment, the largest number of gates on any given bond is 12, for an estimated bond dimension of $4^{12} \approx 1.7\times 10^7$.  

However, these are slight overestimates for two main reasons. (1) The first layer of gates only increases the bond dimension by a factor of 2, rather than 4, since they are applied to the all-0 state. (2) If our goal is to reproduce the experiment, we do not need a numerically exact value for the OTOC.  Instead, we only need to match the estimated SNR achieved by the experiment, on the order of 3.5.  The latter effect is taken into account in our empirical study of cost scaling in 1D, where we find that for an SNR of 5, the required bond dimension is reduced relative to $2^\text{[\# bonds]}$ by a factor of approximately $0.1$.  In a large 2D system, the bond dimension will be reduced less for the same overall wavefunction fidelity.  This is because the overall fidelity is approximately a product of the fidelity from bond dimension truncation on each bond in the lattice, and a large 2D system has more bonds.  We expect that, likewise, achieving a certain target SNR in 2D will allow less reduction in bond dimension than achieving the same target in 1D.  Thus we can lower-bound the true bond dimension required to achieve accuracy comparable to that of the experiment if we multiply our bond dimension estimates by a factor of about $1/10$.  Also taking into account the factor of 1/2 from (1), we find required bond dimensions of $D\approx 3.2 \times 10^3$ for OTOC on $N=95$ qubits, and $D\approx 8.4 \times 10^5$ for OTOC$^{(2)}$ on $N=65$ qubits.

These bond dimensions are far beyond what is practical in PEPS simulation, so it is already clear that reproducing the experiments is out of reach for any realistic TNBP simulation.  However, it is also informative to see whether such a simulation is beyond classical even if we had access to the largest classical computing systems in the world.  To this end, we consider the memory requirements for TNBP relative to the available resources on the full Frontier supercomputer.

A single PEPS tensor with $D\approx 8.4 \times 10^5$ contains $2D^4$ complex numbers, and in single precision each uses 8 bytes.  This gives a total memory requirement for a single tensor for OTOC$^{(2)}$ on 65 qubits of $7.9\times 10^{24}$ bytes, or more than $10^9$ petabytes.  This is orders of magnitude larger than the full storage space on Frontier, about 700 petabytes, and in fact is larger than the storage needed for the full 65-qubit wavefunction.  Even for OTOC on 95 qubits, with $D\approx 3.2 \times 10^3$, a single tensor uses 1.8 petabytes.  While this is in principle possible to store even in RAM on Frontier, the larger tensors that appear during the extraction step again exceed the cluster's total storage space.  

\paragraph{TNBP is more expensive than optimized exact contraction:}
It is generally true that in cases where compression is not possible, tensor network state methods are simply more expensive than full state vector simulation.  For example, an MPS representing a typical Haar-random state on $N$ qubits has bond dimension $2^{N/2}$ at the center of the system.  The cost of contracting the MPS then scales as $\mathcal{O}(2^{3N/2})$, higher than the $\mathcal{O}(2^N)$ cost for performing operations on the full state-vector representation of the same state.  

The same is true in two dimensions.  If the circuit for $|\phi\rangle$ were completely incompressible, the cost of computing the state by approximate PEPS evolution with BP-based truncations, as in TNBP, would be strictly higher than the cost of evolving the state exactly.  In turn, the cost of finding $|\phi\rangle$ exactly, then computing the expectation value $\langle\phi|M|\phi\rangle$, is strictly higher than performing an optimized exact contraction of the full tensor network for $\langle\phi|M|\phi\rangle$.  This exact contraction is precisely what was found using TNCO in~\cite{google2025observation} to be more than 13,000$\times$ slower to compute on the Frontier supercomputer than to measure on the Willow chip.  

One possible flaw in this argument is that, as discussed in Sec.~\ref{subsec:scaling_with_compressibility} above, the OTOC circuits actually \emph{are} somewhat compressible.  Specifically, the gates outside the dressed physical lightcones of $B$ and $M$ have little effect on the final value of the OTOC and the corresponding parts of the circuit can be compressed during TNBP simulation.  

Taking this compressibility into account, we can still use exact contraction costs to lower bound the cost of TNBP simulation.  Specifically, the most extreme version of taking into account compressibility in certain parts of the circuit would be to simply remove the gates in the compressible regions from the circuit.  The cost of TNBP simulation of the full circuit, with compressibility taken into account, is higher than the cost of TNBP simulation of the circuit with the gates in the compressible parts of the circuit removed entirely.  That simulation has higher computational cost than exact contraction of the circuit with those gates removed. 
Finally, the strategy of removing gates from the circuit to reduce cost, while maintaining an accurate evaluation of the OTOC or OTOC$^{(2)}$, was already tried in~\cite{google2025observation}.  The conclusion there was that exact contraction with gate removal, if keeping enough gates to achieve SNR comparable to what was achieved in the experiment, would still take much longer on the full Frontier supercomputer than the time spent on the experiment.  

In short, TNBP simulation has higher computational cost than exact contraction with gate removal, which was already found to be more expensive than the experiment on the Willow chip.  
We thus conclude that the largest experiments from~\cite{google2025observation} are indeed beyond the capabilities of TNBP simulation, or indeed of any classical PEPS-evolution-based simulation in the Schr\"{o}dinger picture, with any feasible level of computational resources.  


\section{Conclusions}
\label{sec:conclusions}

We have demonstrated that the random-circuit OTOC measured in Google's quantum echoes experiment~\cite{google2025observation} cannot be feasibly simulated using tensor networks with belief propagation, specifically with the Schr\"{o}dinger picture approach described in Sec.~\ref{sec:bp}.  Even for relatively small system sizes, with just 23 qubits, TNBP already fails for the same PEPS and bMPS bond dimensions, $D$ and $\chi$, used to successfully simulate other state-of-the-art quantum experiments~\cite{tindall2024efficient,tindall2025dynamics}.  For the 23-qubit OTOC circuits, TNBP underperforms both the experiment and another classical method demonstrated in \cite{google2025observation}, TNMC.

To address the feasibility of simulating much larger systems, including the 95-qubit OTOC experiment and 65-qubit OTOC$^{(2)}$ experiment in~\cite{google2025observation}, we argued based on incompressibility of highly entangling random evolution and based on the geometry of OTOC circuits that the cost of TNBP simulation is exponential in linear system size.  While portions of the OTOC circuits that lie outside of the physical lightcones of the measurement and butterfly operators actually \emph{can} be compressed, we argued that this compression does not affect the exponential scaling.  
We observed precisely the predicted exponential scaling in numerical simulations in 1D, while in 2D we confirmed the incompressibility of the OTOC circuits both for the aforementioned 23-qubit circuits from~\cite{google2025observation} and from analogous OTOC circuits we constructed on a range of system sizes up to 28 qubits.  Additionally, we showed that when the 2D circuits have their highly entangling (iSWAP-like) 2-qubit gates replaced with weakly entangling gates, the circuit evolution \emph{is} compressible, and TNBP can efficiently simulate the circuits and can accurately compute the OTOC with low computational cost; we therefore confirm that TNBP \emph{can} exploit compressibility when it is present in a circuit.

From the finding of incompressibility, we conclude that an accurate PEPS representation of the output state of the circuit run in the 65-qubit OTOC$^{(2)}$ experiment requires a bond dimension on the order of $4^{12}$, well over a million.  With this bond dimension, a single tensor would already take up more storage space than is available on the entirety of the Frontier supercomputer.  Furthermore, for incompressible circuits, approximate tensor network state-based simulations are more computationally expensive than exact tensor network contraction, and it was already shown in~\cite{google2025observation} that exact contraction takes far longer for this circuit than does the experiment on the Willow chip.  Thus we conclude that TNBP does not challenge the claim that quantum echoes experiment is beyond classical.

Here we have primarily focused on the TNBP method as outlined in Sec.~\ref{sec:bp}.  It is natural to ask the extent to which our key result, that TNBP cannot feasibly simulate the quantum echoes experiment, also holds for related tensor network-based simulation methods.  If the evolution step in TNBP is replaced by PEPS evolution with a different truncation scheme, such as full update, simple update, or truncations based on BP with loop corrections~\cite{evenbly2024loop,
gray2025tensor, midha2025belief, midha2026belief}, we expect that simulation of the largest OTOC experiments will remain infeasible.  Our argument for exponential cost scaling (in Sec.~\ref{subsec:IIB_scaling_theory}) rests only on incompressibility of the circuit (in the region inside the physical lightcones) and is independent of the algorithm used to attempt such a compression.  

On the other hand, our arguments do not directly apply to simulations in the Heisenberg picture, in which we evolve the $M$ operator, the $B$ operator, or both.  
As we argued in App.~\ref{sec:Heis_picture}, if we evolve just one of these operators (\textit{e.g.} $B$) in the Heisenberg picture, the scaling remains exponential in linear system size, and the cost of the simulations is even higher than in the Schr\"{o}dinger picture due to evolving in a doubled Hilbert space.  
On the other hand, we show that a weaker exponential scaling, with a square root of the linear system size, is possible for a simulation that evolves both $M$ and $B$ in the Heisenberg picture.  Further investigation of this approach, including numerical investigations of the simulation cost in practice, is a promising avenue for future work.

\section*{Acknowledgements}
We thank Sergio Boixo, Paolo Braccia, Kostya Kechedzhi, Salvatore Mandr\`a, Manuel Rudolph, Vadim Smelyanskiy, and Adam Zalcman for helpful discussions and ideas.  P.B. acknowledges constant support from Donostia International Physics Center.

\begin{figure*}[t]
    \centering
    \includegraphics[width=\linewidth]{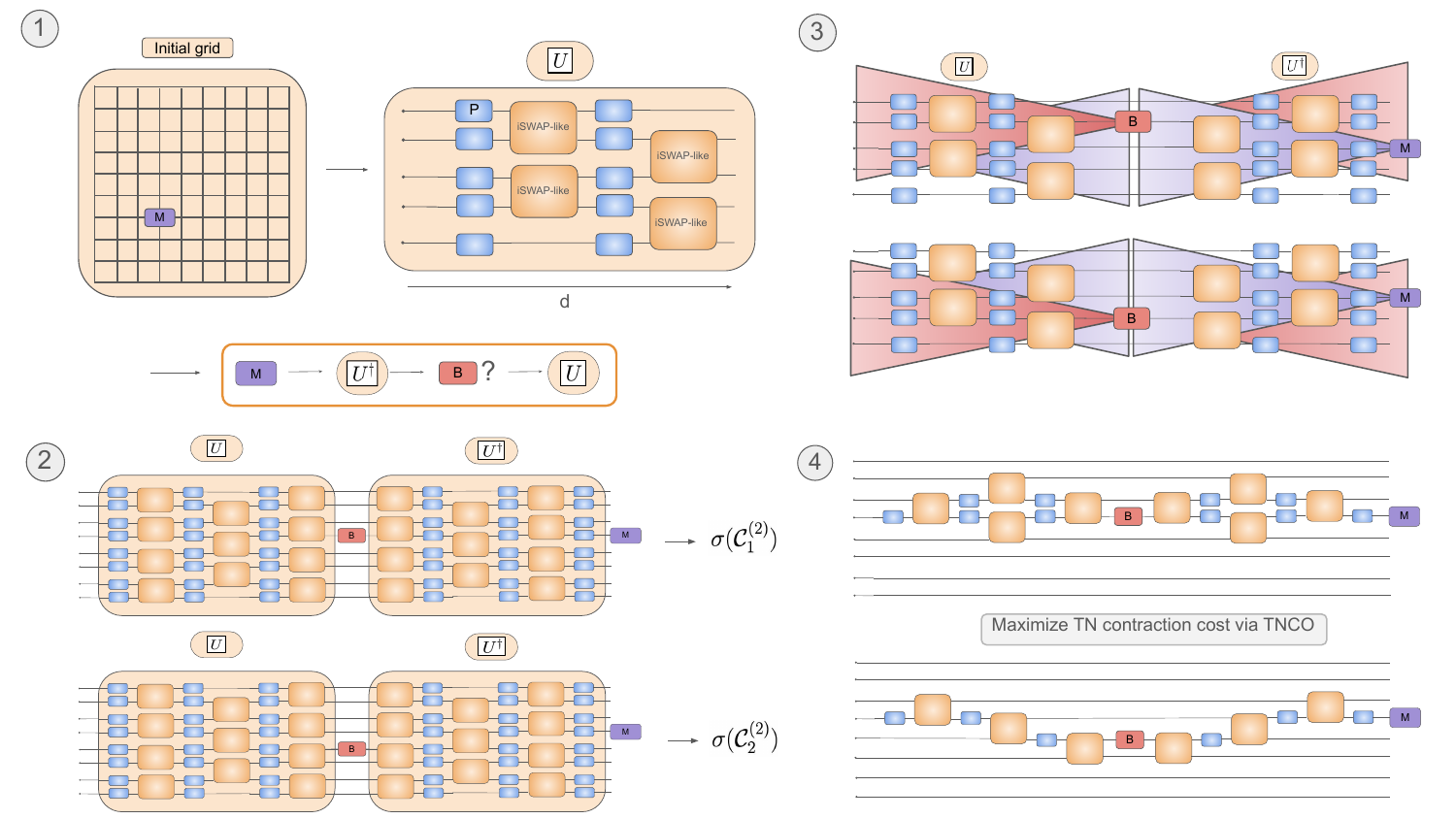}
    \caption{\textbf{Procedure to generate the light cones simulated in the scaling analysis}.  
1) Fix a $10 \times 10$ grid and place $M$, then choose the layout for random time-evolution circuits $U$ in order to define an OTOC ensemble as a function of the position of $B$.  
2) Compute $\sigma(\mathcal{C}^{(2)})$ for all possible positions of $B$, and select the locations of $B$ with $\sigma$ above 30\% of the maximum.  
3) Simplify circuits by pruning gates, exploiting gate cancellations in $U B U^\dagger$ and removing gates outside the lightcone of $M$.  
4) Select the location of $B$ that gives the pruned circuit with the most challenging TN contraction based on TNCO.}
    \label{fig:lightcone}
\end{figure*}

\bibliography{biblio}

@article{google2025observation,
  author={{Google Quantum AI and Collaborators}},
  title={Observation of constructive interference at the edge of quantum ergodicity},
  journal={Nature},
  volume={646},
  number={8086},
  pages={825--830},
  year={2025},
  publisher={Nature Publishing Group UK London},
  url = {https://www.nature.com/articles/s41586-025-09526-6}
}

@misc{OTOC_NMR,
      title={Quantum computation of molecular geometry via many-body nuclear spin echoes}, 
      author={C. Zhang and R. G. Cortiñas and others},
      fullauthorlist={C. Zhang and R. G. Cortiñas and A. H. Karamlou and N. Noll and J. Provazza and J. Bausch and S. Shirobokov and A. White and M. Claassen and S. H. Kang and A. W. Senior and N. Tomašev and J. Gross and K. Lee and T. Schuster and W. J. Huggins and H. Celik and A. Greene and B. Kozlovskii and F. J. H. Heras and A. Bengtsson and A. Grajales Dau and I. Drozdov and B. Ying and W. Livingstone and V. Sivak and N. Yosri and C. Quintana and D. Abanin and A. Abbas and R. Acharya and L. Aghababaie Beni and G. Aigeldinger and R. Alcaraz and S. Alcaraz and T. I. Andersen and M. Ansmann and F. Arute and K. Arya and W. Askew and N. Astrakhantsev and J. Atalaya and B. Ballard and J. C. Bardin and H. Bates and M. Bigdeli Karimi and A. Bilmes and S. Bilodeau and F. Borjans and A. Bourassa and J. Bovaird and D. Bowers and L. Brill and P. Brooks and M. Broughton and D. A. Browne and B. Buchea and B. B. Buckley and T. Burger and B. Burkett and J. Busnaina and N. Bushnell and A. Cabrera and J. Campero and H. -S. Chang and S. Chen and Z. Chen and B. Chiaro and L. -Y. Chih and A. Y. Cleland and B. Cochrane and M. Cockrell and J. Cogan and R. Collins and P. Conner and H. Cook and W. Courtney and A. L. Crook and B. Curtin and S. Das and M. Damyanov and D. M. Debroy and L. De Lorenzo and S. Demura and L. B. De Rose and A. Di Paolo and P. Donohoe and A. Dunsworth and V. Ehimhen and A. Eickbusch and A. M. Elbag and L. Ella and M. Elzouka and D. Enriquez and C. Erickson and V. S. Ferreira and M. Flores and L. Flores Burgos and E. Forati and J. Ford and A. G. Fowler and B. Foxen and M. Fukami and A. W. L. Fung and L. Fuste and S. Ganjam and G. Garcia and C. Garrick and R. Gasca and H. Gehring and R. Geiger and É. Genois and W. Giang and C. Gidney and D. Gilboa and J. E. Goeders and E. C. Gonzales and R. Gosula and S. J. de Graaf and D. Graumann and J. Grebel and J. Guerrero and J. D. Guimarães and T. Ha and S. Habegger and T. Hadick and A. Hadjikhani and M. P. Harrigan and S. D. Harrington and J. Hartshorn and S. Heslin and P. Heu and O. Higgott and R. Hiltermann and J. Hilton and H. -Y. Huang and M. Hucka and C. Hudspeth and A. Huff and E. Jeffrey and S. Jevons and Z. Jiang and X. Jin and C. Joshi and P. Juhas and A. Kabel and H. Kang and K. Kang and R. Kaufman and K. Kechedzhi and T. Khattar and M. Khezri and S. Kim and R. King and O. Kiss and P. V. Klimov and C. M. Knaut and B. Kobrin and F. Kostritsa and J. M. Kreikebaum and R. Kudo and B. Kueffler and A. Kumar and V. D. Kurilovich and V. Kutsko and N. Lacroix and D. Landhuis and T. Lange-Dei and B. W. Langley and P. Laptev and K. -M. Lau and L. Le Guevel and J. Ledford and J. Lee and B. J. Lester and W. Leung and L. Li and W. Y. Li and M. Li and A. T. Lill and M. T. Lloyd and A. Locharla and D. Lundahl and A. Lunt and S. Madhuk and A. Maiti and A. Maloney and S. Mandra and L. S. Martin and O. Martin and E. Mascot and P. Masih Das and D. Maslov and M. Mathews and C. Maxfield and J. R. McClean and M. McEwen and S. Meeks and K. C. Miao and R. Molavi and S. Molina and S. Montazeri and C. Neill and M. Newman and A. Nguyen and M. Nguyen and C. -H. Ni and M. Y. Niu and L. Oas and R. Orosco and K. Ottosson and A. Pagano and S. Peek and D. Peterson and A. Pizzuto and E. Portoles and R. Potter and O. Pritchard and M. Qian and A. Ranadive and M. J. Reagor and R. Resnick and D. M. Rhodes and D. Riley and G. Roberts and R. Rodriguez and E. Ropes and E. Rosenberg and E. Rosenfeld and D. Rosenstock and E. Rossi and D. A. Rower and M. S. Rudolph and R. Salazar and K. Sankaragomathi and M. C. Sarihan and K. J. Satzinger and M. Schaefer and S. Schroeder and H. F. Schurkus and A. Shahingohar and M. J. Shearn and A. Shorter and N. Shutty and V. Shvarts and S. Small and W. C. Smith and D. A. Sobel and R. D. Somma and B. Spells and S. Springer and G. Sterling and J. Suchard and A. Szasz and A. Sztein and M. Taylor and J. P. Thiruraman and D. Thor and D. Timucin and E. Tomita and A. Torres and M. M. Torunbalci and H. Tran and A. Vaishnav and J. Vargas and S. Vdovichev and G. Vidal and C. Vollgraff Heidweiller and M. Voorhees and S. Waltman and J. Waltz and S. X. Wang and B. Ware and J. D. Watson and Y. Wei and T. Weidel and T. White and K. Wong and B. W. K. Woo and C. J. Wood and M. Woodson and C. Xing and Z. J. Yao and P. Yeh and J. Yoo and E. Young and G. Young and A. Zalcman and R. Zhang and Y. Zhang and N. Zhu and N. Zobrist and Z. Zou and G. Bortoli and S. Boixo and J. Chen and Y. Chen and M. Devoret and M. Hansen and C. Jones and J. Kelly and P. Kohli and A. Korotkov and E. Lucero and J. Manyika and Y. Matias and A. Megrant and H. Neven and W. D. Oliver and G. Ramachandran and R. Babbush and V. Smelyanskiy and P. Roushan and D. Kafri and R. Sarpong and D. W. Berry and C. Ramanathan and X. Mi and C. Bengs and A. Ajoy and Z. K. Minev and N. C. Rubin and T. E. O'Brien},
      year={2025},
      eprint={2510.19550},
      archivePrefix={arXiv},
      primaryClass={quant-ph},
      url={https://arxiv.org/abs/2510.19550}, 
}

@article{Arute2019,
    author={{Google Quantum AI and Collaborators}},
    fullAuthorList={Arute, Frank
    and Arya, Kunal
    and Babbush, Ryan
    and Bacon, Dave
    and Bardin, Joseph C.
    and Barends, Rami
    and Biswas, Rupak
    and Boixo, Sergio
    and Brandao, Fernando G. S. L.
    and Buell, David A.
    and Burkett, Brian
    and Chen, Yu
    and Chen, Zijun
    and Chiaro, Ben
    and Collins, Roberto
    and Courtney, William
    and Dunsworth, Andrew
    and Farhi, Edward
    and Foxen, Brooks
    and Fowler, Austin
    and Gidney, Craig
    and Giustina, Marissa
    and Graff, Rob
    and Guerin, Keith
    and Habegger, Steve
    and Harrigan, Matthew P.
    and Hartmann, Michael J.
    and Ho, Alan
    and Hoffmann, Markus
    and Huang, Trent
    and Humble, Travis S.
    and Isakov, Sergei V.
    and Jeffrey, Evan
    and Jiang, Zhang
    and Kafri, Dvir
    and Kechedzhi, Kostyantyn
    and Kelly, Julian
    and Klimov, Paul V.
    and Knysh, Sergey
    and Korotkov, Alexander
    and Kostritsa, Fedor
    and Landhuis, David
    and Lindmark, Mike
    and Lucero, Erik
    and Lyakh, Dmitry
    and Mandr{\`a}, Salvatore
    and McClean, Jarrod R.
    and McEwen, Matthew
    and Megrant, Anthony
    and Mi, Xiao
    and Michielsen, Kristel
    and Mohseni, Masoud
    and Mutus, Josh
    and Naaman, Ofer
    and Neeley, Matthew
    and Neill, Charles
    and Niu, Murphy Yuezhen
    and Ostby, Eric
    and Petukhov, Andre
    and Platt, John C.
    and Quintana, Chris
    and Rieffel, Eleanor G.
    and Roushan, Pedram
    and Rubin, Nicholas C.
    and Sank, Daniel
    and Satzinger, Kevin J.
    and Smelyanskiy, Vadim
    and Sung, Kevin J.
    and Trevithick, Matthew D.
    and Vainsencher, Amit
    and Villalonga, Benjamin
    and White, Theodore
    and Yao, Z. Jamie
    and Yeh, Ping
    and Zalcman, Adam
    and Neven, Hartmut
    and Martinis, John M.},
    title={Quantum supremacy using a programmable superconducting processor},
    journal={Nature},
    year={2019},
    month={Oct},
    day={01},
    volume={574},
    number={7779},
    pages={505-510},
    issn={1476-4687},
    doi={10.1038/s41586-019-1666-5},
    url={https://doi.org/10.1038/s41586-019-1666-5}
}

@article{USTC2020,
    author = {Han-Sen Zhong  and Hui Wang  and others},
    fullauthorlist = {Han-Sen Zhong  and Hui Wang  and Yu-Hao Deng  and Ming-Cheng Chen  and Li-Chao Peng  and Yi-Han Luo  and Jian Qin  and Dian Wu  and Xing Ding  and Yi Hu  and Peng Hu  and Xiao-Yan Yang  and Wei-Jun Zhang  and Hao Li  and Yuxuan Li  and Xiao Jiang  and Lin Gan  and Guangwen Yang  and Lixing You  and Zhen Wang  and Li Li  and Nai-Le Liu  and Chao-Yang Lu  and Jian-Wei Pan },
    title = {Quantum computational advantage using photons},
    journal = {Science},
    volume = {370},
    number = {6523},
    pages = {1460-1463},
    year = {2020},
    doi = {10.1126/science.abe8770},
    URL = {https://www.science.org/doi/abs/10.1126/science.abe8770},
}

@article{USTC2021a,
  title = {Strong Quantum Computational Advantage Using a Superconducting Quantum Processor},
  author = {Wu, Yulin and Bao, Wan-Su and others},
  fullauthorlist = {Wu, Yulin and Bao, Wan-Su and Cao, Sirui and Chen, Fusheng and Chen, Ming-Cheng and Chen, Xiawei and Chung, Tung-Hsun and Deng, Hui and Du, Yajie and Fan, Daojin and Gong, Ming and Guo, Cheng and Guo, Chu and Guo, Shaojun and Han, Lianchen and Hong, Linyin and Huang, He-Liang and Huo, Yong-Heng and Li, Liping and Li, Na and Li, Shaowei and Li, Yuan and Liang, Futian and Lin, Chun and Lin, Jin and Qian, Haoran and Qiao, Dan and Rong, Hao and Su, Hong and Sun, Lihua and Wang, Liangyuan and Wang, Shiyu and Wu, Dachao and Xu, Yu and Yan, Kai and Yang, Weifeng and Yang, Yang and Ye, Yangsen and Yin, Jianghan and Ying, Chong and Yu, Jiale and Zha, Chen and Zhang, Cha and Zhang, Haibin and Zhang, Kaili and Zhang, Yiming and Zhao, Han and Zhao, Youwei and Zhou, Liang and Zhu, Qingling and Lu, Chao-Yang and Peng, Cheng-Zhi and Zhu, Xiaobo and Pan, Jian-Wei},
  journal = {Phys. Rev. Lett.},
  volume = {127},
  issue = {18},
  pages = {180501},
  numpages = {7},
  year = {2021},
  month = {Oct},
  publisher = {American Physical Society},
  doi = {10.1103/PhysRevLett.127.180501},
  url = {https://link.aps.org/doi/10.1103/PhysRevLett.127.180501}
}

@article{USTC2021b,
  title = {Phase-Programmable Gaussian Boson Sampling Using Stimulated Squeezed Light},
  author = {Zhong, Han-Sen and Deng, Yu-Hao and others},
  fullauthorlist = {Zhong, Han-Sen and Deng, Yu-Hao and Qin, Jian and Wang, Hui and Chen, Ming-Cheng and Peng, Li-Chao and Luo, Yi-Han and Wu, Dian and Gong, Si-Qiu and Su, Hao and Hu, Yi and Hu, Peng and Yang, Xiao-Yan and Zhang, Wei-Jun and Li, Hao and Li, Yuxuan and Jiang, Xiao and Gan, Lin and Yang, Guangwen and You, Lixing and Wang, Zhen and Li, Li and Liu, Nai-Le and Renema, Jelmer J. and Lu, Chao-Yang and Pan, Jian-Wei},
  journal = {Phys. Rev. Lett.},
  volume = {127},
  issue = {18},
  pages = {180502},
  numpages = {9},
  year = {2021},
  month = {Oct},
  publisher = {American Physical Society},
  doi = {10.1103/PhysRevLett.127.180502},
  url = {https://link.aps.org/doi/10.1103/PhysRevLett.127.180502}
}

@article{USTC2021c,
    title = {Quantum computational advantage via 60-qubit 24-cycle random circuit sampling},
    journal = {Science Bulletin},
    volume = {67},
    number = {3},
    pages = {240-245},
    year = {2022},
    issn = {2095-9273},
    doi = {https://doi.org/10.1016/j.scib.2021.10.017},
    url = {https://www.sciencedirect.com/science/article/pii/S2095927321006733},
    author = {Qingling Zhu and Sirui Cao and others},
    fullauthorlist = {Qingling Zhu and Sirui Cao and Fusheng Chen and Ming-Cheng Chen and Xiawei Chen and Tung-Hsun Chung and Hui Deng and Yajie Du and Daojin Fan and Ming Gong and Cheng Guo and Chu Guo and Shaojun Guo and Lianchen Han and Linyin Hong and He-Liang Huang and Yong-Heng Huo and Liping Li and Na Li and Shaowei Li and Yuan Li and Futian Liang and Chun Lin and Jin Lin and Haoran Qian and Dan Qiao and Hao Rong and Hong Su and Lihua Sun and Liangyuan Wang and Shiyu Wang and Dachao Wu and Yulin Wu and Yu Xu and Kai Yan and Weifeng Yang and Yang Yang and Yangsen Ye and Jianghan Yin and Chong Ying and Jiale Yu and Chen Zha and Cha Zhang and Haibin Zhang and Kaili Zhang and Yiming Zhang and Han Zhao and Youwei Zhao and Liang Zhou and Chao-Yang Lu and Cheng-Zhi Peng and Xiaobo Zhu and Jian-Wei Pan},
}

@article{Xanadu2022,
    author={Madsen, Lars S. and Laudenbach, Fabian and others},
    fullauthorlist={Madsen, Lars S. and Laudenbach, Fabian and Askarani, Mohsen Falamarzi. and Rortais, Fabien and Vincent, Trevor and Bulmer, Jacob F. F. and Miatto, Filippo M. and Neuhaus, Leonhard and Helt, Lukas G. and Collins, Matthew J. and Lita, Adriana E. and Gerrits, Thomas and Nam, Sae Woo and Vaidya, Varun D. and Menotti, Matteo and Dhand, Ish and Vernon, Zachary and Quesada, Nicol{\'a}s and Lavoie, Jonathan},
    title={Quantum computational advantage with a programmable photonic processor},
    journal={Nature},
    year={2022},
    month={Jun},
    day={01},
    volume={606},
    number={7912},
    pages={75-81},
    issn={1476-4687},
    doi={10.1038/s41586-022-04725-x},
    url={https://doi.org/10.1038/s41586-022-04725-x}
}

@misc{USTC2023,
      title={Gaussian Boson Sampling with Pseudo-Photon-Number Resolving Detectors and Quantum Computational Advantage}, 
      author={Yu-Hao Deng and Yi-Chao Gu and others},
      fullauthorlist={Yu-Hao Deng and Yi-Chao Gu and Hua-Liang Liu and Si-Qiu Gong and Hao Su and Zhi-Jiong Zhang and Hao-Yang Tang and Meng-Hao Jia and Jia-Min Xu and Ming-Cheng Chen and Jian Qin and Li-Chao Peng and Jiarong Yan and Yi Hu and Jia Huang and Hao Li and Yuxuan Li and Yaojian Chen and Xiao Jiang and Lin Gan and Guangwen Yang and Lixing You and Li Li and Han-Sen Zhong and Hui Wang and Nai-Le Liu and Jelmer J. Renema and Chao-Yang Lu and Jian-Wei Pan},
      year={2023},
      eprint={2304.12240},
      archivePrefix={arXiv},
      primaryClass={quant-ph},
      url={https://arxiv.org/abs/2304.12240}, 
}

@article{IBM2023,
    author={Kim, Youngseok
    and Eddins, Andrew
    and Anand, Sajant
    and Wei, Ken Xuan
    and van den Berg, Ewout
    and Rosenblatt, Sami
    and Nayfeh, Hasan
    and Wu, Yantao
    and Zaletel, Michael
    and Temme, Kristan
    and Kandala, Abhinav},
    title={Evidence for the utility of quantum computing before fault tolerance},
    journal={Nature},
    year={2023},
    month={Jun},
    day={01},
    volume={618},
    number={7965},
    pages={500-505},
    issn={1476-4687},
    doi={10.1038/s41586-023-06096-3},
    url={https://doi.org/10.1038/s41586-023-06096-3}
}

@article{DWave2024,
    author = {Andrew D. King  and Alberto Nocera  and others},
    fullauthorlist = {Andrew D. King  and Alberto Nocera  and Marek M. Rams  and Jacek Dziarmaga  and Roeland Wiersema  and William Bernoudy  and Jack Raymond  and Nitin Kaushal  and Niclas Heinsdorf  and Richard Harris  and Kelly Boothby  and Fabio Altomare  and Mohsen Asad  and Andrew J. Berkley  and Martin Boschnak  and Kevin Chern  and Holly Christiani  and Samantha Cibere  and Jake Connor  and Martin H. Dehn  and Rahul Deshpande  and Sara Ejtemaee  and Pau Farre  and Kelsey Hamer  and Emile Hoskinson  and Shuiyuan Huang  and Mark W. Johnson  and Samuel Kortas  and Eric Ladizinsky  and Trevor Lanting  and Tony Lai  and Ryan Li  and Allison J. R. MacDonald  and Gaelen Marsden  and Catherine C. McGeoch  and Reza Molavi  and Travis Oh  and Richard Neufeld  and Mana Norouzpour  and Joel Pasvolsky  and Patrick Poitras  and Gabriel Poulin-Lamarre  and Thomas Prescott  and Mauricio Reis  and Chris Rich  and Mohammad Samani  and Benjamin Sheldan  and Anatoly Smirnov  and Edward Sterpka  and Berta Trullas Clavera  and Nicholas Tsai  and Mark Volkmann  and Alexander M. Whiticar  and Jed D. Whittaker  and Warren Wilkinson  and Jason Yao  and T. J. Yi  and Anders W. Sandvik  and Gonzalo Alvarez  and Roger G. Melko  and Juan Carrasquilla  and Marcel Franz  and Mohammad H. Amin },
    title = {Beyond-classical computation in quantum simulation},
    journal = {Science},
    volume = {388},
    number = {6743},
    pages = {199-204},
    year = {2025},
    doi = {10.1126/science.ado6285},
    URL = {https://www.science.org/doi/abs/10.1126/science.ado6285},
}

@article{GoogleAnalog2025,
    author={Andersen, T. I. and Astrakhantsev, N. and others},
    fullauthorlist={Andersen, T. I.
    and Astrakhantsev, N.
    and Karamlou, A. H.
    and Berndtsson, J.
    and Motruk, J.
    and Szasz, A.
    and Gross, J. A.
    and Schuckert, A.
    and Westerhout, T.
    and Zhang, Y.
    and Forati, E.
    and Rossi, D.
    and Kobrin, B.
    and Paolo, A. Di
    and Klots, A. R.
    and Drozdov, I.
    and Kurilovich, V.
    and Petukhov, A.
    and Ioffe, L. B.
    and Elben, A.
    and Rath, A.
    and Vitale, V.
    and Vermersch, B.
    and Acharya, R.
    and Beni, L. A.
    and Anderson, K.
    and Ansmann, M.
    and Arute, F.
    and Arya, K.
    and Asfaw, A.
    and Atalaya, J.
    and Ballard, B.
    and Bardin, J. C.
    and Bengtsson, A.
    and Bilmes, A.
    and Bortoli, G.
    and Bourassa, A.
    and Bovaird, J.
    and Brill, L.
    and Broughton, M.
    and Browne, D. A.
    and Buchea, B.
    and Buckley, B. B.
    and Buell, D. A.
    and Burger, T.
    and Burkett, B.
    and Bushnell, N.
    and Cabrera, A.
    and Campero, J.
    and Chang, H.-S.
    and Chen, Z.
    and Chiaro, B.
    and Claes, J.
    and Cleland, A. Y.
    and Cogan, J.
    and Collins, R.
    and Conner, P.
    and Courtney, W.
    and Crook, A. L.
    and Das, S.
    and Debroy, D. M.
    and Lorenzo, L. De
    and Barba, A. Del Toro
    and Demura, S.
    and Donohoe, P.
    and Dunsworth, A.
    and Earle, C.
    and Eickbusch, A.
    and Elbag, A. M.
    and Elzouka, M.
    and Erickson, C.
    and Faoro, L.
    and Fatemi, R.
    and Ferreira, V. S.
    and Burgos, L. Flores
    and Fowler, A. G.
    and Foxen, B.
    and Ganjam, S.
    and Gasca, R.
    and Giang, W.
    and Gidney, C.
    and Gilboa, D.
    and Giustina, M.
    and Gosula, R.
    and Dau, A. Grajales
    and Graumann, D.
    and Greene, A.
    and Habegger, S.
    and Hamilton, M. C.
    and Hansen, M.
    and Harrigan, M. P.
    and Harrington, S. D.
    and Heslin, S.
    and Heu, P.
    and Hill, G.
    and Hoffmann, M. R.
    and Huang, H.-Y.
    and Huang, T.
    and Huff, A.
    and Huggins, W. J.
    and Isakov, S. V.
    and Jeffrey, E.
    and Jiang, Z.
    and Jones, C.
    and Jordan, S.
    and Joshi, C.
    and Juhas, P.
    and Kafri, D.
    and Kang, H.
    and Kechedzhi, K.
    and Khaire, T.
    and Khattar, T.
    and Khezri, M.
    and Kieferov{\'a}, M.
    and Kim, S.
    and Kitaev, A.
    and Klimov, P.
    and Korotkov, A. N.
    and Kostritsa, F.
    and Kreikebaum, J. M.
    and Landhuis, D.
    and Langley, B. W.
    and Laptev, P.
    and Lau, K.-M.
    and Guevel, L. Le
    and Ledford, J.
    and Lee, J.
    and Lee, K. W.
    and Lensky, Y. D.
    and Lester, B. J.
    and Li, W. Y.
    and Lill, A. T.
    and Liu, W.
    and Livingston, W. P.
    and Locharla, A.
    and Lundahl, D.
    and Lunt, A.
    and Madhuk, S.
    and Maloney, A.
    and Mandr{\`a}, S.
    and Martin, L. S.
    and Martin, O.
    and Martin, S.
    and Maxfield, C.
    and McClean, J. R.
    and McEwen, M.
    and Meeks, S.
    and Miao, K. C.
    and Mieszala, A.
    and Molina, S.
    and Montazeri, S.
    and Morvan, A.
    and Movassagh, R.
    and Neill, C.
    and Nersisyan, A.
    and Newman, M.
    and Nguyen, A.
    and Nguyen, M.
    and Ni, C.-H.
    and Niu, M. Y.
    and Oliver, W. D.
    and Ottosson, K.
    and Pizzuto, A.
    and Potter, R.
    and Pritchard, O.
    and Pryadko, L. P.
    and Quintana, C.
    and Reagor, M. J.
    and Rhodes, D. M.
    and Roberts, G.
    and Rocque, C.
    and Rosenberg, E.
    and Rubin, N. C.
    and Saei, N.
    and Sankaragomathi, K.
    and Satzinger, K. J.
    and Schurkus, H. F.
    and Schuster, C.
    and Shearn, M. J.
    and Shorter, A.
    and Shutty, N.
    and Shvarts, V.
    and Sivak, V.
    and Skruzny, J.
    and Small, S.
    and Smith, W. Clarke
    and Springer, S.
    and Sterling, G.
    and Suchard, J.
    and Szalay, M.
    and Sztein, A.
    and Thor, D.
    and Torres, A.
    and Torunbalci, M. M.
    and Vaishnav, A.
    and Vdovichev, S.
    and Villalonga, B.
    and Heidweiller, C. Vollgraff
    and Waltman, S.
    and Wang, S. X.
    and White, T.
    and Wong, K.
    and Woo, B. W. K.
    and Xing, C.
    and Yao, Z. Jamie
    and Yeh, P.
    and Ying, B.
    and Yoo, J.
    and Yosri, N.
    and Young, G.
    and Zalcman, A.
    and Zhu, N.
    and Zobrist, N.
    and Neven, H.
    and Babbush, R.
    and Boixo, S.
    and Hilton, J.
    and Lucero, E.
    and Megrant, A.
    and Kelly, J.
    and Chen, Y.
    and Smelyanskiy, V.
    and Vidal, G.
    and Roushan, P.
    and L{\"a}uchli, A. M.
    and Abanin, D. A.
    and Mi, X.},
    title={Thermalization and criticality on an analogue--digital quantum simulator},
    journal={Nature},
    year={2025},
    month={Feb},
    day={01},
    volume={638},
    number={8049},
    pages={79-85},
    issn={1476-4687},
    doi={10.1038/s41586-024-08460-3},
    url={https://doi.org/10.1038/s41586-024-08460-3}
}

@article{QuantinuumRuszlan2025,
    author={Liu, Minzhao and Shaydulin, Ruslan and others},
    fullauthorlist={Liu, Minzhao and Shaydulin, Ruslan and Niroula, Pradeep and DeCross, Matthew and Hung, Shih-Han and Kon, Wen Yu and Cervero-Mart{\'i}n, Enrique and Chakraborty, Kaushik and Amer, Omar and Aaronson, Scott and Acharya, Atithi and Alexeev, Yuri and Berg, K. Jordan and Chakrabarti, Shouvanik and Curchod, Florian J. and Dreiling, Joan M. and Erickson, Neal and Foltz, Cameronand Foss-Feig, Michael and Hayes, David and Humble, Travis S. and Kumar, Niraj and Larson, Jeffrey and Lykov, Danylo and Mills, Michael and Moses, Steven A. and Neyenhuis, Brian and Eloul, Shaltiel and Siegfried, Peter and Walker, James and Lim, Charles and Pistoia, Marco},
    title={Certified randomness using a trapped-ion quantum processor},
    journal={Nature},
    year={2025},
    month={Apr},
    day={01},
    volume={640},
    number={8058},
    pages={343-348},
    issn={1476-4687},
    doi={10.1038/s41586-025-08737-1},
    url={https://doi.org/10.1038/s41586-025-08737-1}
}

@misc{QuantinuumIsing2025,
      title={Digital quantum magnetism at the frontier of classical simulations}, 
      author={Reza Haghshenas and Eli Chertkov and others},
      fullauthorlist={Reza Haghshenas and Eli Chertkov and Michael Mills and Wilhelm Kadow and Sheng-Hsuan Lin and Yi-Hsiang Chen and Chris Cade and Ido Niesen and Tomislav Begušić and Manuel S. Rudolph and Cristina Cirstoiu and Kevin Hemery and Conor Mc Keever and Michael Lubasch and Etienne Granet and Charles H. Baldwin and John P. Bartolotta and Matthew Bohn and Julia Cline and Matthew DeCross and Joan M. Dreiling and Cameron Foltz and David Francois and John P. Gaebler and Christopher N. Gilbreth and Johnnie Gray and Dan Gresh and Alex Hall and Aaron Hankin and Azure Hansen and Nathan Hewitt and Ross B. Hutson and Mohsin Iqbal and Nikhil Kotibhaskar and Elliot Lehman and Dominic Lucchetti and Ivaylo S. Madjarov and Karl Mayer and Alistair R. Milne and Steven A. Moses and Brian Neyenhuis and Gunhee Park and Boris Ponsioen and Michael Schecter and Peter E. Siegfried and David T. Stephen and Bruce G. Tiemann and Maxwell D. Urmey and James Walker and Andrew C. Potter and David Hayes and Garnet Kin-Lic Chan and Frank Pollmann and Michael Knap and Henrik Dreyer and Michael Foss-Feig},
      year={2025},
      eprint={2503.20870},
      archivePrefix={arXiv},
      primaryClass={quant-ph},
      url={https://arxiv.org/abs/2503.20870}, 
}

@misc{USTC2025,
      title={Robust quantum computational advantage with programmable 3050-photon Gaussian boson sampling}, 
      author={Hua-Liang Liu and Hao Su and others},
      fullauthorlist={Hua-Liang Liu and Hao Su and Si-Qiu Gong and Yi-Chao Gu and Hao-Yang Tang and Meng-Hao Jia and Qian Wei and Yukun Song and Dongzhou Wang and Mingyang Zheng and Faxi Chen and Libo Li and Siyu Ren and Xuezhi Zhu and Meihong Wang and Yaojian Chen and Yanfei Liu and Longsheng Song and Pengyu Yang and Junshi Chen and Hong An and Lei Zhang and Lin Gan and Guangwen Yang and Jia-Min Xu and Yu-Ming He and Hui Wang and Han-Sen Zhong and Ming-Cheng Chen and Xiao Jiang and Li Li and Nai-Le Liu and Yu-Hao Deng and Xiao-Long Su and Qiang Zhang and Chao-Yang Lu and Jian-Wei Pan},
      year={2025},
      eprint={2508.09092},
      archivePrefix={arXiv},
      primaryClass={quant-ph},
      url={https://arxiv.org/abs/2508.09092}, 
}

@misc{larose2024history,
      title={A brief history of quantum vs classical computational advantage}, 
      author={Ryan LaRose},
      year={2024},
      eprint={2412.14703},
      archivePrefix={arXiv},
      primaryClass={quant-ph},
      url={https://arxiv.org/abs/2412.14703}, 
}

@article{lubasch2014algorithms,
  title={Algorithms for finite projected entangled pair states},
  author={Lubasch, Michael and Cirac, J Ignacio and Banuls, Mari-Carmen},
  journal={Physical Review B},
  volume={90},
  number={6},
  pages={064425},
  year={2014},
  publisher={APS},
  url ={https://journals.aps.org/prb/abstract/10.1103/PhysRevB.90.064425}
}

@article{jiang2008accurate,
  title={Accurate determination of tensor network state of quantum lattice models in two dimensions},
  author={Jiang, Hong-Chen and Weng, Zheng-Yu and Xiang, Tao},
  journal={Physical review letters},
  volume={101},
  number={9},
  pages={090603},
  year={2008},
  publisher={APS},
  url = {https://journals.aps.org/prl/abstract/10.1103/PhysRevLett.101.090603}
}

@article{verstraete2004renormalization,
  title={Renormalization algorithms for quantum-many body systems in two and higher dimensions},
  author={Verstraete, Frank and Cirac, J Ignacio},
  journal={arXiv preprint cond-mat/0407066},
  year={2004},
  url = {https://arxiv.org/abs/cond-mat/0407066}
}

@inbook{yedidia2003understanding,
author = {Yedidia, Jonathan S. and Freeman, William T. and Weiss, Yair},
title = {Understanding belief propagation and its generalizations},
year = {2003},
isbn = {1558608117},
publisher = {Morgan Kaufmann Publishers Inc.},
address = {San Francisco, CA, USA},
booktitle = {Exploring Artificial Intelligence in the New Millennium},
pages = {239–269},
numpages = {31},
url = {https://dl.acm.org/doi/abs/10.5555/779343.779352}
}

@article{alkabetz2021tensor,
  title={Tensor networks contraction and the belief propagation algorithm},
  author={Alkabetz, Roy and Arad, Itai},
  journal={Physical Review Research},
  volume={3},
  number={2},
  pages={023073},
  year={2021},
  publisher={APS},
  url={https://journals.aps.org/prresearch/abstract/10.1103/PhysRevResearch.3.023073}
}

@article{tindall2024efficient,
  title={Efficient tensor network simulation of {IBM}’s {E}agle kicked {I}sing experiment},
  author={Tindall, Joseph and Fishman, Matthew and Stoudenmire, E Miles and Sels, Dries},
  journal={PRX Quantum},
  volume={5},
  number={1},
  pages={010308},
  year={2024},
  publisher={APS},
  url={https://journals.aps.org/prxquantum/abstract/10.1103/PRXQuantum.5.010308}
}

@article{tindall2025dynamics,
  title={Dynamics of disordered quantum systems with two-and three-dimensional tensor networks},
  author={Tindall, Joseph and Mello, Antonio and Fishman, Matt and Stoudenmire, Miles and Sels, Dries},
  journal={arXiv preprint arXiv:2503.05693},
  year={2025},
  url={https://arxiv.org/abs/2503.05693}
}

@article{tindall2023gauging,
  title={Gauging tensor networks with belief propagation},
  author={Tindall, Joseph and Fishman, Matthew},
  journal={SciPost Physics},
  volume={15},
  number={6},
  pages={222},
  year={2023},
  url={https://scipost.org/SciPostPhys.15.6.222}
}

@article{zhao2025leapfrogging,
  title={Leapfrogging {S}ycamore: {H}arnessing 1432 {GPU}s for 7$\times$ faster quantum random circuit sampling},
  author={Zhao, Xian-He and Zhong, Han-Sen and Pan, Feng and Chen, Zi-Han and Fu, Rong and Su, Zhongling and Xie, Xiaotong and Zhao, Chaoxing and Zhang, Pan and Ouyang, Wanli and others},
  journal={National Science Review},
  volume={12},
  number={3},
  pages={nwae317},
  year={2025},
  publisher={Oxford University Press},
  URL={https://academic.oup.com/nsr/article/12/3/nwae317/7756427}
}

@article{
beguvsic2024fast,
author = {Begu{\v{s}}i{\'c}, Tomislav and Gray, Johnnie and Chan, Garnet Kin-Lic},
title = {Fast and converged classical simulations of evidence for the utility of quantum computing before fault tolerance},
journal = {Science Advances},
volume = {10},
number = {3},
pages = {eadk4321},
year = {2024},
doi = {10.1126/sciadv.adk4321},
URL = {https://www.science.org/doi/abs/10.1126/sciadv.adk4321},
}

@article{gray2025tensor,
  title={Tensor Network Loop Cluster Expansions for Quantum Many-Body Problems},
  author={Gray, Johnnie and Park, Gunhee and Evenbly, Glen and Pancotti, Nicola and Chan, Garnet Kin},
  journal={arXiv preprint arXiv:2510.05647},
  year={2025},
  url={https://arxiv.org/abs/2510.05647}
}

@article{evenbly2024loop,
  title={Loop Series Expansions for Tensor Networks},
  author={Evenbly, Glen and Pancotti, Nicola and Milsted, Ashley and Gray, Johnnie and Chan, Garnet Kin},
  journal={arXiv preprint arXiv:2409.03108},
  year={2024},
  url={https://arxiv.org/abs/2409.03108}
}

@article{gonzalez2024random,
  title={Random insights into the complexity of two-dimensional tensor network calculations},
  author={Gonz{\'a}lez-Garc{\'\i}a, Sof{\'\i}a and Sang, Shengqi and Hsieh, Timothy H and Boixo, Sergio and Vidal, Guifr{\'e} and Potter, Andrew C and Vasseur, Romain},
  journal={Physical Review B},
  volume={109},
  number={23},
  pages={235102},
  year={2024},
  publisher={APS},
  url={https://journals.aps.org/prb/abstract/10.1103/PhysRevB.109.235102}
}

@misc{google2025tnco,
  author = {Google Research},
  title  = {{TNCO}},
  note   = {\url{https://github.com/google-research/tnco/tree/main/tnco}},
  year   = {2025},


}

@article{rudolph2025simulating,
  title={Simulating and sampling from quantum circuits with 2D tensor networks},
  author={Rudolph, Manuel S and Tindall, Joseph},
  journal={arXiv preprint arXiv:2507.11424},
  year={2025},
  url={https://arxiv.org/abs/2507.11424}

}

@article{fishman2022itensor,
  title={The {ITensor} software library for tensor network calculations},
  author={Fishman, Matthew and White, Steven and Stoudenmire, Edwin Miles},
  journal={SciPost Physics Codebases},
  pages={004},
  year={2022},
  url={https://scipost.org/SciPostPhysCodeb.4}

}

@misc{tnqs,
  author    = {Joseph Tindall and Manuel Rudolph},
  title     = {{TensorNetworkQuantumSimulator}.jl},
  year      = {2025},
  publisher = {GitHub},
  note      = {\url{https://github.com/JoeyT1994/TensorNetworkQuantumSimulator.jl}},
}

@article{hashimoto2017out,
  title={Out-of-time-order correlators in quantum mechanics},
  author={Hashimoto, Koji and Murata, Keiju and Yoshii, Ryosuke},
  journal={Journal of High Energy Physics},
  volume={2017},
  number={10},
  pages={1--31},
  year={2017},
  publisher={Springer},
  URL={https://link.springer.com/article/10.1007/JHEP10(2017)138}
}

@article{haehl2019classification,
  title={Classification of out-of-time-order correlators},
  author={Haehl, Felix and Loganayagam, R and Narayan, Prithvi and Rangamani, Mukund},
  journal={SciPost Physics},
  volume={6},
  number={1},
  pages={001},
  year={2019},
  URL={https://scipost.org/SciPostPhys.6.1.001}
}

@article{decross2406computational,
  title={Computational power of random quantum circuits in arbitrary geometries},
  author={DeCross, Matthew and Haghshenas, Reza and Liu, Minzhao and Rinaldi, Enrico and Gray, Johnnie and Alexeev, Yuri and Baldwin, Charles H and Bartolotta, John P and Bohn, Matthew and Chertkov, Eli and others},
  journal={Physical Review X},
  volume={15},
  number={2},
  pages={021052},
  year={2025},
  publisher={APS},
  URL={https://journals.aps.org/prx/abstract/10.1103/PhysRevX.15.021052}
}

@article{morvan2024phase,
  title={Phase transitions in random circuit sampling},
  author={Morvan, Alexis and Villalonga, B and Mi, X and Mandr{\`a}, S and Bengtsson, A and Klimov, PV and Chen, Z and Hong, S and Erickson, C and Drozdov, IK and others},
  journal={Nature},
  volume={634},
  number={8033},
  pages={328--333},
  year={2024},
  publisher={Nature Publishing Group UK London},
  URl={https://www.nature.com/articles/s41586-024-07998-6}
}

@article{garcia2022out,
  title={Out-of-time-order correlators and quantum chaos},
  author={Garc{\'\i}a-Mata, Ignacio and Jalabert, Rodolfo A and Wisniacki, Diego A},
  journal={arXiv preprint arXiv:2209.07965},
  year={2022},
  URL={https://arxiv.org/abs/2209.07965}
}

@article{mi2021information,
  title={Information scrambling in quantum circuits},
  author={Mi, Xiao and Roushan, Pedram and Quintana, Chris and Mandra, Salvatore and Marshall, Jeffrey and Neill, Charles and Arute, Frank and Arya, Kunal and Atalaya, Juan and Babbush, Ryan and others},
  journal={Science},
  volume={374},
  number={6574},
  pages={1479--1483},
  year={2021},
  publisher={American Association for the Advancement of Science},
  URL={https://www.science.org/doi/10.1126/science.abg5029}
}

@article{khemani2018operator,
  title={Operator spreading and the emergence of dissipative hydrodynamics under unitary evolution with conservation laws},
  author={Khemani, Vedika and Vishwanath, Ashvin and Huse, David A},
  journal={Physical Review X},
  volume={8},
  number={3},
  pages={031057},
  year={2018},
  publisher={APS},
  URL={https://journals.aps.org/prx/abstract/10.1103/PhysRevX.8.031057}
}

@article{jordan2008classical,
  title={Classical simulation of infinite-size quantum lattice systems in two spatial dimensions},
  author={Jordan, Jacob and Or{\'u}s, Roman and Vidal, Guifre and Verstraete, Frank and Cirac, J Ignacio},
  journal={Physical review letters},
  volume={101},
  number={25},
  pages={250602},
  year={2008},
  publisher={APS},
  URL={https://journals.aps.org/prl/abstract/10.1103/PhysRevLett.101.250602}
}

@article{nahum2018operator,
  title = {Operator Spreading in Random Unitary Circuits},
  author = {Nahum, Adam and Vijay, Sagar and Haah, Jeongwan},
  journal = {Phys. Rev. X},
  volume = {8},
  issue = {2},
  pages = {021014},
  numpages = {30},
  year = {2018},
  month = {Apr},
  publisher = {American Physical Society},
  doi = {10.1103/PhysRevX.8.021014},
  url = {https://link.aps.org/doi/10.1103/PhysRevX.8.021014}
}

@article{xu2020accessing,
  title={Accessing scrambling using matrix product operators},
  author={Xu, Shenglong and Swingle, Brian},
  journal={Nature Physics},
  volume={16},
  number={2},
  pages={199--204},
  year={2020},
  publisher={Nature Publishing Group UK London},
  URL={https://www.nature.com/articles/s41567-019-0712-4}
}

@article{zhou2020entanglement,
  title={Entanglement membrane in chaotic many-body systems},
  author={Zhou, Tianci and Nahum, Adam},
  journal={Physical Review X},
  volume={10},
  number={3},
  pages={031066},
  year={2020},
  publisher={APS},
  URL={https://journals.aps.org/prx/abstract/10.1103/PhysRevX.10.031066}
}

@misc{sahu2022efficienttensornetworksimulation,
      title={Efficient tensor network simulation of quantum many-body physics on sparse graphs}, 
      author={Subhayan Sahu and Brian Swingle},
      year={2022},
      eprint={2206.04701},
      archivePrefix={arXiv},
      primaryClass={quant-ph},
      url={https://arxiv.org/abs/2206.04701}, 
}

@article{midha2026belief,
  title={Belief Propagation and Tensor Network Expansions for Many-Body Quantum Systems: Rigorous Results and Fundamental Limits},
  author={Midha, Siddhant and Sommers, Grace M. and Tindall, Joseph and Abanin, Dmitry A.},
  journal={arXiv preprint arXiv:2604.03228},
  year={2026},
  url ={https://arxiv.org/abs/2604.03228}
}

@misc{midha2025belief,
      title={Beyond Belief Propagation: Cluster-Corrected Tensor Network Contraction with Exponential Convergence}, 
      author={Siddhant Midha and Yifan F. Zhang},
      year={2025},
      eprint={2510.02290},
      archivePrefix={arXiv},
      primaryClass={quant-ph},
      url={https://arxiv.org/abs/2510.02290}, 
}

@article{Xu2020otocOutside,
    author={Xu, Shenglong
    and Swingle, Brian},
    title={Accessing scrambling using matrix product operators},
    journal={Nature Physics},
    year={2020},
    month={Feb},
    day={01},
    volume={16},
    number={2},
    pages={199-204},
    issn={1745-2481},
    doi={10.1038/s41567-019-0712-4},
    url={https://doi.org/10.1038/s41567-019-0712-4}
}

@article{Bertini2019dualUnitaries,
  title = {Exact Correlation Functions for Dual-Unitary Lattice Models in $1+1$ Dimensions},
  author = {Bertini, Bruno and Kos, Pavel and Prosen, Toma\ifmmode \check{z}\else \v{z}\fi{}},
  journal = {Phys. Rev. Lett.},
  volume = {123},
  issue = {21},
  pages = {210601},
  numpages = {6},
  year = {2019},
  month = {Nov},
  publisher = {American Physical Society},
  doi = {10.1103/PhysRevLett.123.210601},
  url = {https://link.aps.org/doi/10.1103/PhysRevLett.123.210601}
}

@misc{mandra2025,
      title={A Heuristic for Matrix Product State Simulation of Out-of-Equilibrium Dynamics of Two-Dimensional Transverse-Field {I}sing Models}, 
      author={Salvatore Mandr\`{a} and Nikita Astrakhantsev and Sergei Isakov and Benjamin Villalonga and Brayden Ware and Tom Westerhout and Kostyantyn Kechedzhi},
      year={2025},
      eprint={2511.23438},
      archivePrefix={arXiv},
      primaryClass={quant-ph},
      url={https://arxiv.org/abs/2511.23438}, 
}

@article{Kechedzhi2024effective,
    title = {Effective quantum volume, fidelity and computational cost of noisy quantum processing experiments},
    journal = {Future Generation Computer Systems},
    volume = {153},
    pages = {431-441},
    year = {2024},
    issn = {0167-739X},
    doi = {https://doi.org/10.1016/j.future.2023.12.002},
    url = {https://www.sciencedirect.com/science/article/pii/S0167739X23004569},
    author = {K. Kechedzhi and S.V. Isakov and S. Mandrà and B. Villalonga and X. Mi and S. Boixo and V. Smelyanskiy}
}


\appendix

\section{Details of construction of circuits for scaling analysis}
\label{section:obtain_lightcone}

\begin{figure*}
    \centering
    \includegraphics[width=0.8\linewidth]{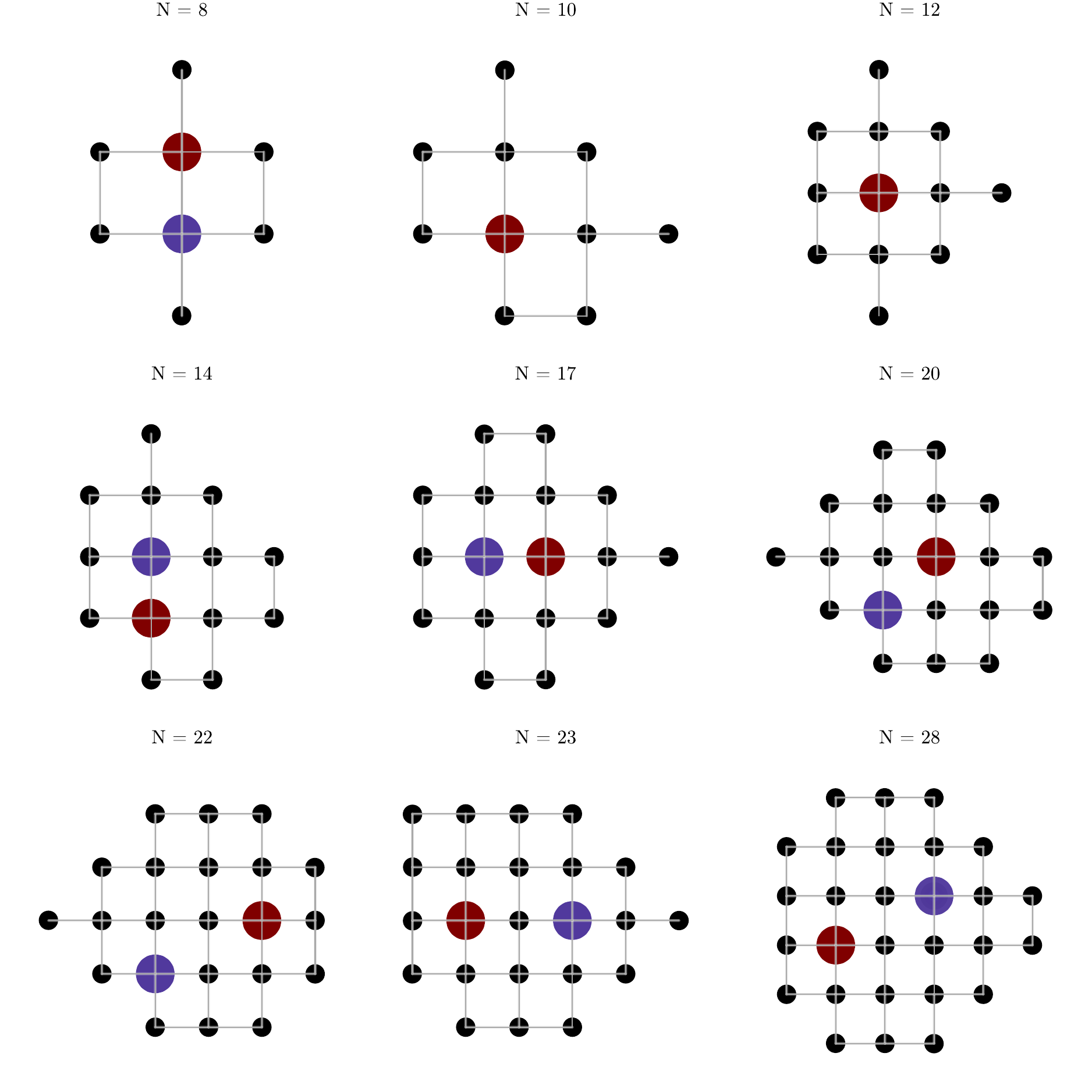}
    \caption{\textbf{Circuit footprint for each system size studied in the scaling analysis}.  For each system size (equivalently, circuit depth) considered in the scaling analysis, we illustrate the layout of the qubits in space and indicate the locations of the measurement operator $M$ (purple) and butterfly operator $B$ (red).  If only the red circle is visible, $M$ and $B$ are located on the same qubit.}
    \label{fig:circuits_last_t}
\end{figure*}

\subsubsection{Procedure to determine the locations of measurement and butterfly operators}

In the main text, we study the performance of TNBP in two different scenarios. First, we test the method on 23-qubit quantum circuits taken directly from the quantum echoes experiment presented in Ref.~\cite{google2025observation}. Second, we construct analogous quantum circuits for OTOC ensembles for a range of sizes and depths to characterize the scaling of the cost for TNBP. In this section, we describe the steps taken to generate the quantum circuits employed for the scaling analysis. This procedure for constructing an OTOC ensemble, which is designed to follow the key steps in \cite{google2025observation}, is illustrated in Figure \ref{fig:lightcone}.  The steps can be summarized as follows:
\begin{enumerate}
    \item[\circled{1}] On a large square qubit grid, determine the layout of the circuit for the time evolution $U$, and choose a location for the measurement operator $M$.
    \item[\circled{2}] Find the locations for the butterfly operator $B$ that give a \emph{large signal}, \textit{i.e.} large instance-to-instance fluctuations in the value of the OTOC. 
    \item[\circled{3}] For each of these locations for $B$, prune gates that do not contribute to the value of the OTOC, and then remove from the system any qubits that no longer have gates acting on them.
    \item[\circled{4}] Among all the locations for $B$ that give a large signal, select the one that maximizes exact tensor network contraction cost.  Choosing this location for $B$ is intended to make computation of the OTOC \emph{classically hard}.
\end{enumerate}
We now elaborate on each of these steps.

$\circled{1}$: We start with an initial square grid whose width is large enough that, within the depth of the time evolution circuit $U$, the geometric lightcones of $M$ and $B$ do not reach the edge in most parts of the grid.  In practice, we use a $10\times10$ grid, which is large enough for the circuit depths we consider.  (In contrast, for the circuits presented in \cite{google2025observation} the initial grid is smaller, so the lightcones do reach the edges of the grid.)

We place the measurement operator $M$ at a specific position near the center of the grid (one could instead choose to fix $B$; the important feature for the computation of the OTOC will be the relative position between $M$ and $B$). We set the depth $d$ of the unitary $U$, which determines the number of single- and two-qubit gates in the circuit. The gates in $U$ follow the pattern shown in Fig.~\ref{fig:OTOC circuit}, consisting of alternating layers of $\iSWAP$-like gates and single-qubit Pauli gates.

\circled{2}: Once the circuit layout of the unitaries $U$ (\textit{i.e.} the distribution of unitaries) is specified, and we have chosen the location of $M$, the next step is to find which locations of $B$ give a large signal.  Each possible position of the butterfly operator $B$ defines an OTOC ensemble, with fixed $M$ and $B$ and a fixed layout for the random unitaries $U$.  For each such ensemble, we compute the standard deviation in the OTOC, $\sigma\left(\mathcal{C}^{(2)}\right)$.  Importantly, computing this standard deviation \emph{does not} require computing the OTOC itself for many instances and taking the standard deviation of those values; rather, $\sigma$ can be computed directly and efficiently for the full ensemble~\cite{google2025observation, mi2021information, khemani2018operator, nahum2018operator}.  

In order for a quantum experiment to reliably detect variations in $\mathrm{OTOC}^{(1)}$ across different circuit instances, \textit{i.e.} to have a signal large, we need to pick an ensemble that has a large standard deviation.  To this end, let $\Sigma$ be the maximal value of $\sigma$ over all positions for the butterfly operator $B$.  We take as our remaining candidate locations for $B$ all the locations with $\sigma_B \geq 0.3\,\Sigma$, in other words the locations where the signal size is at least 30\% of the maximum possible.  

\begin{figure}[h]
    \centering
    \includegraphics[width=0.7\linewidth]{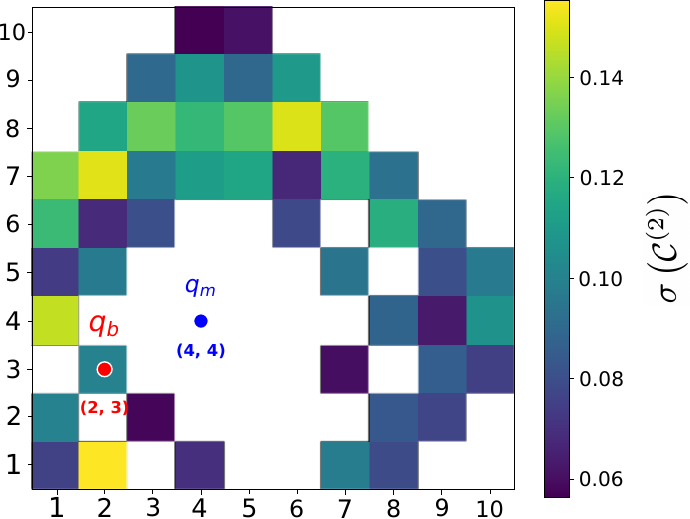}
    \caption{\textbf{OTOC standard deviation for $N = 28$ qubits}.  The measurement operator is placed at qubit $q_m$ in the center of a $10\times10$ grid. We then compute the standard deviation $\sigma$ for the OTOC ensemble we would have for each possible location of the butterfly operator $B$; the color in each square indicates the value of $\sigma$ when $B$ is placed at that location.  We leave blank all locations where $\sigma$ is less than 30\% of its maximum value over all possible locations of $B$, as the corresponding ensembles have instance-to-instance fluctuations that are too small to reliably measure in experiment.  Of the remaining locations for $B$ (colored squares), we choose $q_b=(2,3)$ because it has the highest estimated classical cost for exact tensor network contraction.  Note that $\sigma$ does not have symmetry around $q_m$, which is expected because the brickwall circuit structure of the evolution circuit $U$ breaks the mirror and discrete rotation symmetries. }
    \label{fig:standard deviation}
\end{figure}

This procedure is illustrated in Fig.~\ref{fig:standard deviation} for the case $N = 28$ qubits. The color in each square shows the size of $\sigma$ if $B$ were to be located in that square;\footnote{The quantity plotted in the figure is not exactly equal to the standard deviation $\sigma(\mathcal{C}^{(2)})$. Instead, we compute a simpler approximation obtained via Monte Carlo sampling. This quantity corresponds to $\sigma(\mathcal{C}^{(2)})$ computed using the diagonal approximation of the evolved operator $M$. As shown in \cite{google2025observation}, this method provides a cheaper and reliable estimate of $\sigma(\mathcal{C}^{(2)})$.} squares left blank show the locations for $B$ where $\sigma < 0.3\Sigma$, while the colored squares show the locations where $\sigma$ surpasses this threshold.  The measurement operator $M$ is at position (4,4) near the center of the grid.  The standard deviation is too small to allow for a reliable experiment when $B$ is close to $M$ because the OTOC is very close to 0 for all instances, hence the large blank region around $M$.  The standard deviation is also too small to allow for a reliable experiment when $B$ is very far from $M$ because the OTOC value is equal to or close to 1 for all instances, leading to the blank regions in the corners.  The remaining candidate locations for $B$ roughly form a ring around the location of $M$.

\circled{3}: Next, for each remaining possible location for $B$, we remove all gates that are outside the lightcones of the measurement and butterfly operators.  The lightcones are schematically illustrated in Fig.~\ref{fig:lightcone}, and the step-by-step pruning of gates is illustrated in Fig.~\ref{fig:OTOC circuit} in the main text.  We also remove from each circuit layout any qubits that no longer have gates acting on them.

\begin{figure*}[t]
    \centering
    \subfloat[N=10]{
        \includegraphics[width=0.22\textwidth]{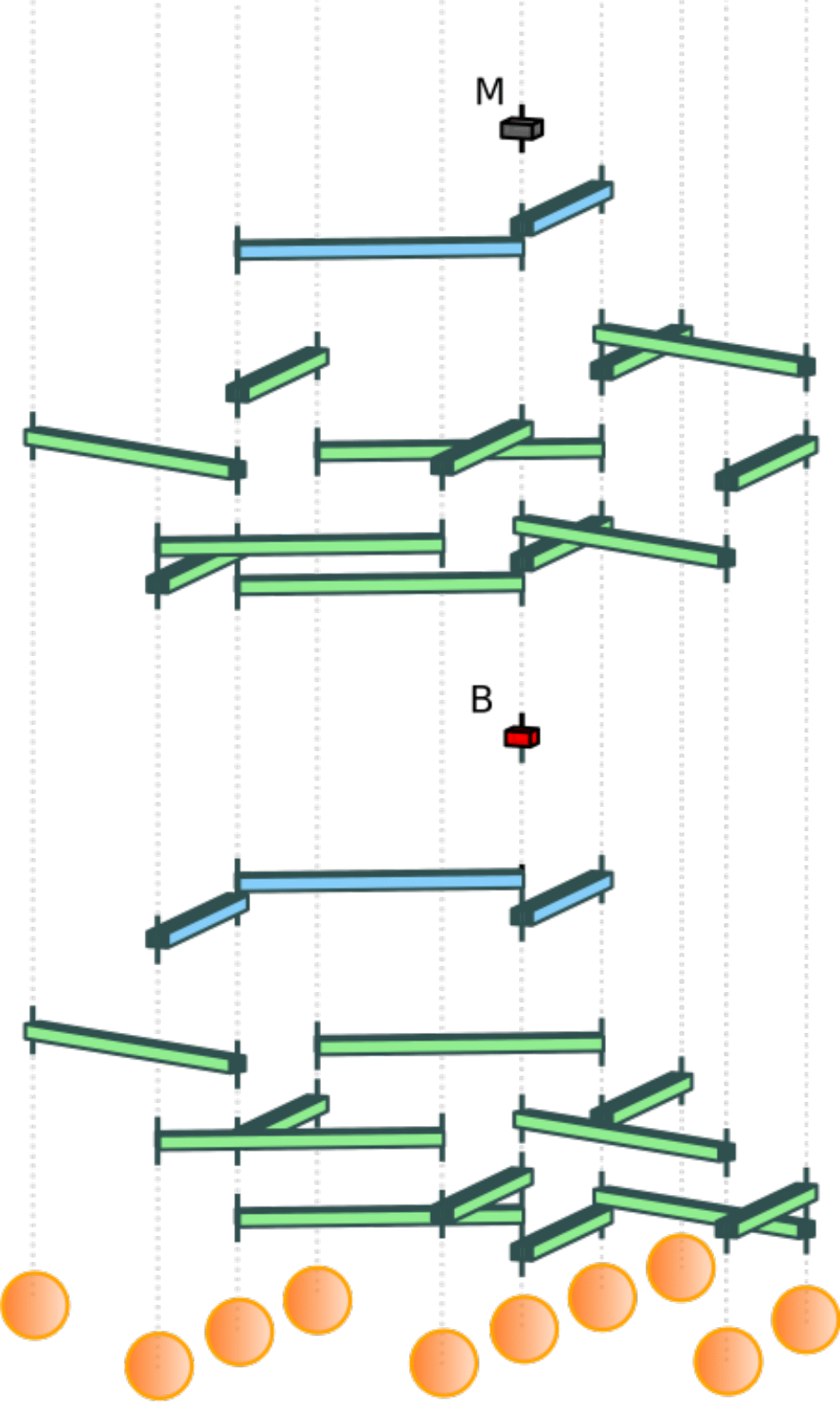}
        \label{fig:circuit-n10}
    }
    \hspace{0.03\linewidth} 
    \subfloat[N=20]{
        \includegraphics[width=0.24\textwidth]{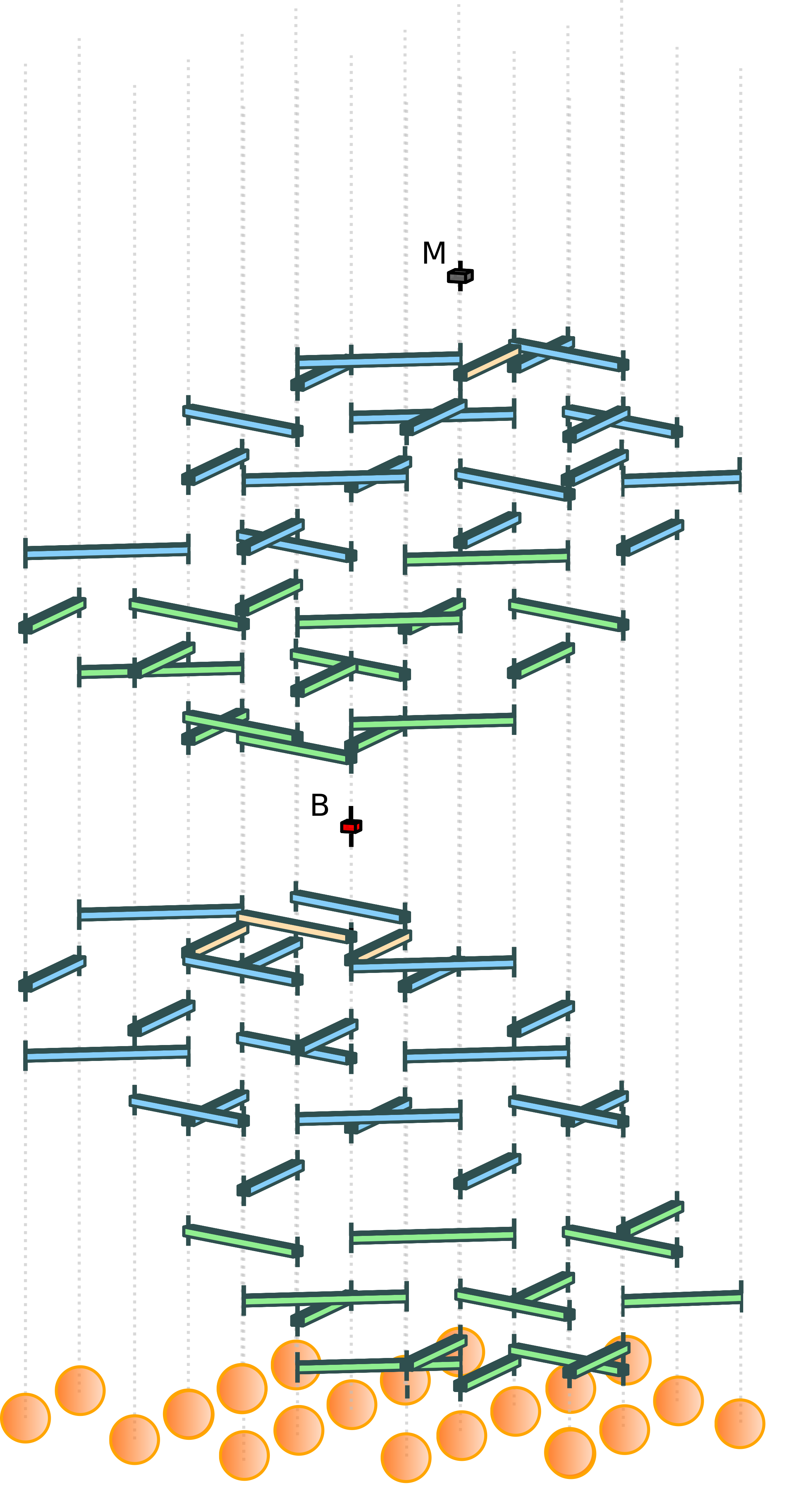}
        \label{fig:circuit-n20}
    }
    \hspace{0.03\linewidth}
    \subfloat[N=28]{
        \includegraphics[width=0.27\textwidth]{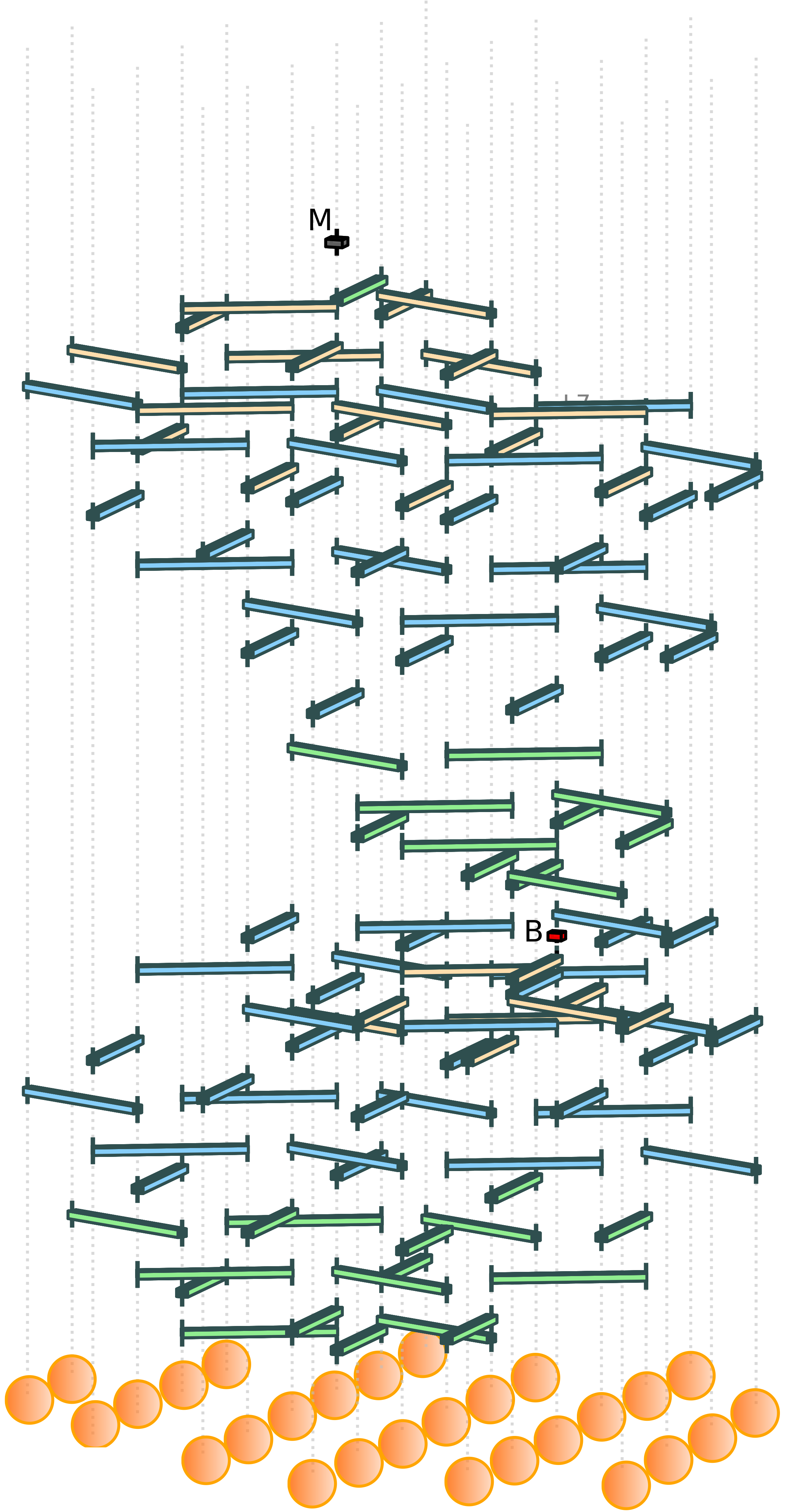}
        \label{fig:circuit-n28}
    }
    \caption{\textbf{Diagrams of the quantum circuits from the scaling analysis for $\text{N} = 10, 20, 28$}. Single qubit gates have been absorbed into two-qubit gates. Different colors denote different layers of the quantum circuit, where each layer is defined as having a single two-qubit gate on a given edge of the network. Once a second two-qubit acts upon a given link, that gate defines the start for a new layer.
}
    \label{fig:circuits-diagram}
\end{figure*}

\circled{4}: We select the final location of $B$, to define the OTOC ensemble we will study with TNBP, by maximizing exact tensor network contraction cost (as a proxy for classical hardness more generally).  To be precise, for each remaining candidate location of $B$, we evaluate the cost of exact tensor network contraction of the pruned circuit (from step \circled{3}).  We estimate the cost using a tensor network contraction optimizer (TNCO) based on simulated annealing, as described in \cite{google2025observation}, with code available in \cite{google2025tnco}. One detail is that we estimate contraction cost not for the actual unitary circuits $U$, where some angles in the random single-qubit gates take discrete values as described in Sec.~\ref{subsec:circuit_design}, but rather for fully Haar-random single-qubit gates; this avoids bias in the contraction path.  In the example shown in Fig.~\ref{fig:standard deviation}, the result is to place $B$ at the location labeled $q_b$.

The end result of steps \circled{1}\, - \,\circled{4} is an OTOC ensemble with fixed positions for $M$ and $B$ as a function of the depth of the random time evolution circuit $U$.  

\subsubsection{Relationship between circuit depth and system size}

Furthermore, this procedure makes the system size $N$ also a function of the circuit depth.  When the initial qubit grid in \circled{1} is large enough that the lightcones of $M$ and $B$ do not yet reach the edge of the grid within the depth of $U$, all qubits around the edges of the initial grid are pruned away during step \circled{3}.  Then the size of the initial grid does not affect the number of qubits in the final circuit.  Instead, the final system size only depends on how large the lightcones of $M$ and $B$ can become during the time evolution $U$, giving a linear size approximately proportional to the circuit depth (note, however, that finite size effects can make the scaling deviate substantially from direct proportionality for the small systems/low depths that we study in this paper).  

The circuits generated for the scaling analysis in this paper are close to this case, where the initial qubit grid is effectively infinite.  However, in some of the larger circuits, small effects from the boundaries of the initial qubit grid do remain.  For example, in the case of $N=23$, the lightcone of the butterfly operator does reach the edge; this can be seen from the final qubit layout in Fig.~\ref{fig:circuits_last_t}, where for $N=23$ we would expect the circuit to include two qubits to the left of the butterfly operator (red circle), but the second qubit to the left would be beyond the edge of the grid.

In~\cite{google2025observation} the circuits were generated using exactly the same steps \circled{1} - \circled{4} except that (a) more general butterfly operators were considered, including multi-qubit operators; and (b) the initial qubit grids were small enough relative to the depth of $U$ that the lightcones of $M$ and $B$ could quickly reach the edge of the grid within the evolution, hence the system size was not directly a function of depth.  In particular, (b) explains why in the experiment there could be two distinct circuits, with different circuit depths, both with $N=23$ qubits.  In contrast, with our construction, using a very large initial qubit grid relative to the depth of $U$, we find a 1-1 relationship between system size and circuit depth.

\subsubsection{Illustrations of the derived circuits}

We follow steps \circled{1}\,-\,\circled{4} for a variety of depths of the time evolution $U$.  After pruning (step \circled{3}), the resulting circuits have system sizes ranging from 8 to 28 qubits.  We show the final layout of qubits in each case, marking the locations of the measurement (purple) and butterfly (red) operators in Fig.~\ref{fig:circuits_last_t}.

We also show the full layout of entangling two-qubit gates in Fig.~\ref{fig:circuits-diagram} for three representative system sizes, $N = 10, 20, 28$.  Single-qubit gates have been absorbed into two-qubit gates for convenience.


\section{Details of tensor networks with belief propagation method}
\label{app:TNBP}

\begin{figure*}[t]
    \centering
    \includegraphics[width=1\linewidth]{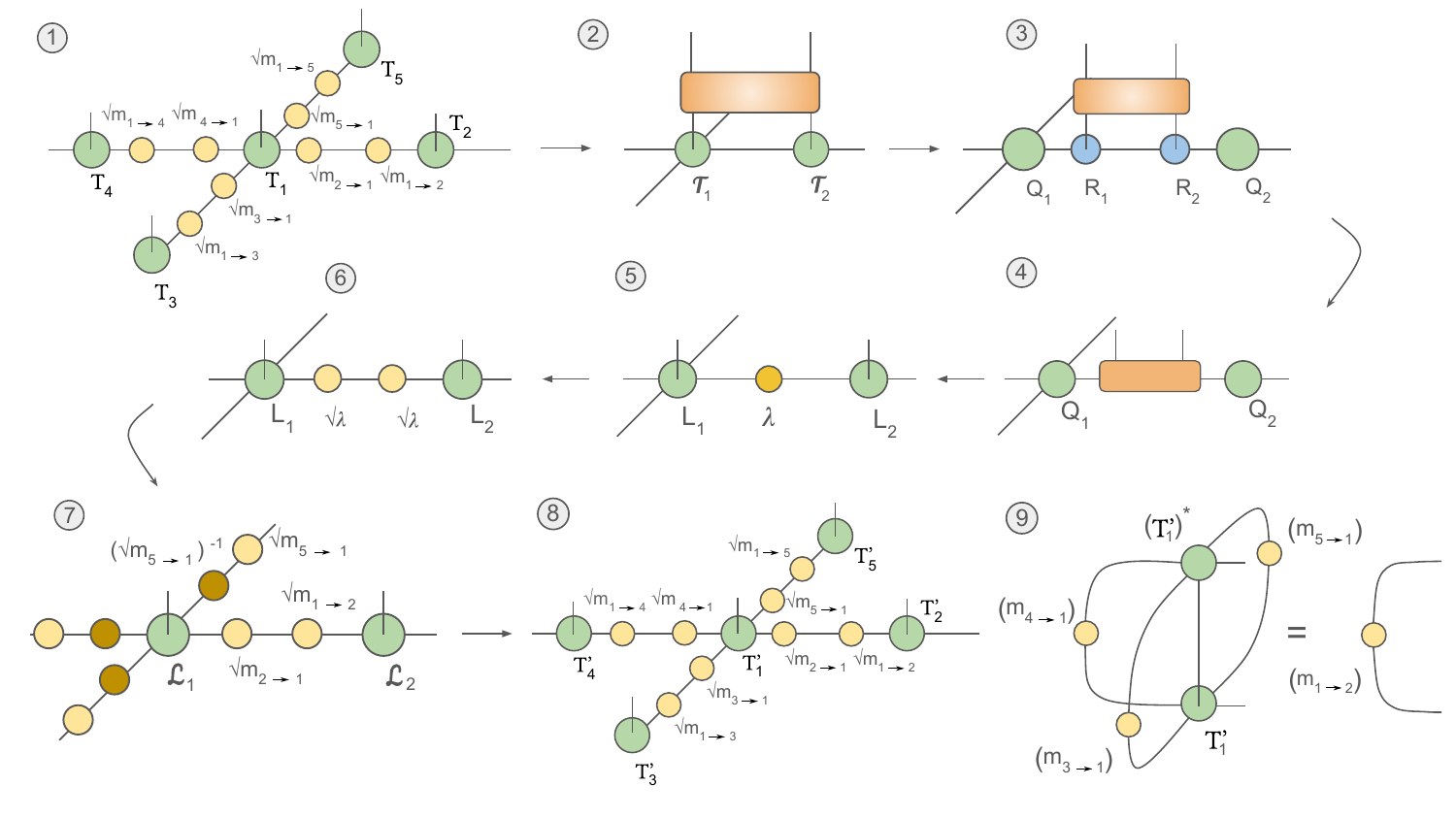}
    \caption{\textbf{Time evolution update via Belief Propagation}. 
(1) Representation of the PEPS in which the square roots of the messages are placed on the edges between tensors.
(2) During the time evolution, a gate is applied to the PEPS after incoming messages are absorbed into target tensors.
(3) The target tensors are QR-decomposed to reduce the computational cost of applying the gate.
(4) After contracting the gate, we perform a singular value decomposition of the resulting tensor, absorb the left and right matrices into the target tensors, and place the diagonal matrix of singular values on the edge between them.
(5) The singular values are split symmetrically.
(6) These split factors are promoted to incoming messages.
(7) We also insert resolutions of the identity so that the network reaches a form equivalent to step~(1), where tensors have been updated after one time step.
(8) We repeat this procedure for the remaining tensors in the PEPS.
(9) Finally, we run a self-consistency belief-propagation algorithm until the messages converge.
}
    \label{fig:BP}
\end{figure*}

To evolve a quantum state represented as a PEPS under a quantum circuit, while keeping computational costs manageable by limiting the bond dimension $D$, we must perform truncations. In particular, when a two-qubit entangling gate is applied to two adjacent PEPS tensors, the immediate result is a larger two-site tensor. This tensor can then be exactly decomposed (via singular value decomposition (SVD) or another factorization) back into two separate PEPS tensors, but with an increased bond dimension connecting them. For a system of qubits ($d=2$), the new bond dimension $D'$ is generically $4D$. To maintain computational efficiency, we truncate the bond back to $D$, ideally discarding as little information as possible.

The method used to perform this truncation is called an update scheme. Its precision and computational cost depend strongly on features such as the connectivity of the tensor network and the amount of entanglement shared between the subsystems.

For PEPS, the high connectivity makes truncations costly. Additionally, the highly entangling nature of OTOC evolution causes the bond dimension $D$ to grow rapidly, further increasing truncation complexity.

A variety of methods have been explored in the literature, ranging from computationally cheaper schemes, like simple update~\cite{jiang2008accurate} (whose cost scales as $\mathcal{O}(D^5)$) to more precise and demanding ones, as in the case of full update (with a cost scaling as $\mathcal{O} (N (D^4\chi^3 + D^6\chi^2))$ where we assume a boundary MPS rank of  $\chi$)~\cite{jordan2008classical, lubasch2014algorithms}. In this work, we focus on belief propagation (BP), an update scheme adapted from graphical models to the language of tensor networks~\cite{alkabetz2021tensor}. We choose BP due to (a) its favorable computational scaling ($\mathcal{O}(\,D^5)$), and (b) its successful application in the classical simulation of recent quantum experiments~\cite{tindall2024efficient, tindall2025dynamics}.

Belief propagation itself has several variants, differing mostly in implementation details. Overall, BP is a message-passing algorithm designed to compute the action of the corresponding environment on each node of the network. It does so by analyzing the state of a node conditioned on the states of other nodes. In tensor-network language, this means that (1) information flows between qubits in the form of messages, and (2) the state of each qubit is influenced by incoming messages from its neighbors. In tree-like networks, BP finds a gauge where local truncations are globally optimal and these messages provide the exact environments. In loopy networks, BP is approximate, and the truncation errors cannot be predicted \textit{a priori}. 

We implement BP following the frameworks of Refs.~\cite{tindall2024efficient, tindall2025dynamics}, using the libraries~\cite{fishman2022itensor, tnqs} to realize a particular variant of BP. The main differences between variants lie in the message-update strategy. Here, we employ a ``symmetric update'', known as the ``symmetric gauge''~\cite{tindall2023gauging}, which ensures that for a given edge, the information flowing toward one node is equal to that flowing toward the other.

The full procedure is illustrated in Fig.~\ref{fig:BP}. We begin by representing the wavefunction as a PEPS with one tensor $T$ per qubit. For numerical stability, the message-passing step is performed in the ``double-layer picture,'' where each node is promoted to a positive semi-definite matrix. We denote the double-layer messages by $m_{i \rightarrow j}$, which reside on the edges and encode the effective environment that tensor $T_i$ and its conjugate $T_i^*$ induce on tensors $T_j$ and $T_j^*$.

After introducing the double-layer picture, we return to the single-layer (ket) picture, as shown in \circled{1} of Fig.~\ref{fig:BP}, which is convenient for applying quantum gates. In this representation, each message is factorized as $\sqrt{m_{i \rightarrow j}}$. These single-layer messages also reside on the edges, carrying information from tensor $T_i$ to tensor $T_j$.

Consider a two-qubit gate acting on tensors $T_1$ and $T_2$. Before contracting the gate tensor, we absorb the effect of incoming messages on the target tensors. For instance, for $T_1$ we absorb the messages $\sqrt{m_{3 \rightarrow 1}}, \sqrt{m_{4 \rightarrow 1}}, \sqrt{m_{5 \rightarrow 1}}$ to form the effective tensor $\mathcal{T}_1$ (see \circled{2}).

To reduce the computational cost of the subsequent SVD, we first perform a QR decomposition of $\mathcal{T}_1$ and $\mathcal{T}_2$ and contract the resulting $R$ tensors with the gate. The intermediate tensor (shown in \circled{4}) is then decomposed via an SVD. The unitary matrices $U$ and $V$ are absorbed into $Q_1$ and $Q_2$, producing updated tensors $L_1$ and $L_2$. The diagonal matrix $\lambda$ is truncated to the maximum bond dimension $D$. BP guarantees that this truncation is globally optimal in loopless networks, but in loopy networks it is approximate, and errors accumulate over time.

The diagonal matrix $\lambda$ is symmetrically distributed between the tensors by taking its square root, yielding $\mathcal{L}_1$ and $\mathcal{L}_2$. In the double-layer picture, $\lambda^2$ defines the updated messages $m_{1 \rightarrow 2}$ and $m_{2 \rightarrow 1}$, while in the single-layer picture, $\lambda$ directly becomes the edge messages $\sqrt{m_{1 \rightarrow 2}}$ and $\sqrt{m_{2 \rightarrow 1}}$. This symmetric distribution is the essence of the symmetric gauge.  
Next, the influence of previously absorbed neighboring messages is removed by multiplying with their inverses. 

The procedure is repeated for a collection of quantum gates—typically arranged in layers so that at most one two-qubit gate acts on each edge—until step \circled{8} in Fig.~\ref{fig:BP}. 

Convergence is then enforced according to the BP message-passing criterion shown in step \circled{9}. 
In the double-layer formulation, the incoming messages $m_{k \rightarrow i}$ are absorbed into a tensor to form an effective environment, which determines the outgoing messages $m_{i \rightarrow j}$ for neighboring tensors. This iterative update continues until a chosen convergence threshold is reached.

Throughout the time evolution, the bond dimension of the messages is bounded by $D \times D$, where $D$ is the maximum bond dimension retained. Additionally, a numerical cutoff is applied during the SVD at step \circled{5}: singular values below $10^{-14}$ are discarded.  


\section{Defining incompressibility of a PEPS or quantum circuit}
\label{app:def_incompressibility}

\begin{figure}
    \centering
    \includegraphics[width=\linewidth]{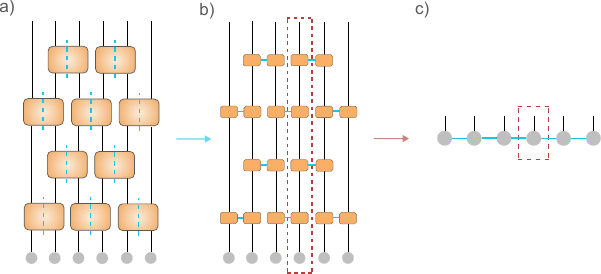}
    \caption{\textbf{Construction of an exact PEPS from a quantum circuit.} We illustrate a procedure to exactly convert any quantum circuit with only single-qubit or nearest-neighbor gates to a PEPS.  We illustrate the procedure in 1D for clarity, but it naturally extends to higher dimensions.  Beginning with the circuit in (a), we perform an SVD on each two-qubit gate, getting two tensors of size $2\times 2\times 4$, where the larger dimension is the one linking the two tensors.  The result of performing all these decompositions is shown in (b).  If we contracted adjacent tensors along the cyan bonds, we would recover the original circuit from (a).  Then for each qubit we contract along the physical legs all the single-qubit tensors on that site that were produced by the SVDs; we also include any single-qubit gates acting on that qubit (not shown).  The contracted tensors for one example qubit are those inside the dashed red box.  The end result is one large tensor for each qubit, giving an exact tensor network representation of the final state, shown in (c).}
    \label{fig:PEPS_no_trunc_construction}
\end{figure}

A central concept in our argument that random-circuit OTOCs are hard to simulate classically is that the random-circuit evolution is incompressible.  In this appendix, we define what we mean by incompressibility.

We start by defining incompressibility for a PEPS representing a wavefunction.  First we say that a PEPS bond index can be compressed if it can be reduced without losing significant fidelity and without needing to increase the bond dimensions elsewhere in the PEPS.  Otherwise, we say the bond index is incompressible.  Then we say a PEPS is incompressible if all its bond indices are incompressible.  In practice, a PEPS may be compressible in some spatial regions and not in others.

Finally, we use this definition for incompressibility of a PEPS to define incompressibility of a quantum circuit.  Our definition is based on construction of an exact PEPS representation of the output state of the quantum circuit when applied to an initial product state.  We say that the circuit is incompressible if the exact PEPS is incompressible, in the sense defined in the previous paragraph.

We now provide an explicit construction of the exact PEPS, which we illustrate for a 1D system in Fig.~\ref{fig:PEPS_no_trunc_construction}.  First, we perform a singular value decomposition (SVD) of all two-qubit gates in the circuit, illustrated by the cyan dashed lines in Fig.~\ref{fig:PEPS_no_trunc_construction}(a).  Because the gates act on qubits, the maximum number of singular values for each gate is four.  Thus each gate is decomposed into two $2\times2\times 4$ tensors, one on each of the two qubits acted on by the gates; contracting along the auxiliary leg created by the SVD, of dimension 4, would produce the original two-qubit gate.  The rank-three tensors are shown in Fig.~\ref{fig:PEPS_no_trunc_construction}(b), with the auxiliary leg shown in cyan.  
Finally, we contract all of the tensors acting on a given qubit, indicated by the dashed red box in Fig.~\ref{fig:PEPS_no_trunc_construction}(b), along their physical legs.  In the direction connecting to each neighboring qubit, the auxiliary legs of dimension 4 are merged, giving a single large leg or bond index of dimension [4 raised to the power of the number of gates acting on that bond].\footnote{In fact, the final dimensions of some bond indices in the exact PEPS may be lower because some gates, for example those acting directly on the initial product state, can only increase the bond dimension by a factor of 2 rather than 4. See the discussion of bond dimensions in 1D below.}  The final tensor network with these exponentially large bond dimensions (indicated by thick cyan lines) is shown in Fig.~\ref{fig:PEPS_no_trunc_construction}(c).

The construction in 2D works exactly the same way.  For each two-qubit gate, we decompose it with an SVD, then contract the resulting rank-three tensors onto the PEPS tensors for the respective qubits along the physical legs.  This increases the PEPS bond dimension by a factor of 4 on the bond where the gate is applied. Thus we can explicitly construct a PEPS representation for the output state that has, on each bond between adjacent qubits, a bond dimension equal to 4 raised to the power of the number of two-qubit gates applied on that bond.\footnote{Note that the exact PEPS does not necessarily have uniform bond dimension throughout the system, as some bonds may have had more gates applied on them than others.}  We refer to this PEPS as the exact PEPS for the output state.

Note that there is a subtlety in 1D.  While the construction of an exact tensor network for the output state of a circuit, described above and illustrated in Fig.~\ref{fig:PEPS_no_trunc_construction}, also gives a bond dimension $4^{\text{\# gates}}$ in 1D, in fact for our OTOC circuits it is always possible to compress to $D_\text{max}=2^{\text{\# gates}}$.  Imagine that a gate is applied to two adjacent sites of a matrix product state with uniform bond dimension $D$.  An SVD of the gate still produces two rank-three tensors connected by an auxiliary leg of dimension 4, so it seems that the dimension of the bond on which the gate is applied should indeed by multiplied by 4.  However, consider the composite tensor formed by applying the gate onto the two MPS sites: it has dimensions $2D\times 2D$, and hence when decomposed back into two MPS tensors via an SVD, it can have at most $2D$ nonzero singular values, and thus the bond dimension is multiplied by only $2$.\footnote{In contrast, if we consider applying a gate to two adjacent PEPS tensors in 2D, the composite tensor has dimensions $2D^3 \times 2D^3$, and hence as long as $D>1$ the number of singular values after decomposing is not limited by the size of this composite tensor but rather by the 4 singular values of applied gate.}

In some 1D evolutions, bond dimensions will still grow by a factor of 4 with each gate.  Starting from the product state $|0\rangle^{\otimes N}$, imagine applying a uniform layer of two-qubit gates on all even bonds.  This makes the MPS bond dimensions 2 on even bonds and 1 on odd bonds.  Now consider a gate applied to an odd bond.  The composite tensor after applying the gate has dimensions $4\times 4$ or $2D_\text{even}\times 2D_\text{even}$.  This tensor is larger than $2D_\text{odd}\times 2D_\text{odd}$, because the MPS bond dimension is $2\times$ higher on even bonds.  Thus the bond dimension on the odd bonds can grow by a factor of $2D_\text{even}/D_\text{odd}=4$.  The same logic applies for all subsequent gate layers.

In our 1D OTOC circuits, the increase in bond dimension from each gate is a factor of 2 rather than a factor of 4.  Because we prune gates outside the geometric lightcones, we are not precisely in either of the cases discussed in the previous two paragraphs: the MPS bond dimension is not uniform across the system after a fixed number of gate layers have been applied, but we also do not apply uniform gate layers across the system.  However, as can be checked explicitly by considering the first few layers of the circuit, the fact that gates are applied starting from $M$ and in a region of increasing size in subsequent layers leads to the growth by a factor of 2 with each gate.  As in the case of gates applied to an MPS with uniform $D$, the slower growth comes from the maximum number of singular values of the composite tensor when each gate is applied.


\section{Geometry of two-dimensional lightcones and resulting cost scaling}
\label{app:2D_lightcone_geometry}

\begin{figure*}
    \centering
    \includegraphics[width=\linewidth]{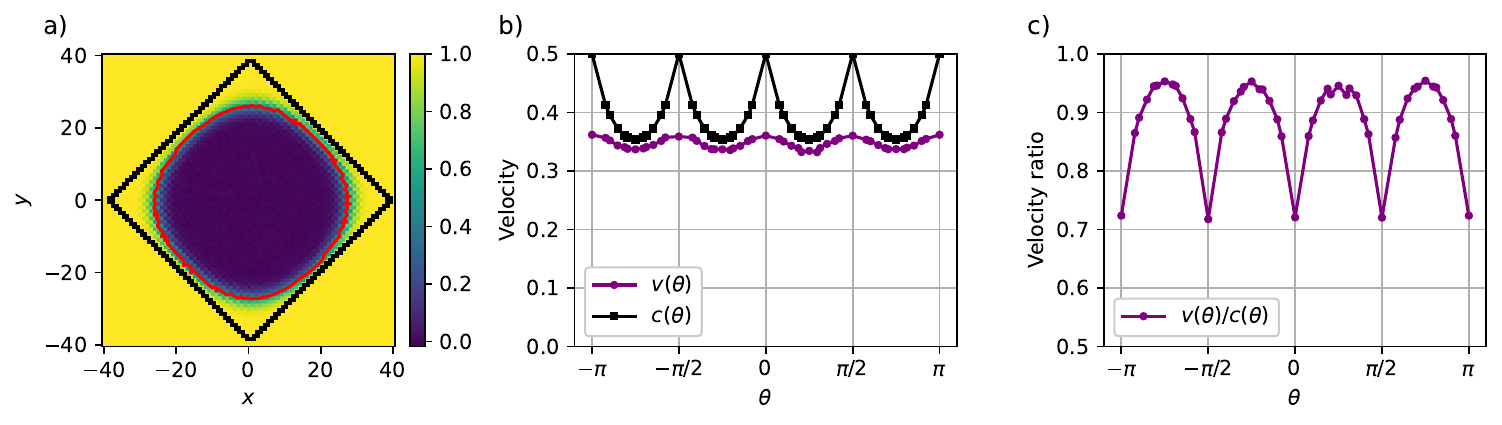}
    \caption{\textbf{2D geometric and physical lightcones.} 
    Using Monte Carlo simulations of the ensemble averaged OTOC values~\cite{google2025observation}, we find the how the physical lightcone grows for large systems with the random circuit ensemble as described in Sec.~\ref{subsec:circuit_design}.  This physical lightcone has a non-circular shape described here using an angle-dependent butterfly velocity $v(\theta)$, which does not match the square shape of the geometric lightcone -- described here with a corresponding velocity $c(\theta)$. The physical lightcone is sensitive to the choice of the gate ensemble and the precise pattern of how entangling gates are laid out in the the circuit; our circuit uses gate layers as shown in Fig.~S7a in the Supplementary Information of~\cite{google2025observation}. (a) The lightcone cross-section after 80 layers of two-qubit gates.  The measurement operator $M$ is placed at the center of the system, and the color shows the value of the OTOC for each location of $B$.  The black diamond shows the edge of the geometric lightcone; the OTOC outside of this region is exactly 1.  The red line shows the physical lightcone edge, where the average OTOC crosses 1/2.  (b)  The velocity at which the physical lightcone spreads for each angle, $v(\theta)$, is extracted using a linear fit of distance to the contour where OTOC crosses 1/2 versus depth [red line in (a)]. The velocity of the geometric lightcone [black line in (a)] for each angle, $c(\theta)$, is shown for comparison. Fits for the velocity use depths from 30 to 80. (c) Speed of the physical lightcone edge relative to the geometric lightcone edge, computed from the ratio of the curves in (b).  In the directions of the primitive lattice vectors, the ratio is $v/c\approx 0.7$, while in the diagonal directions the ratio is $v/c\approx 0.95$.
    }
    \label{fig:2D_lightcones_MC}
\end{figure*}

\begin{figure}
    \centering
    \includegraphics[width=\linewidth]{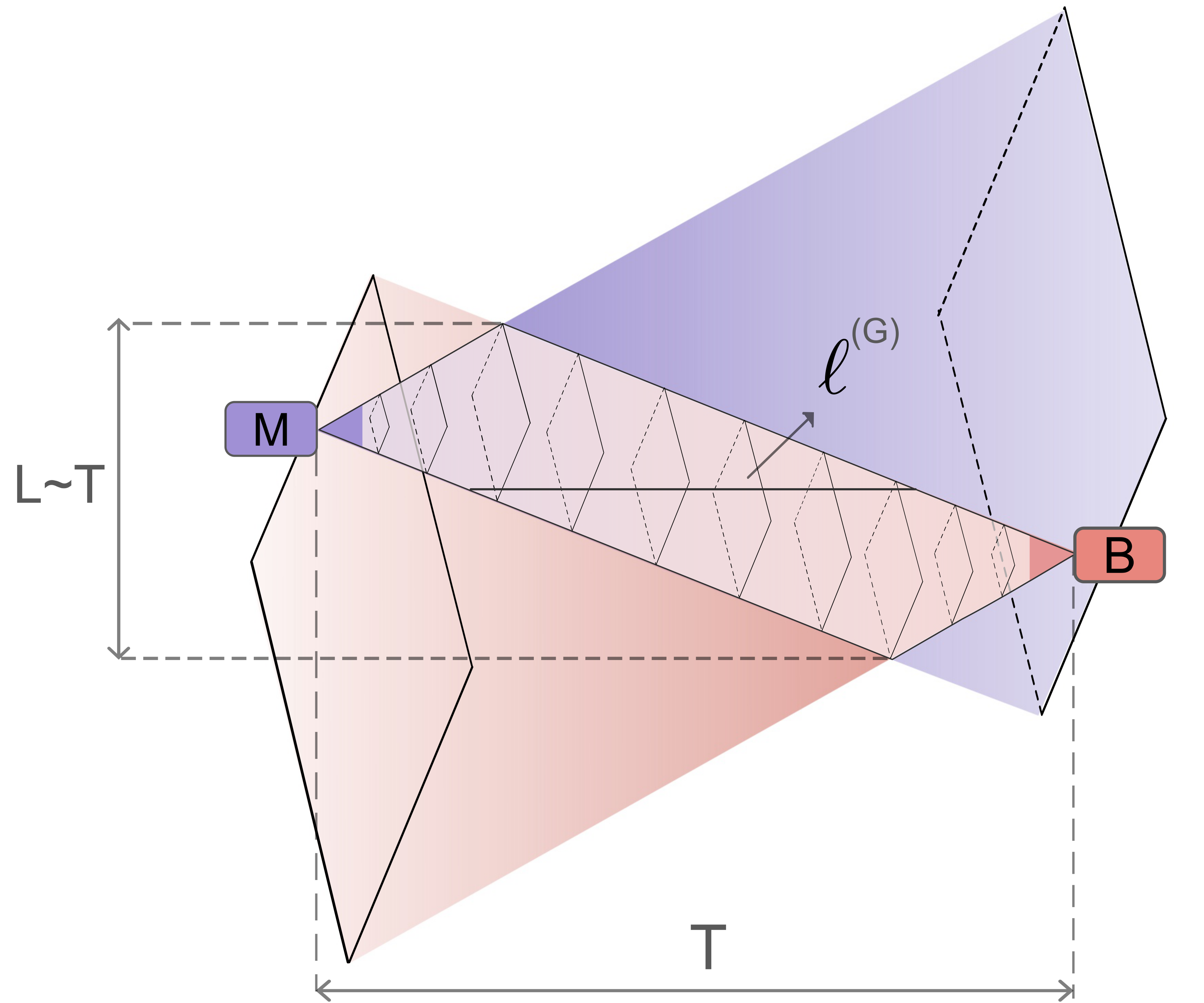}
    \caption{\textbf{2D geometric lightcones, ``horizontal'' case.} In Fig.~\ref{fig:depth_scaling}(c) in the main text, we show the geometric lightcones in two spatial dimensions, in the case that the separation between the $M$ and $B$ operators is at angle of $\pi/4$ to a primitive lattice vector, what we call the ``diagonal'' case.  Here we show what the lightcones and their intersections look like if the direction from $M$ to $B$ is along a primitive lattice vector.}
    \label{fig:2D_geometric_lightcones_horizontal}
\end{figure}

Here we provide further details on the geometric reasoning used to derive the OTOC cost scaling in 2D, as summarized in Eq.~\eqref{eq:expected_scaling_2D_sep_x}.

There are two main reasons why the cost scaling derivation is more involved in 2D than in 1D.  First, for a random brickwall circuit evolution, the geometric lightcone in 2D has a square rather than circular cross-section at any given time $t$.  Since rotational symmetry around, say, $M$ is then lost, the required bond dimension will depend on the relative orientation of $M$ and $B$ relative to the primitive lattice vectors of the square lattice.  Second, the physical lightcone has a different shape from the geometric one --- it is not just a square cross-section scaled down by a constant factor $v/c$. The shape of the physical lightcone of a typical circuit from our ensemble can be measured by using the ensemble-averaged OTOC, which is efficiently computable by virtue of the invariance of the ensemble under single qubit Pauli gate insertions, as shown in Ref.~\cite{google2025observation}. We compute the ensemble-averaged OTOC with the stochastic Monte Carlo algorithm of Ref.~\cite{google2025observation}, using $50,000$ total Pauli trajectories from randomly chosen circuits in our ensemble, and measure the extent of the physical lightcone by determining the locations where the OTOC crosses the value $\frac12$. The result is shown in Fig.~\ref{fig:2D_lightcones_MC} --- the physical lightcone has a cross-section which is like a square with rounded-off corners. We additionally extract velocities for a number of angular rays from linear fits to these distances versus time, showing that the physical lightcone size is an angle-dependent fraction of the geometric lightcone, where the fraction  varies from $v/c \approx 0.7$ along lattice directions to $v/c \approx 0.95$ along diagonal directions.

We will explicitly find the scaling of the required bond dimension with system size for two orientations of the relative position of $M$ and $B$.  To be precise, let $\hat{x}$ and $\hat{y}$ be the directions of the primitive lattice vectors of the square lattice.  Then we consider the cases: (1) ``horizontal,'' where the line from $M$ to $B$ lies along $\hat{x}$, $\theta=0$ in Fig.~\ref{fig:2D_lightcones_MC}; and (2) ``diagonal,'' where the line lies along $\hat{d}\equiv(\hat{x}+\hat{y})/2$, $\theta=\pi/4$ in the figure.  We will show that the required bond dimension is higher in the horizontal case than in the diagonal case.  The required bond dimension for all other orientations falls between these two limiting cases.

Let us start with a brief summary of the geometry of the lightcones.  The geometric lightcone is a square, with sides along the diagonal directions $\hat{d}$ and $\hat{d}'\equiv(\hat{x}-\hat{y})/2$.  The lightcone edge therefore travels fastest along the primitive lattice vectors, at a speed $c_h$ ($h$ for ``horizontal''), and more slowly in diagonal directions.  The speed along $\hat{d}$, which we call $c_d$ ($d$ for ``diagonal'') is slower by a factor of $\sqrt{2}$.  The physical lightcone is closer to circular.  As shown in Fig.~\ref{fig:2D_lightcones_MC}(b), the physical lightcone velocity varies by only about 10\% from one direction to another.  The result is that the physical lightcone spreads nearly as fast as the geometric lightcone along the diagonal directions ($v_d/c_d\approx 0.95$), and more slowly relative to the geometric lightcone along $\hat{x}$ and $\hat{y}$ ($v_h/c_h\approx 0.7$); this is shown in Fig.~\ref{fig:2D_lightcones_MC}(c).  

\paragraph{Scaling in diagonal case:} 
First consider the case where $M$ and $B$ are separated along $\hat{d}$, as illustrated in Fig.~\ref{fig:depth_scaling}(c).  Then as in 1D the maximum number of gates applied on any given bond is proportional to $\ell^{(G)} = (1-v_{MB}/c)T$ where $v_{MB}$ gives the speed of propagation needed to reach $B$ from $M$ in time $T$.  However, here we have to be more specific about what we mean by $c$ since the geometric lightcone spreads at different speeds in different directions.  The correct expression is $\ell^{(G)} = (1-v_{MB}/c_d)T$ where $c_d$ is the speed that the geometric lightcone spreads in the direction $\hat{d}$, at $\theta=\pi/4$.  

Unlike in 1D, it is now the linear system size $L$, rather than the full size $N\sim L^2$, that is proportional to $T$.  More precisely, the full set of qubits after pruning gates outside the geometric lightcones is given by an octagon with regular angles and alternating sides of length $(c_d-v_{MB})T$ and $v_{MB}T/\sqrt{2}$, which has area $N = (c_d^2 - v_{MB}^2/2)T^2$.  Then as claimed in Sec.~\ref{subsec:IIB_scaling_theory}, $\ell^{(G)}$ scales as $L\sim\sqrt{N}$, and thus the required PEPS bond dimension scales exponentially in the square root of system size, $D\sim\exp(\sqrt{N})$.

To find the precise asymptotic scaling for our brickwall circuits, we need to translate from this geometric picture to one in which we take into account the discrete gate layers.  We explicitly define $T$ to be the number of 2-qubit gate layers in $U$.  From this definition, in 2D we get $c_d=1/(2\sqrt{2})\approx 0.35$ because along the directions of the primitive lattice vectors (\textit{e.g.} along $\hat{x}$), only half of the layers have gates in that direction, while the other half propagate information in a perpendicular direction, giving $c_h=1/2$, and the lightcone speed along the diagonal is reduced by a factor of $\sqrt{2}$, $c_d = c_h/\sqrt{2}$.  

We also translate the geometric length $\ell^{(G)}$ to number of gate layers applied on a bond.  If $T$ counts total number of gate layers, then $\ell^{(G)}$ counts the maximum number of local gate layers around any given bond, including those gate layers that do not include a gate on the specific bond of interest.  Only one fourth of the layers in a 2D brickwall circuit apply a gate on any particular bond, so the actual number of gates applied to a specific bond during $U$ is $\ell^{(G)}/4$.  In finding the OTOC we evolve by both $U$ and $U^\dagger$, making the maximum number of gates per bond $\ell^{(G)}/2$.

Then the final expression for the expected number of maximum number of gates applied on a given bond in the OTOC circuit is
\begin{equation}
    \text{\# gates} = \sqrt{N}\cdot \frac{1}{2}\cdot\frac{1-v_{MB}/c_d}{\sqrt{c_d^2-v_{MB}^2/2}}. \label{eq:2D_scaling_diagonal}
\end{equation}
For OTOC$^{(2)}$ the number of gates is doubled.  
Specializing to OTOC circuits that have $B$ right on the edge of the physical lightcone of $M$ in order to maximize the size of instance-to-instance fluctuations in the OTOC so that $v_{MB}$ is equal to the speed at which the physical lightcone spreads, we can substitute the values $c_d = 1/(2\sqrt{2})$ and $v_{MB}/c_d \approx 0.95$ in Eq.~\eqref{eq:2D_scaling_diagonal}.  This gives the maximum number of gates on a given bond as approximately $0.1\sqrt{N}$, and hence the final bond dimension scaling for OTOC in the diagonal case is $D = A_2\cdot 4^{0.1\sqrt{N}}$, for some constant $A_2$ that depends on the desired level of accuracy.  
$A_2$ is at most 0.5, since it would be 1 if every gate multiplied the bond dimension by 4 and the very first layer only multiplies $D$ by 2 because it acts on a product state.

\paragraph{Scaling in horizontal case:} 
Next we consider the case where $M$ and $B$ are separated along $\hat{x}$, as illustrated in Fig.~\ref{fig:2D_geometric_lightcones_horizontal}.  Again the local circuit depth is given by $\ell^{(G)} = (1-v_{MB}/c)T$, but now we use $c=c_h$, the speed that the geometric lightcone spreads along $\hat{x}$.   
The total system size after pruning is $N = T^2(c_h^2 - v_{MB}^2)/2$.  We once again find that $\ell^{(G)}$ scales as $L\sim\sqrt{N}$ and hence $D\sim\exp(\sqrt{N})$, but the details of the scaling are different from the diagonal case.

Again using the fact that the maximum number of gates per bond is $\ell^{(G)}/2$, and solving for $T$ as a function of $N$, we find
\begin{equation}
    \text{\# gates} = \sqrt{N}\cdot \frac{1}{\sqrt{2}}\cdot\frac{1-v_{MB}/c_h}{\sqrt{c_h^2-v_{MB}^2}}
\end{equation}
for OTOC, and double that for OTOC$^{(2)}$. 
Again assuming that we choose $v_{MB}$ to be the speed of propagation of the physical lightcone edge, we can substitute in $c_h=1/2$ and $v_{MB}/c\approx 0.7$, giving a maximum number of gates on any given bond that scales as approximately $0.6\sqrt{N}$.  Hence the final predicted bond dimension scaling for OTOC in the horizontal case is $D = \tilde{A}_2\cdot 4^{0.6\sqrt{N}}$ for $\tilde{A}_2 \leq 0.5$ another constant depending on desired accuracy.  This is precisely Eq.~\eqref{eq:expected_scaling_2D_sep_x}.  

\paragraph{Which scaling to use in cost estimates:} 
Evidently, when the butterfly operator $B$ is placed on the edge of the physical lightcone of the measurement operator $M$ in order to maximize the signal, \textit{i.e.} the instance-to-instance fluctuations, the simulations are much harder classically in the horizontal case, where the operators are separated along a primitive lattice vector, than in the diagonal case.  If the relative orientation of $M$ and $B$ is somewhere in-between these limits, at an angle $\theta < \pi/4$ from $\hat{x}$, the cost scaling will be somewhere in-between as well.

As described in App.~\ref{section:obtain_lightcone}, we consider OTOC circuits which are constructed first by choosing relative positions for $M$ and $B$ that give a large signal, then by selecting among these positions the one that has the highest projected cost for exact tensor network contraction.  Thus the geometric reasoning above suggests that this process would select the horizontal case, with $M$ and $B$ separated along $\hat{x}$, to maximize cost.  
We therefore use the predicted scaling from the horizontal case in Eq.~\eqref{eq:expected_scaling_2D_sep_x} in the main text. 

It is an important note, however, that this scaling \emph{can} underestimate the true cost for small systems, since the circuit depth may be determined more by edge effects than by the geometry used here.


\section{Heisenberg-picture approaches to simulating the OTOC}
\label{sec:Heis_picture}

\begin{figure*}[ht]
    \centering
    \includegraphics[width=0.8\linewidth]{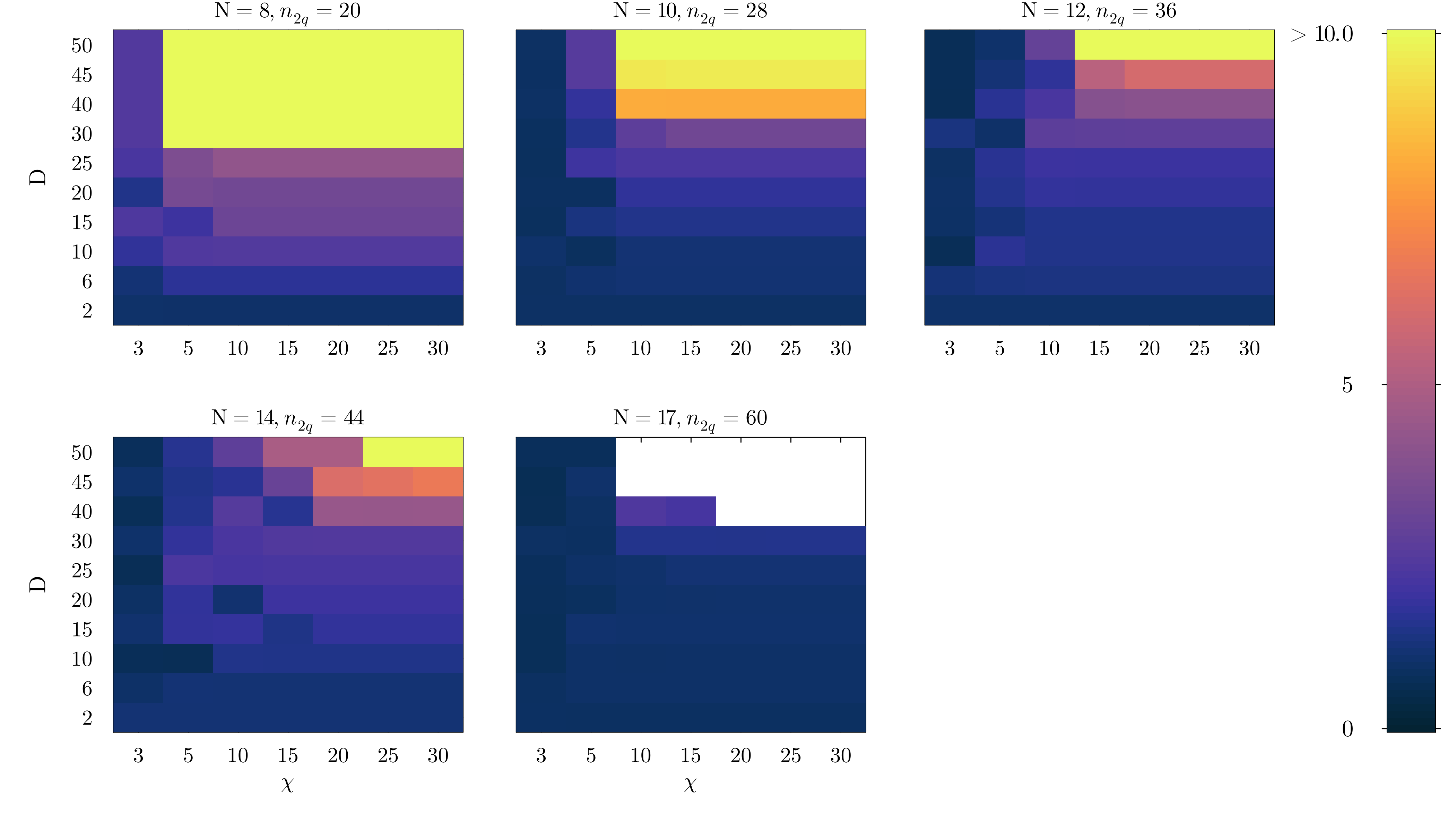}
    \caption{\textbf{SNR for Heisenberg-evolved operator in the vectorized picture}
    We show the same analysis as in Fig.~\ref{fig:heatmap_scaling}, and for the same OTOC ensembles, but with evolution performed in the Heisenberg picture, evolving the butterfly operator $B$ in a doubled Hilbert space. $D$ is the PEPS bond dimension and controls the accuracy of time evolution; $\chi$ is the boundary MPS bond dimension and controls the accuracy of extraction of the final OTOC observable.  We find that, compared with the Schr\"{o}dinger picture approach discussed in the main text, this Heisenberg picture approach requires a larger $D$ and $\chi$ to achieve the same SNR for the same OTOC ensemble.  The white region for $N=17$ indicates where bMPS contraction would exceed the available memory on one a2-ultragpu-8g node on Google Cloud.
    }
    \label{fig:heatmap_scaling_vectorized}
\end{figure*}

In the main text, we consider the cost of TNBP simulation of OTOCs, which consists of: an evolution step, in which we evolve the initial state $|0\rangle^{\otimes N}$ to get $|\phi\rangle$ in the case of the first-order OTOC, or $|\varphi\rangle$ in the case of OTOC$^{(2)}$; and the extraction step, in which we compute the expectation value of the measurement operator $M$ in this state.

In this Schr\"{o}dinger picture approach, as discussed in Sec.~\ref{subsec:scaling_with_compressibility} it is possible to take advantage of the fact that gates outside the dressed physical lightcones of $M$ and $B$, even if still inside the geometric lightcones, have little effect on the OTOC.  However, as shown also in Sec.~\ref{subsec:scaling_with_compressibility}, we still need a PEPS bond dimension that scales exponentially in linear system size.

In this Appendix, we argue that simulations in the Heisenberg picture could further exploit compressibility outside the dressed physical lightcones to achieve an improved cost scaling.  

\paragraph{Heisenberg evolution of $B$:} 
We begin with a Heisenberg picture approach that does not, in fact, improve the cost scaling.  In this approach, instead of reaching the state $|\phi\rangle$ by evolving the initial state $|0\rangle^{\otimes N}$ under $U$, then $B$, then $U^\dagger$, in the evolution step we evolve the operator $B$ as a vector in a doubled Hilbert space, $|B\rangle\!\rangle$.  In the doubled Hilbert space, the evolved operator $U B U^\dagger$ becomes $(U\otimes U^\ast\!) |B\rangle\!\rangle$.  Then in the extraction step we return to the single Hilbert space by taking index 0 for half the tensor legs to select the initial state $|0\rangle^{\otimes N}$ before proceeding to extract the OTOC value as in the Schr\"{o}dinger picture.

In Sec.~\ref{subsec:scaling_with_compressibility} we argued that in the Schr\"{o}dinger picture the OTOC circuit can be partially compressed during the evolution under $U^\dagger$ because gates outside the dressed physical lightcone of $B$ partially cancel between $U$ and $U^\dagger$.  However, we had to evolve under $U$ with no compression first.  If we had tried to compress prior to finishing evolution by $U$, the gates outside the physical lightcone of $B$ in $U^\dagger$ would have no longer approximately canceled with the truncated version of the evolution in $U$.

In the Heisenberg picture, we can better exploit the partial cancellation of gates outside the dressed physical lightcone.  Because we simultaneously apply each gate in $U$ with the corresponding gate in $U^\dagger$, we never need to temporarily increase the bond dimension before being able to compress it back down again later.  

However, the maximum bond dimension $D$ that we need still scales exponentially in $T$.  To see this, consider bonds near the location of $B$.  As indicated in Fig.~\ref{fig:geometric_compressibility}(a), the number of gates applied on these bonds still scales as $\ell^{(G)}\sim T$; the scaling is not reduced because, while we eliminate the effect of gates outside of the dressed physical lightcone of $B$, we have not made any such simplification using the dressed physical lightcone of $M$ (bounded by the purple lines in Fig.~\ref{fig:geometric_compressibility}(a)).  
Thus this first approach to Heisenberg evolution will have the same scaling as Schr\"{o}dinger picture evolution, with the additional cost of a doubled Hilbert space.

We empirically test this approach on small systems up to $N=17$ qubits.  We perform an analysis of how SNR scales with bond dimensions $D$ and $\chi$, analogous to the analysis in Sec.~\ref{sec:scaling_experiment} for the Schr\"{o}dinger picture; the results are shown in Fig.~\ref{fig:heatmap_scaling_vectorized}.  Simulations were run on an a2-ultragpu-8g system (A100 and H100 GPUs) with 80~GB of RAM. 

We find that this approach requires more computational resources than Schr\"{o}dinger picture simulation to achieve the same SNR for the same OTOC ensembles.  This can be understood from two complementary perspectives.

First, the evolution takes place in a doubled Hilbert space with local dimension $d=4$ instead of $d=2$. This increases the bond dimension required to capture correlations in a scrambling circuit, leading to higher memory usage and contraction cost.

Second, the method does not mitigate entanglement growth. Although gates outside the dressed lightcone of $B$ contribute weakly, bonds near $B$ are still acted on by a number of gate applications that scales linearly with circuit depth $T$. Consequently, the bond dimension required to accurately represent the operator grows exponentially in $T$, as in the Schr\"{o}dinger picture.

\begin{figure*}[ht!]
    \centering
    \includegraphics[width=1\linewidth]{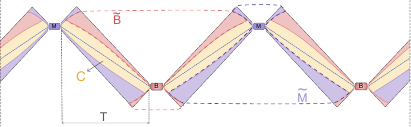 }
    \caption{\textbf{Heisenberg picture contraction scheme for computing the OTOC.}  We consider the OTOC as defined in Eq~\ref{eq:OTOC_def_original}, with a trace.  Then the OTOC can be computed by contracting the circuit shown here: two copies of $M$ (purple boxes), two copies of $B$ (red boxes), time evolutions for time $T$ between them (black-bordered parallelograms, after pruning gates outside the geometric lightcones), and a trace (taken at the dashed black lines at the edges).  Within each section of time evolution, only gates within the yellow-shaded region, which we call the irreducible core of the evolution, significantly contribute to the value of the OTOC; we call the operator defined by the circuit in the yellow region $C$.  The gates outside the irreducible core approximately (but not exactly) cancel between $U$ and $U^\dagger$.  We can therefore efficiently evolve each copy of $M$ in the Heisenberg picture to include the adjacent purple-shaded regions in the time evolution; the operator $M$ evolved under the part of $U\otimes U^\dagger$ in the purple-shaded region defines $\tilde{M}$.  We likewise define $\tilde{B}$ as the partially Heisenberg-evolved $B$, incorporating the part of the time evolution shaded red.  Then the OTOC can be computed according to Eq.~\eqref{eq:OTOC_Heis_efficient}, which can be computed as an exact tensor network contraction scaling as $\exp{\sqrt{T}}$.
    }
    \label{fig:efficient_Heisenberg_OTOC}
\end{figure*}

\paragraph{Simulation via dual Heisenberg evolution of $M$ and $B$:} 
We now describe a Heisenberg picture approach that \emph{does} improve the cost scaling with system size relative to simulation in the Schr\"{o}dinger picture. 
To improve the cost scaling, we must perform Heisenberg picture evolution not only of $B$ but also of $M$.  (The approach described here is valid when the OTOC is defined as in Eq.~\eqref{eq:OTOC_def_original}, with a trace and two copies of $M$.)  We illustrate this bi-directional Heisenberg picture approach in Fig.~\ref{fig:efficient_Heisenberg_OTOC}, in the case that $M$ and $B$ are chosen to lie on each other's physical lightcone edges in order to maximize instance-to-instance fluctuations in the OTOC.  

We evolve $|M\rangle\!\rangle$ (the doubled Hilbert space vector representing the operator $M$) under the portion of $(U^\ast\! \otimes U)$ that lies in the geometric lightcone of $M$ but outside of its dressed physical lightcone, \textit{i.e.} the gates in the region shaded purple in Fig.~\ref{fig:efficient_Heisenberg_OTOC}, to get $|\tilde{M}\rangle\!\rangle$.  We likewise evolve $|B\rangle\!\rangle$ under the portion of $(U\otimes U^\ast\!)$ that lies in the geometric lightcone of $B$ but outside of its dressed physical lightcone, namely the gates in the region shaded red in the figure, to get $|\tilde{B}\rangle\!\rangle$.  Both of these evolutions are expected to be efficient when simulated with a PEPS: because the incorporated gates are outside of the dressed physical lightcones, they almost cancel between $U$ and $U^\dagger$ and therefore do not substantially increase the bond dimension of the PEPS.  

The intersection of the two dressed physical lightcones, shaded yellow in Fig.~\ref{fig:efficient_Heisenberg_OTOC}, we expect to be nearly incompressible.  We refer to this region as the ``incompressible core'' of the tensor network giving the OTOC, and the operator defined by the gates in the incompressible core we call $C$.  All gates in the incompressible core are highly entangling and do not cancel even approximately between $U$ and $U^\dagger$.  Therefore an accurate PEPS simulation of the evolution will require a bond dimension exponentially large in $\ell^{(P)}$, the largest length crossing the incompressible core in the time direction (shown in Fig.~\ref{fig:geometric_compressibility}(a)).  

With these definitions, the OTOC can be found as
\begin{equation}
    \text{OTOC} = \text{Tr}\left[\tilde{M} C \tilde{B} C^\dagger \tilde{M} C \tilde{B} C^\dagger \right]/2^N, \label{eq:OTOC_Heis_efficient}
\end{equation}
which can be computed by an exact tensor network contraction.  The operators $\tilde{M}$ and $\tilde{B}$ are PEPS that have low bond dimension since they only incorporate gates outside the dressed physical lightcones.  OTOC$^{(2)}$ can be found similarly, as
\begin{equation}
    \text{OTOC}^{(2)} = \text{Tr}\left[\left(\tilde{M} C \tilde{B} C^\dagger\right)^4  \right]/2^N.\label{eq:OTOC2_Heis_efficient}
\end{equation}

In 1D, and specifically for the case that $M$ and $B$ lie on each other's physical lightcone edges as shown in Fig.~\ref{fig:efficient_Heisenberg_OTOC}, the remaining operators $C$ have a tensor network width that scales as the square root of the circuit depth.  Specifically, the time-direction width of the irreducible core, $\ell^{(P)}$ (shown in Fig.~\ref{fig:geometric_compressibility}(a)), is given by $\ell^{(P)} \approx (a/2v)^2\left[\sqrt{1+2T(a/2v)^{-2}} - 1\right]$ where $v$ is the speed defining the edge of the physical lightcone and the $a$ gives the width of the diffusive front, $W(t)=a\sqrt{t}$, as shown in Fig.~\ref{fig:OTOC_vs_B_loc}.  Thus when $T$ is large, $\ell^{(P)} \sim \sqrt{T}$.  The diagonal length $\ell^{(P)}_D$ (again shown in Fig.~\ref{fig:geometric_compressibility}(a)) also scales as $\sqrt{T}$.

Thus in 1D the width of the tensor network to be contracted according to Eq.~\eqref{eq:OTOC_Heis_efficient} scales as $\sqrt{T}$, coming almost entirely from the four copies of $C$.  The total contraction cost is exponential in the width, giving a final computation cost that scales as $\exp(\sqrt{T})$.  The total number of qubits scales as $N \sim T$, so the cost of computing the OTOC scales as $\exp(\sqrt{N})$.

In 2D this approach also leads to a reduced computational complexity.  However, the geometry of the physical lightcones and their broadened fronts with width $W(t)\sim t^{1/3}$ is more complicated, and we have not worked out the precise scaling.

We conclude that, both in 1D and in 2D, a careful Heisenberg picture approach as described here will achieve an improved asymptotic scaling relative to the Schr\"{o}dinger picture approach. 
A practical demonstration of this bi-directional Heisenberg approach is outside the scope of the current paper, but it would be an interesting target for future work.  While we expect this method to have a lower cost for very large systems compared with Schr\"{o}dinger picture approaches, simulations of the 65-qubit OTOC$^{(2)}$ experiment are likely still out of reach.  Furthermore, a system of 65 qubits may be small enough that the improved asymptotic scaling in the bi-directional Heisenberg approach does not yet overcome the larger prefactors incurred in this method, so that this method could be even more expensive than Schr\"{o}dinger picture simulations.


\section{Intermediate BP truncations vs final BP truncation}
\label{section:final_truncation}

\begin{figure}[ht]
        \centering
        \includegraphics[width=\linewidth]{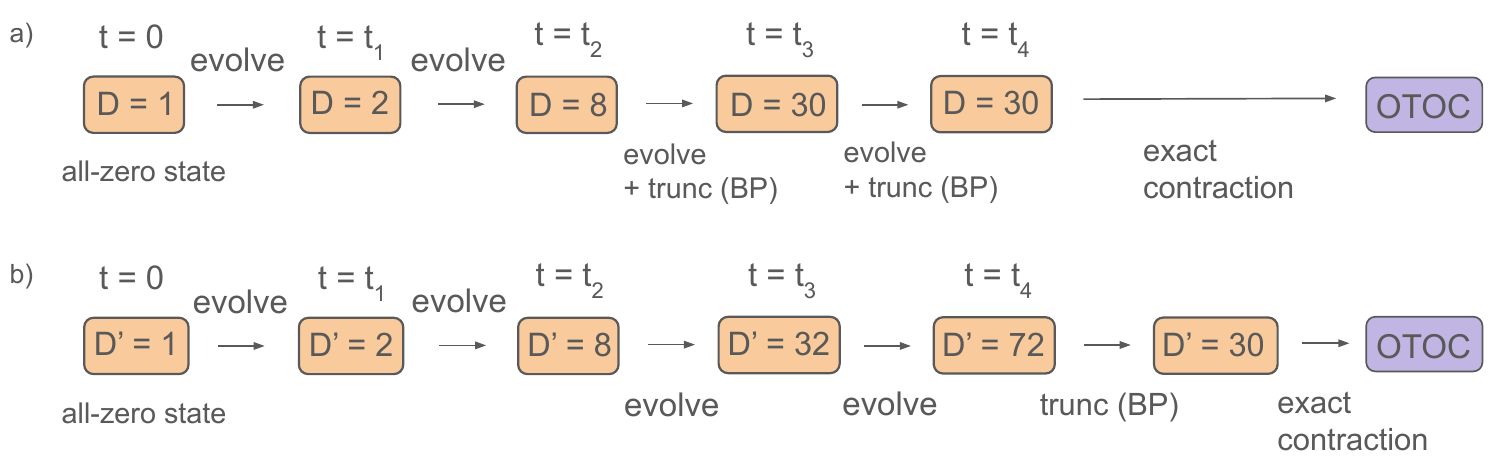}
       \caption{\textbf{Intermediate vs.\ final BP truncations.} Panel (a) shows the time-evolution procedure used in the main text. Here, the wavefunction is truncated to bond dimension $D$ throughout the time evolution; $D$ must be small enough (around 30) that the extraction step to compute the OTOC is computationally feasible. The SNR computed with this scheme is shown in \textit{blue} in Fig.~\ref{fig:snr_truncated}. Panel (b) shows the alternative evolution procedure, in which no truncation is performed during the time evolution beyond a singular-value cutoff of $10^{-14}$, leading to a potentially very large bond dimension $D'$.  To make extraction of the OTOC feasible, BP is applied once at the end of the time evolution to reduce the maximum bond dimension in the PEPS from $D'$ to $D$. The SNR computed with this scheme is shown in \textit{red} in Fig.~\ref{fig:snr_truncated}.}

        \label{flow_chart_truncations}
\end{figure} 

\begin{figure*}[ht]
    \centering
    \begin{minipage}{0.48\linewidth}
        \centering
        \includegraphics[width=\linewidth]{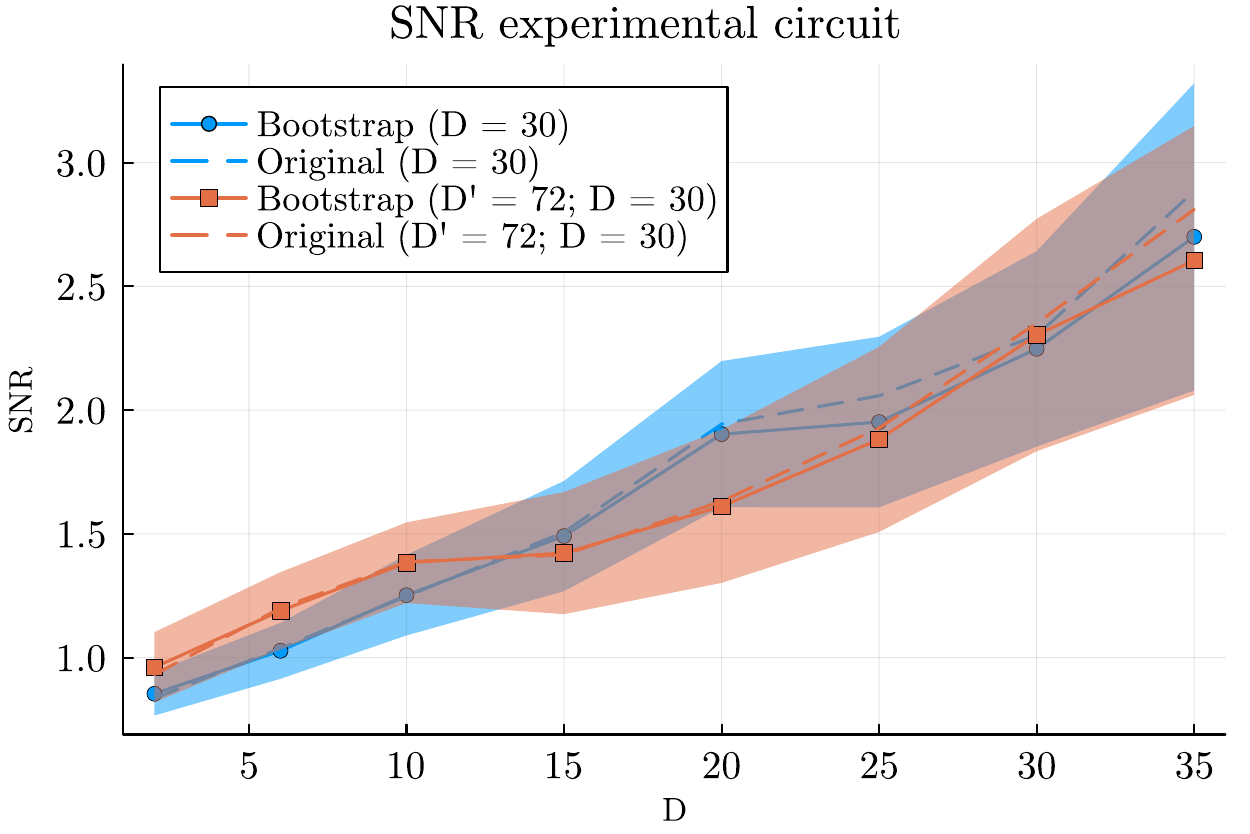}
    \end{minipage}\hfill
    \begin{minipage}{0.48\linewidth}
        \centering
        \includegraphics[width=\linewidth]{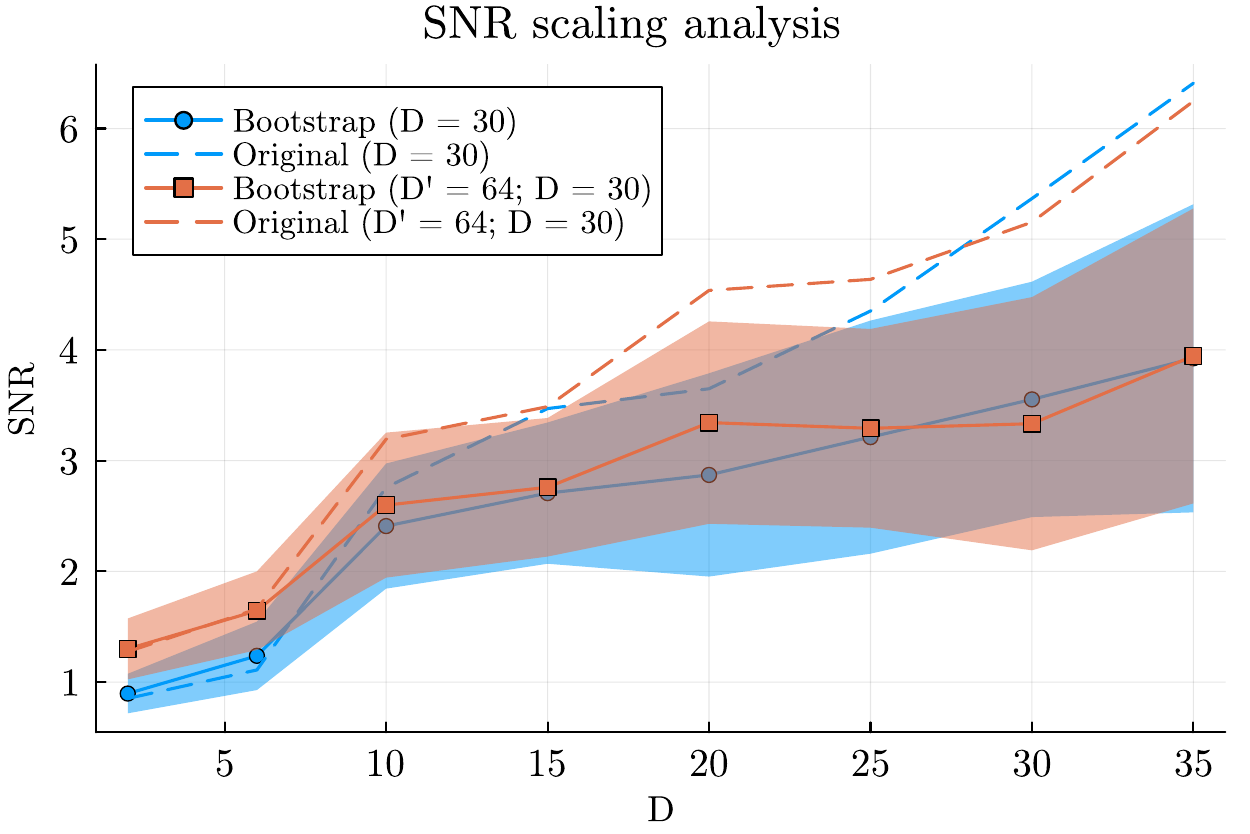}
    \end{minipage}
    \caption{\textbf{SNR}. Left panel: experimental 23-qubit quantum circuit from \cite{google2025observation} ($n_{2q}=108$). Right panel: new 23-qubit quantum circuit studied in Sec.~\ref{sec:scaling_experiment} ($n_{2q}=94$). The blue curve shows the SNR when the wavefunction is truncated to $D = 30$ during the time evolution. The red curve shows the SNR when the wavefunction is evolved without significant truncation, reaching a large bond dimension $D'$, which is 72 in the experimental circuit and 64 in the scaling analysis circuit. In both cases, the wavefunction is truncated to $D = 30$ once the time evolution is over. In all cases, we compute the SNR from OTOC values computed via exact contraction of the final PEPS with bond dimension $D$. The dashed line shows the SNR for 50 instances, while the solid line shows the mean (and shaded region the standard deviation) of a bootstrap process with 100 batches of size $s=30$, where the samples are taken with replacement.}
    \label{fig:snr_truncated}
\end{figure*} 

In the main text, we showed that when BP is used to perform truncation during time evolution, a large bond dimension $D$ is needed to obtain a final PEPS that faithfully represents the state $|\phi\rangle$.  When $D$ is too small, important information about correlations is lost during the evolution, and then even an exact contraction of the final PEPS cannot recover an accurate value for the OTOC.

However, these data alone do not disambiguate between several possible explanations for needing a large bond dimension, $D$:
\begin{enumerate}
    \item No PEPS wavefunction can accurately represent the state $|\phi\rangle$ without a large $D$.
    \item There exists a PEPS with a smaller $D$ that could faithfully represent the state, but performing even a single BP truncation, which does not accurately take into account correlations around loops in the lattice, makes the bond dimension $D$ insufficient.  
    \item The relevant information in the state is only lost because we perform many BP truncations throughout the evolution.  With only a single large BP truncation after exact evolution, the final state would still contain enough information to extract the OTOC.
\end{enumerate}
If 3.\ were the case, the OTOC would be easy to simulate classically.  We can afford to run time evolution with a large PEPS bond dimension $D'$, on the order of hundreds.  Then if a single truncation to much lower $D$, in order to allow for an affordable bMPS contraction, would not lose too much information, we could compute the OTOC accurately at quite moderate cost even for large systems.

In this Appendix, we show that 3.\ is not the case.  We show empirically that performing even a single significant BP truncation on a numerically exact PEPS is enough to reduce the information about the OTOC to the same level as if we had truncated to the final $D$ at every step (as we do in the simulations reported in the main text). This result is in contrast with the analysis performed in \cite{tindall2025dynamics}, where final truncations were shown to have a better performance than intermediate truncations to capture the value of two-point correlators. Note that we do not distinguish between case 1.\ and case 2.\ here, though our general arguments throughout the paper on cost scaling and incompressibility suggest that case 1.\ is correct.

The numerical experiment we consider here is summarized in Fig.~\ref{flow_chart_truncations}.  Panel (a) illustrates the procedure used in the main simulations in the paper: with every gate applied in the OTOC circuit, we use BP to immediately truncate back down to some fixed PEPS bond dimension, $D$.  Panel (b) illustrates the alternative procedure: during the time evolution, we impose no restriction on the final bond dimension\footnote{We do still truncate singular values smaller than \(10^{-14}\), which are likely only different from 0 due to limitations of numerical precision.}, but we truncate the final state to bond dimension $D$ using BP.  
In both cases, we extract the value of the OTOC using an exact ($\chi=\infty$) contraction, to isolate the effect of the truncation of the state as the only approximation.

We perform this numerical experiment both for the 23-qubit circuit with $n_{2q}=108$ from \cite{google2025observation}, studied in Sec.~\ref{sec:comparison with experiment} above, and for the 23-qubit circuit from our analysis of cost for a range of small system sizes, in Sec.~\ref{sec:scaling_experiment}.  When we do not perform truncation during the evolution, the final states have bond dimensions $D'=72$ and $D'=64$, respectively.  These states are then truncated to a variety of different final PEPS bond dimensions $D$, in the range of 5 to 35, from which the OTOC is extracted by exact contraction.  We compare the resulting SNR with the values obtained in the original simulation, with truncation to $D$ after applying each gate.  

The results are shown in Fig.~\ref{fig:snr_truncated}.  Data from the original simulations, with truncation throughout the evolution, are shown in blue; data from the procedure with a single large BP truncation are shown in red.  We see that, as claimed, both procedures lead to comparable SNR as a function of the final PEPS bond dimension $D$.  

We also confirm that instance-to-instance fluctuations in the accuracy of the OTOC are comparable between the two approaches.  The dashed lines in the figure show SNR computed using 50 random instances.  We use a bootstrap resampling procedure to see the statistical error in the SNR.  We take $p=100$ different batches of $s=30$ instances with replacement, compute the SNR for each, and then plot the mean (solid line) and standard deviation (shaded region).  For the circuit from the scaling analysis (right panel), the SNR with 30 instances is much lower than with 50,\footnote{For \(s = 30\), only about 54.5\% of the samples are expected to be independent. This suggests that the SNR of the full dataset may be affected by outlier instances with unusually high performance. This finding is consistent with the preparation of shallow circuits following the procedure in App.~\ref{section:obtain_lightcone}, where no monotonic increase in complexity is guaranteed and high variability across instances is expected even within the same system size.} but the red and blue data are still on top of each other.  Thus the comparable accuracy between truncating repeatedly and truncating once holds even on an instance-by-instance basis.

To summarize, these results show that, even when a high bond dimension is allowed during time evolution, truncations performed via belief propagation erase relevant features needed to compute the OTOC accurately. Therefore, regardless of when the wavefunction is truncated using belief propagation, important information is lost. The numerical experiments reported in this appendix are not sufficient to determine whether more computationally demanding truncation schemes could allow the PEPS to be reduced to a lower $D$ without losing information (case 2) or whether the loss of information when truncating to $D$ is a fundamental feature of the state $|\phi\rangle$ and the PEPS ansatz (case 1).  However, in the main text we argue that the circuits are incompressible, which is case 1.


\section{Empirical cost scaling for small 2D systems}
\label{app:empirical_scaling_2D}

\begin{figure*}[ht]
    \centering
    \includegraphics[width=1\linewidth]{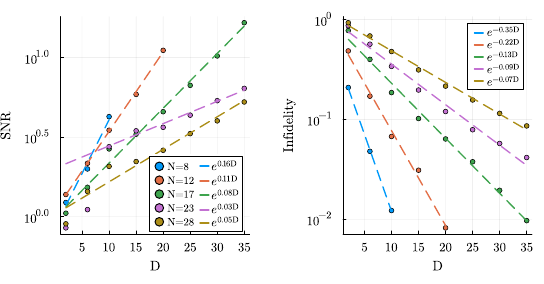}
    \caption{\textbf{Scaling of SNR and global wavefunction fidelity with bond dimension.} Left panel: SNR as a function of PEPS bond dimension $D$ for a variety of systems sizes.  OTOC values are computed by exact contraction of the final PEPS, so finite $D$ is the only approximation.  For each system size, SNR scales approximately exponentially in $D$.  Center panel: Infidelity vs SNR for different system sizes and bond dimensions (different points in the scatterplot) are strongly correlated.  Right panel: global wavefunction infidelity vs $D$ decays exponentially for each system size, with an exponent that decreases with increasing $N$.}
    \label{fig:infidelity_snr}
\end{figure*}

\begin{figure}[ht]
    \centering
    \includegraphics[width=1\linewidth]{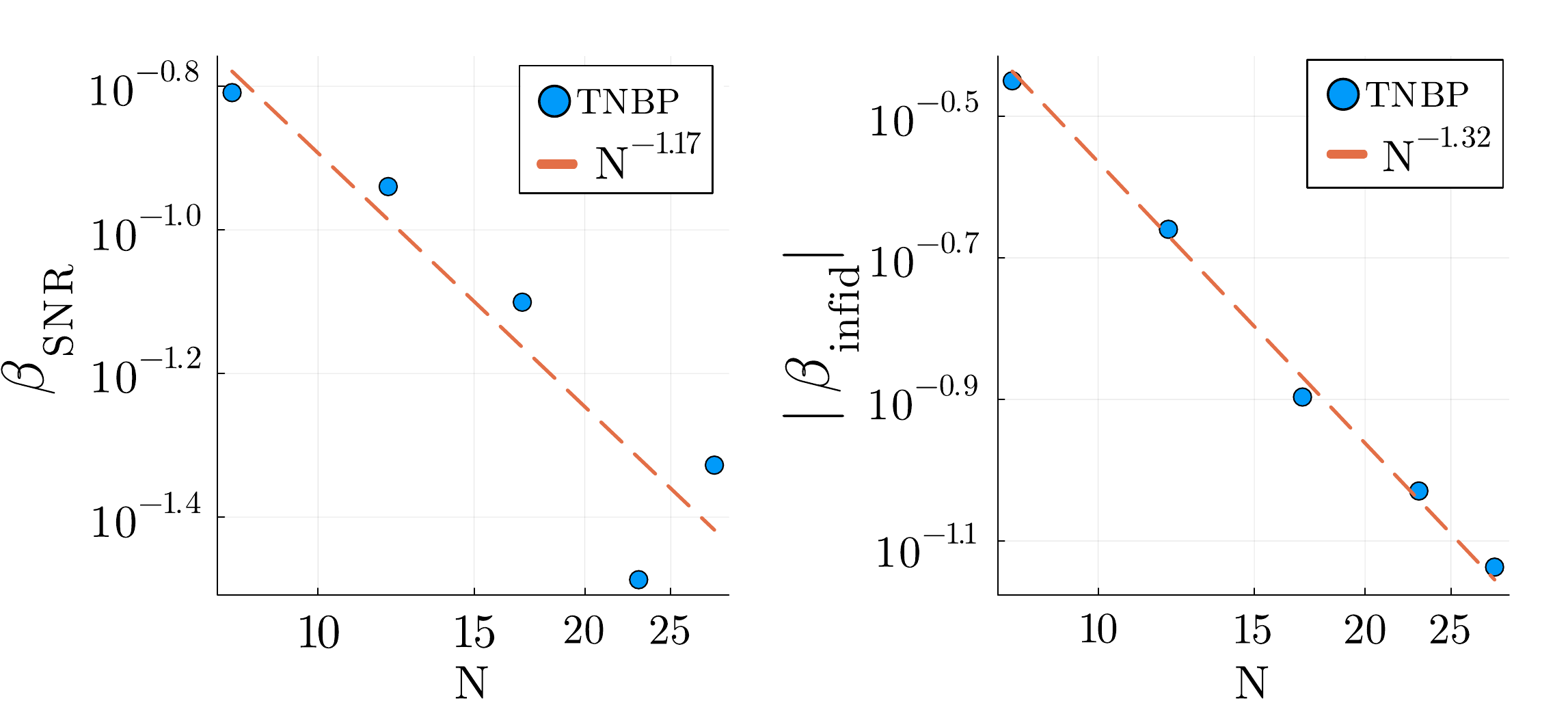}
    \caption{\textbf{Dependence on $N$ of SNR and infidelity exponents.} As shown in Fig.~\ref{fig:infidelity_snr}, SNR grows and infidelity decays exponentially in $D$ for each system size.  Here we show that the SNR growth rate $\beta_\text{SNR}$ (right) and infidelity decay (left) rate $\beta_\text{infid}$ are each approximately given by a power of $N$.  In each case, the rate approximately collapses to a straight line on a log-log plot. }
    \label{fig:exponents_scaling}
\end{figure}

\begin{figure}[ht]
    \centering
    \includegraphics[width=0.8\linewidth]{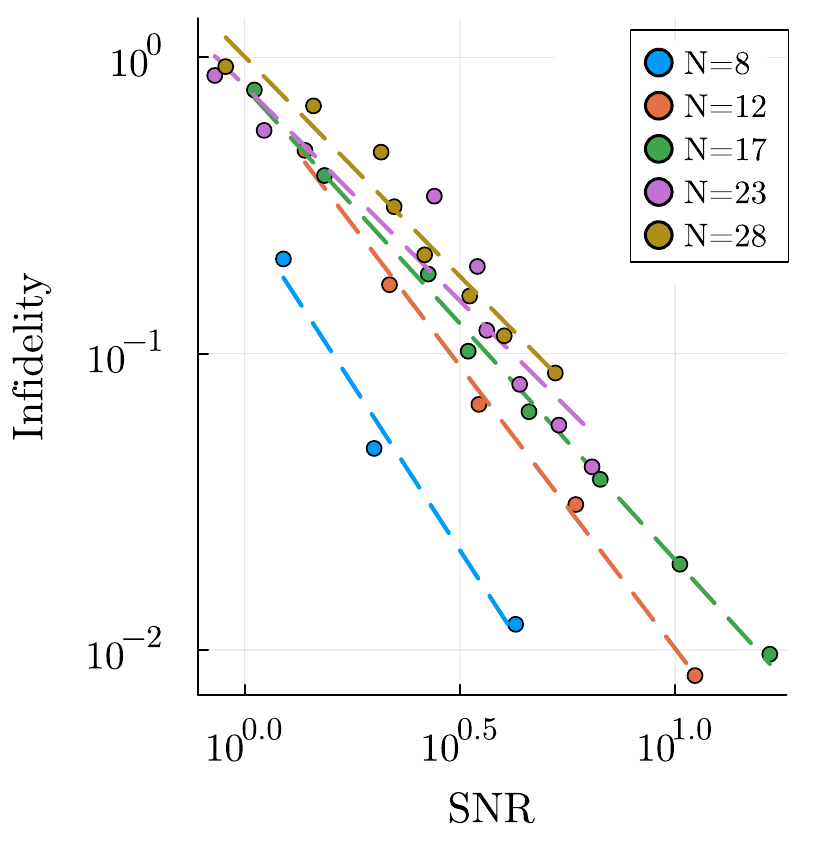}
    \caption{\textbf{Global wavefunction fidelity vs SNR} For OTOC ensembles with a range of system sizes from $N=8$ to $N=28$ qubits, we compute the SNR using a range of PEPS bond dimension cutoffs, $D$, and exact contraction ($\chi=\infty$).  For each $D$, we also compute the fidelity of the final PEPS with the state $|\phi\rangle$ computed by an exact state vector evolution.  We plot infidelity vs SNR, with each point in the figure corresponding to one $(N,D)$ pair.  We observe a high degree of correlation across two orders of magnitude in wavefunction fidelity.}
    \label{fig:infidelity_vs_snr}
\end{figure}

From the TNBP simulations of small 2D OTOC ensembles ranging in size from $N=8$ to $N=28$ qubits, Sec.~\ref{sec:scaling_experiment},
we drew the qualitative conclusion that an accurate TNBP computation of random-circuit OTOCs requires both the PEPS and bMPS bond dimensions, $D$ and $\chi$, to be large and to scale with system size.  Here we ask a more quantitative question: for an OTOC circuit with a given number of qubits, $N$, what PEPS bond dimension $D$ is needed in order to achieve a large SNR with TNBP?  We empirically find that the bond dimension scales only as $D\sim N^{1.3}$, much weaker than the exponential scaling predicted in Sec.~\ref{subsec:IIB_scaling_theory}.  However, we argue that the empirically observed scaling here is dominated by finite size effects, and the predicted exponential scaling would appear for larger systems.

\paragraph{Scaling of $D$ with $N$ based on measured SNR:} 
We first ask: in order to achieve a particular SNR (for example, $\text{SNR}=3.5$ as found for the experiment on $N=95$ qubits in~\cite{google2025observation}), as we increase the system size, how does the required PEPS bond dimension $D$ scale?  To answer this question, in Fig.~\ref{fig:infidelity_snr} (left panel), we plot the SNR as a function of $D$ for five system sizes, using the data from Sec.~\ref{sec:scaling_experiment}.  We want to address only the approximation during time evolution, so we evaluate SNR using an exact contraction, $\chi=\infty$.  We find that for each system size $N$, SNR is roughly exponential in $1/D$, $\text{SNR}\approx\exp(\beta_\text{SNR} D)$.  The exponential rate $\beta$ depends on $N$, and we estimate the dependence as shown in the left panel of Fig.~\ref{fig:exponents_scaling}, finding $\beta_\text{SNR}\approx 1.2\, N^{-1.17}$.  Inverting the equation for SNR as a function of $D$ gives $D\sim N^{1.17} \times \log(\text{SNR})/1.2$, so that to keep a fixed SNR as we scale up the system size, we need to increase the bond dimension $D$ proportionally with $N^{1.17}$.  

However, non-monotonicity and other irregularities in the observed SNR vs $D$ relation at each size $N$ make this scaling unreliable, even for the small system sizes we study here where exponential scaling is not yet expected.  One solution is to estimate the required scaling of $D$ with $N$ using a different measure of simulation quality that is less sensitive to small details of the circuit such as choice of location for the $M$ and $B$ operators.

\paragraph{Correlation between SNR and global fidelity:} 

One possible metric to use to derive the scaling is global wavefunction fidelity.  For our TNBP simulations, global fidelity is indeed a good proxy for the SNR.  

It is not immediately clear that these quantities should be correlated.  Specifically, in Sec.~\ref{subsec:scaling_with_compressibility} we argued that gates outside the dressed physical lightcone of $M$ could be compressed without significantly changing the value of the OTOC for any instance in an ensemble.  However, such compressions \emph{do} change the global wavefunction.  Thus it is possible to achieve high SNR even when global fidelity is low.

However, in the specific approach to TNBP simulation that we use in our numerical simulations, we observe that SNR and global fidelity are strongly correlated: an accurate value for the local observable $\langle M\rangle$ requires the full state $|\phi\rangle$ to be accurate.  In our approach, after applying each gate, we truncate back to a fixed PEPS bond dimension $D$ that is uniform throughout the system.  We therefore do not exploit the compressibility discussed in Sec.~\ref{subsec:scaling_with_compressibility}.  This creates a correlation between the SNR and global fidelity because any degree of truncation that leads to discarding information outside the physical lightcone of $M$ will also discard information inside the lightcone and thus reduce the SNR.

To compute fidelity, we first perform an exact ($\chi=\infty$) contraction of the final PEPS after time evolution, then we compare the resulting state vector with the result of a full state vector simulation of the circuit.  Observed infidelity comes entirely from the BP truncations to bond dimension $D$ during the evolution step, with no approximation during the extraction step. 
We plot infidelity ($1-\text{fidelity}$) vs SNR in Fig.~\ref{fig:infidelity_vs_snr}.  The log-log plot shows a strong correlation across an order of magnitude range in SNR.

\paragraph{Scaling of $D$ with $N$ based on global fidelity:} 
Having now established that the SNR and global wavefunction fidelity are highly correlated in our simulations across a range of system sizes and PEPS bond dimensions, we can find how $D$ needs to scale with $N$ to achieve a fixed fidelity, as a proxy for the scaling of $D$ with $N$ for fixed SNR.  

The observed scaling of infidelity ($1-\text{fidelity}$) vs $D$ for different system sizes is shown in Fig.~\ref{fig:infidelity_snr} (right panel).  The infidelity $I$ decays exponentially in $D$, $I\sim \text{exp}(\beta_\text{infid} D)$; using the fit shown in the right panel of Fig.~\ref{fig:exponents_scaling}, we estimate $\beta_\text{infid}\approx -5.6\, N^{-1.32}$.  As seen in the figures, the exponential decay is a good fit to the data for every system size $N\leq 28$ and the decay rate is given by a clean power law in $N$.  

According to this empirical scaling investigation, the required bond dimension for fixed fidelity (or equivalently, fixed SNR) will scale as $D\sim N^{1.3}$.  

\paragraph{Observed scaling is not exponential because of finite size effects:} 
In light of the arguments in Sec.~\ref{subsec:IIB_scaling_theory} above, this result is surprising.  We argued that an accurate simulation of the OTOC in the limit of large system sizes $N$ should be exponentially expensive, with a Schr\"{o}dinger picture simulation of the OTOC for a 2D system requiring a bond dimension that scales exponentially in $\sqrt{N}$.  Instead, we have empirically found that the required bond dimension scales only as a small power of $N$.  

The disagreement can be explained by finite size effects.  The theoretical prediction of exponential scaling is based on the expectation that the number of gates on some bonds will grow proportionally with $\sqrt{N}$.  However, when we construct the circuits used here in the empirical scaling analysis following the procedure in Sec.~\ref{subsec:circuit_design} and App.~\ref{section:obtain_lightcone}, the circuits for $N=20$, 22, 23, and 28 all have at most 4 gates on any given bond.  Thus our systems are simply too small to check how the cost really scales for large $N$.  We believe that for large systems, to maintain a fixed SNR or global fidelity the bond dimension will have to grow as predicted in Sec.~\ref{subsec:IIB_scaling_theory}, $D\sim\exp(\sqrt{N})$.

\end{document}